\documentclass[a4paper, 12pt, sort&compress]{elsarticle}

\usepackage{amsmath}
\usepackage{amssymb}
\usepackage{subfig}
\usepackage{multirow}
\usepackage{algorithm}
\usepackage{algpseudocode}
\usepackage{setspace}
\usepackage{xcolor}
\usepackage{hyperref}
\usepackage{cleveref}
\usepackage{bm}
\usepackage{soul}
\usepackage{lineno}

\newcommand{\pderiv}[2]{\frac{\partial #1}{\partial #2}}
\newcommand{\bs}[1]{\boldsymbol{#1}}

\newcommand{\gammasD}{\gamma^s_{\text{D}}}
\newcommand{\gammasN}{\gamma^s_{\text{N}}}

\newcommand{\T}{\; \mathrm{T}}

\newcommand{\myequal}{\, {=} \,}

\RequirePackage[margin=20mm]{geometry}

\graphicspath{{./figures/}}

\begin{document}

\fontfamily{ptm}\selectfont

\begin{frontmatter}

\title{A comprehensive assessment of accuracy of adaptive integration of cut cells for laminar fluid-structure interaction problems}

\author[add1]{Chennakesava Kadapa\corref{cor1}}
\ead{c.kadapa@bolton.ac.uk}

\author[add2]{Xinyu Wang}
\author[add2]{Yue Mei\corref{cor1}}
\ead{meiyue@dlut.edu.cn}

\cortext[cor1]{Corresponding author}

\address[add1]{School of Engineering, University of Bolton, Bolton BL3 5AB, United Kingdom}
\address[add2]{Department of Engineering Mechanics, Dalian University of Technology, China}

\begin{abstract}

Finite element methods based on cut-cells are becoming increasingly popular because of their advantages over formulations based on body-fitted meshes for problems with moving interfaces. In such methods, the cells (or elements) which are cut by the interface between two different domains need to be integrated using special techniques in order to obtain optimal convergence rates and accurate fluxes across the interface. The adaptive integration technique in which the cells are recursively subdivided is one of the popular techniques for the numerical integration of cut-cells due to its advantages over tessellation, particularly for problems involving complex geometries in three dimensions. Although adaptive integration does not impose any limitations on the representation of the geometry of immersed solids as it requires only point location algorithms, it becomes computationally expensive for recovering optimal convergence rates. This paper presents a comprehensive assessment of the adaptive integration of cut-cells for applications in computational fluid dynamics and fluid-structure interaction. We assess the effect of the accuracy of integration of cut cells on convergence rates in velocity and pressure fields, and then on forces and displacements for fluid-structure interaction problems by studying several examples in two and three dimensions. By taking the computational cost and the accuracy of forces and displacements into account, we demonstrate that numerical results of acceptable accuracy for FSI problems involving laminar flows can be obtained with only fewer levels of refinement. In particular, we show that three levels of adaptive refinement are sufficient for obtaining force and displacement values of acceptable accuracy for laminar fluid-structure interaction problems.

\end{abstract}


\end{frontmatter}

\newpage
\section{Introduction}
Numerical methods for the solution of partial differential equations encountered in science and engineering can be broadly classified into two major groups: i) methods based on body-fitted meshes and ii) methods based on unfitted/immersed/embedded meshes. While the numerical schemes based on body-fitted meshes are well established and available as commercial and open-source software for the simulation of problems in science and engineering, those based on immersed or embedded meshes are relatively recent. Despite their popularity and commercial success, the fundamental difficulty with body-fitted methods(BFMs) is that they require the generation of body-fitted meshes, which can be cumbersome for complex geometries usually encountered in industrial practice. In addition, numerical schemes based on BFMs have limited applicability for the simulation of fluid-structure interaction (FSI) problems in which solids undergo extensive deformations. This limitation stems from the fact that FSI schemes based on body-fitted meshes require sophisticated mesh-updating and/or remeshing algorithms \cite{JohnsonCMAME1994,TezduyarACMM2001,SaksonoIJNME2007}. Due to these limitations, numerical methods based on immersed or embedded boundaries have become a viable alternative for computational FSI problems over the past couple of decades.

Among the embedded/unfitted methods, those based on \textit{cut cells} have received a considerable amount of attention during the past decade. eXtended Finite Element Method  \cite{BelytschkoMSMSE2009}, Partition of unity finite element method \cite{MelenkCMAME1996}, cut finite element method \cite{BurmanIJNME2014,DettmerCMAME2016,KadapaCMAME2017rigid,KadapaCMAME2018}, Finite Cell Method \cite{ParvizianCM2007}, are a few examples of such methods. The basic idea behind these cut-cell based methods is to enrich the finite element space in the vicinity of an interface between two physical domains. This is achieved by cutting the cell intersected by the interface (or immersed boundary) and using some sophisticated quadrature rules to integrate the active portion(s) of the cut-cell.

The major difficulty associated with cut-cell based numerical schemes is the integration of the portion of a cut cell that belongs to a particular domain. This is due to the fact that accurate resolution of discontinuities along the interfaces/immersed boundaries translates directly into accurate integration of the cut cells, for which several techniques have been proposed in the literature. Subtriangulation or tessellation has been and still is the prominent technique for integrating cut cells in the XFEM community. However, this methodology is limited to geometries modelled with lower-order representations, e.g., first-order triangles. Due to the increase in the use of immersed/embedded methods, especially using higher-order representations of geometry, recent years have seen a considerable increase in research work on developing alternative quadrature methods that are efficient and geometrically accurate for the integration of cut cells, see \cite{MousaviCM2011, SudhakarCMAME2013, VenturaIJNME2015, DuczekCM2015, StavrevCMAME2016} and references therein. A widely-used alternative technique for the integration of cut cells is adaptive integration which is based on the recursive subdivision of the cell of interest, see \cite{DettmerCMAME2016, KudelaAMSES2015, KudelaCMAME2016}. Each of these methods has its relative advantages and limitations.

The method of sub-triangulation, although it yields exact integration of cut cells when the boundary is approximated with linear segments, see \cite{BurmanIJNME2014, KadapaCMAME2017rigid}, becomes quite complicated in three-dimensions (3D) when immersed surfaces/interfaces are discretised with triangular elements, usually obtained from standard mesh generation software or STL files. Such scenarios require sophisticated constrained Delaunay tetrahedralisation algorithms, which are not robust enough to be used as a reliable option for complex geometries that can change their position and shape dynamically during an FSI simulation. Moreover, a linear approximation of geometries represented with implicit surfaces or higher-order discretisations introduces numerical errors between the actual and discretised geometries. 

Furthermore, the associated difficulties will be compounded when the immersed boundaries need to be represented with higher-order discretisations, for example, when using higher-order elements for the solid problem. With sub-triangulation, the particular algorithm to be used for quadrature depends upon the representation of the geometry of the immersed boundary. H\"ollig \cite{HolligBook}, and R\"uberg and Cirak \cite{RubergCMAME2012} use level set presentation for the geometry and modify the basis functions of cut cells using Lagrangian polynomials. Stavrev et al. \cite{StavrevCMAME2016} use trimmed NURBS surfaces as immersed geometries and reparameterise active cut cells using higher-order Lagrange polynomials. Kudela et al. \cite{KudelaCMAME2016} use recursive subdivision in combination with node-mapping to achieve accurate and efficient techniques for the integration of cut cells. However, these techniques require sophisticated algorithms for identifying topologies of cut cells and then calculating the mappings, which can be quite expensive for industrially relevant geometries that involve intricate shapes. Therefore, in order to circumvent these issues, adaptive integration is the preferred choice.

Integration of cut cells using adaptive integration relies on the recursive subdivision of a cut cell and applying the quadrature rule for each relevant cell at finer levels. So far in the literature, adaptive integration has been widely used in the context of the generalised finite element method \cite{StrouboulisIJNME2000, StrouboulisCMAME2001}; the Finite Cell Method (FCM) for solid mechanics \cite{DusterCMAME2008, SchillingerCMAME2011, DusterCM2012, DuczekCM2015, StavrevCMAME2016, VarduhnIJNME2016, ThiagarajanCMAME2016, DusterChapter2020, DiviCMA2020}, fluid flow \cite{XuCandF2016} and wave propagation \cite{DuczekIJNME2014}; FEM for level set functions \cite{MuellerIJNME2012}; and fluid-structure interaction \cite{DettmerCMAME2016,KadapaCMAME2018,XuCandF2021}. Because of the way the adaptive integration technique operates, it is expected that this technique requires a significant number of levels of refinement for accurately integrating the cut-cells. However, published literature on the exact number of levels of refinement required for achieving optimal convergence rates in velocity and pressure fields, the sensitivity of convergence of error norms for incompressible Navier-Stokes, and the associated computational cost is not existing. Furthermore, to the best of the authors' knowledge, literature on the effect of inaccuracies in cut-cell integration on the force and displacement values for FSI problems is also lacking. Given the increasing interest in the use of cut-cell based methods for FSI simulations \cite{KadapaCMAME2017rigid, KadapaCMAME2018,XuCandF2021,BoustaniJCP2021,ThariCandF2021}, it is important to assess the effect of inaccuracies in the integration of cut cells on the numerical results for FSI problems, and to the best of our knowledge, such a study is not yet available. Therefore, this paper aims to address this gap by performing a comprehensive assessment of the adaptive integration of cut cells on the numerical results of FSI problems involving laminar flows. First, we study the sensitivity of convergence rates of error norms using the example of Kovasznay flow. Later, we assess the effect of different levels of adaptive integration on FSI problems using fluid-flexible structure interaction problems in two and three dimensions.

The remaining part of this paper is organised as follows. The governing equations for the fluid-structure interaction problems and the respective finite element formulations are discussed briefly in Section \ref{section-fsi-formulation}. In Section \ref{section-cutcells}, integration cut cells using subtriangulation and adaptive integration techniques is discussed. The accuracy of integration of cut cells using adaptive integration and its effect on the numerical results of laminar FSI problems is evaluated in Section \ref{section-examples}. This paper is concluded by summarising the observations and drawing conclusions in Section \ref{section-conclusion}.

\section{Formulation of the fluid-structure interaction problem} \label{section-fsi-formulation}
In this work, the coupled FSI problem is solved using a staggered scheme proposed in Dettmer and Peri\'c \cite{DettmerIJNME2013} and Kadapa \cite{KadapaOE2020} in which the fluid and solid problems are solved separately, once per time step, using their respective solvers after obtaining the appropriate data from solid and fluid problems, respectively. The CutFEM framework based on hierarchical b-spline grids used in the present work is already published in Dettmer et al. \cite{DettmerCMAME2016} and Kadapa et al. \cite{KadapaCMAME2017rigid,KadapaCMAME2018,KadapaJFS2020,KadapaIJCM2021}. Therefore, only the key details are discussed in the present work, and the reader is referred to \cite{DettmerCMAME2016,KadapaCMAME2017rigid,KadapaCMAME2018,KadapaJFS2020,DettmerIJNME2013,KadapaOE2020,KadapaIJCM2021} for the comprehensive details of the formulations and methods used.

\begin{figure}[H]
\centering
\subfloat[]{\includegraphics[clip,scale=0.6]{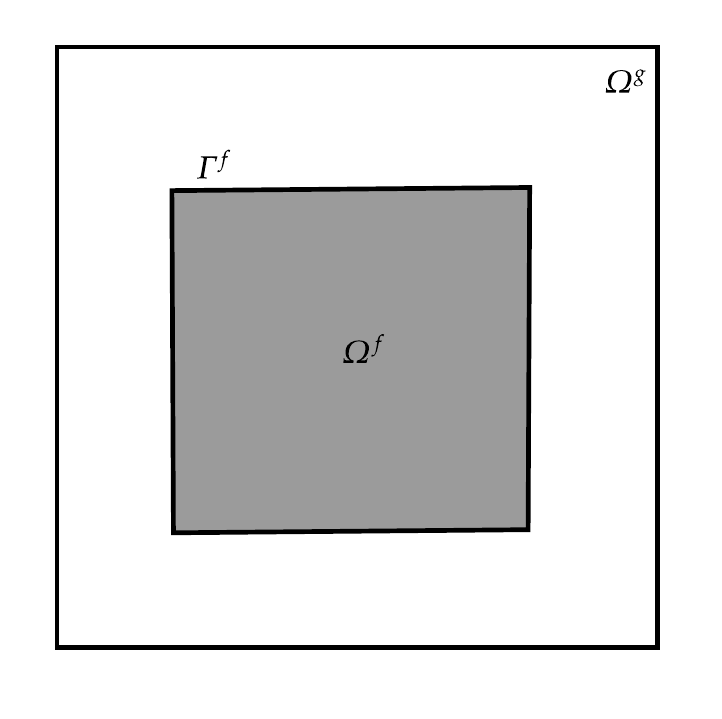} \label{fig-fsi-domains-1} }
\subfloat[]{\includegraphics[clip,scale=0.6]{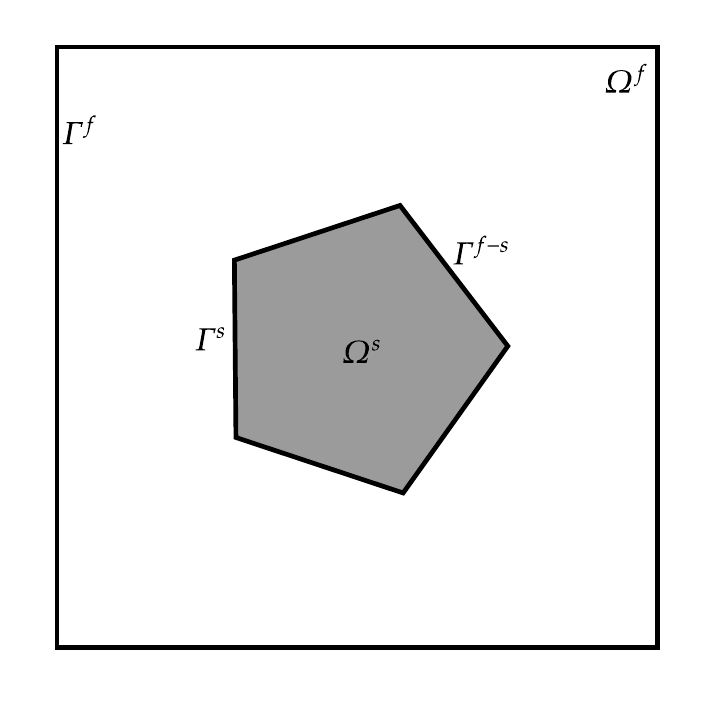} \label{fig-fsi-domains-2} }
\caption{Domains and boundaries in typical CFD and FSI problems.}
\label{fig-fsi-domains}
\end{figure}

Figure \ref{fig-fsi-domains} shows typical scenarios encountered in an embedded boundary framework for fluid flow and FSI problems. A typical fluid flow problem (see Fig. \ref{fig-fsi-domains-1}) consists of a fluid domain $\Omega^f$ with its boundary $\Gamma^f$ embedded in a Cartesian grid $\Omega^g$. A typical FSI problem (see Fig. \ref{fig-fsi-domains-2}) consists of a fluid domain, $\Omega^f$, and solid domain, $\Omega^s$, with their corresponding boundaries denoted as $\Gamma^f$ and $\Gamma^s$ respectively in the original configuration, as shown in Fig. \ref{fig-fsi-domains-1}. The deformed configurations of the fluid and domains are denoted, respectively, as $\omega^f$ and $\omega^s$, along with their corresponding boundaries as $\gamma^f$ and $\gamma^s$. The interface between the two domains is denoted as $\Gamma^{f-s}$ and $\gamma^{f-s}$, respectively, in the original and deformed configurations. For the configuration shown in Fig. \ref{fig-fsi-domains-2}, $\Gamma^s$ and $\Gamma^{f-s}$ are the same. The fluid problem is solved on a fixed Cartesian grid ($\Omega^g$); therefore, $\Gamma^f=\gamma^f$ throughout the simulation.

\subsection{The fluid problem}
For laminar, viscous and incompressible fluid flow problems, the governing equations are the incompressible Navier-Stokes equations, given as

\begin{subequations} \label{eqns-NS-main}
\begin{align}
\rho^f \pderiv{\bm{v}^f}{t} + \rho^f (\bm{v}^f \cdot \bm{\nabla}) \bm{v}^f - \nabla \cdot \bs{\sigma}^f &= \bm{f}^f && \mathrm{in} \quad \Omega^f,  \\
\bm{\nabla} \cdot \bm{v}^f &= 0 && \mathrm{in} \quad \Omega^f, \\
\bm{v}^f &= \bar{\bm{v}}^f && \mathrm{on} \quad \Gamma^f_D, \\
\bm{\sigma}^f \cdot \bm{n}^f &= \bar{\bm{t}}^f &&   \mathrm{in} \quad \Gamma^f_N, \\
\bm{v}^f(t \myequal 0) &= \bm{v}^f_0  && \mathrm{in} \quad \Omega^f, \\
p^f(t \myequal 0) &= p^f_0  && \mathrm{in} \quad \Omega^f,
\end{align}
\end{subequations}
where, $\rho^f$  is the density of the fluid, $\bm{v}^f$ is the velocity of the fluid, $p^f$ is the pressure field in the fluid domain, $t$ is the time variable, $\bm{\nabla}$ is the gradient operator, $\bm{f}^f$ is the body force on the fluid domain, $\bm{\sigma}^f (=\mu^f \bm{\nabla} \bm{v}^f - p^f \, \bm{I})$ is the stress tensor, $\mu ^f$ is the viscosity of the fluid, $\bm{I}$ is the second-order identity tensor, $\bm{n}^f$ is the unit outward normal on the boundary $\Gamma^f$, $\bm{v}^f_0$ is the initial velocity, $p^f_0$ is the initial pressure, $\bar{\bm{v}}^f$ is the prescribed velocity on Dirichlet boundary $\Gamma^f_D$, and $\bar{\bm{t}}^f$ is applied traction on the Neumann boundary $\Gamma^f_N$.

The fluid problem is solved for velocity and pressure fields using the stabilised finite element formulation on hierarchical b-spline grids. The interface constraint is enforced weakly using Nitsche's method, and the ghost-penalty operators are used for circumventing matrix ill-conditioning due to small cut cells. The reader is referred to Dettmer et al. \cite{DettmerCMAME2016} and Kadapa et al. \cite{KadapaCMAME2017rigid,KadapaCMAME2018,KadapaJFS2020} for the comprehensive details of the formulations used in the proposed work. Integration of cut cells is discussed in Section \ref{section-cutcells}.

\subsection{The solid problem}

The flexible solid is modelled using the finite element formulation in the finite strain regime to accurately capture large structural deformations. The governing equations for the elastodynamics of flexible solids in the current configuration can be written as,

\begin{subequations} \label{gov-eqns-solid}
\begin{align}
\rho^s \pderiv{^2 \bm{d}^s}{t^2} - \nabla_{\bm{x}} \cdot \bs{\sigma}^s &= \bm{f}^s \quad\quad \mathrm{in} \quad \omega^s \\
\bm{d}^s &= \bar{\bm{d}}^s \quad\quad \mathrm{on} \quad \gammasD \\
\bm{\sigma}^s \cdot \bm{n}^s &= \bar{\bm{t}}^s \,\quad\quad   \mathrm{on} \quad \gammasN\;,
\end{align}
\end{subequations}
where, $\rho^s$ is the density of solid in the current configuration, $\bm{d}^s$ is the displacement of the solid, $\nabla_{\bm{x}}$ is the gradient operator with respect to the current configuration, $\bm{f}^s$ is the body force, $\bm{n}^s$ is the unit outward normal on the boundary $\gamma^s$, $\bar{\bm{d}}^s$ is the specified displacement on the boundary $\gammasD$, and $\bar{\bm{t}}^s$ is specified traction on the boundary $\gammasN$. The Cauchy stress tensor, $\bs{\sigma}^s$, depends upon the particular constitutive material model employed for the solid. For a compressible hyperelastic material, it is given given by

\begin{align}
\bs{\sigma}^s = \frac{1}{J} \, \pderiv{\Psi}{\bm{F}} \, \bm{F}^{\T},
\end{align}
where $\bm{F}=\bm{I}+\pderiv{\bm{d}^s}{\bm{X}}$ is the deformation gradient and $J=\det(\bm{F})$. In the numerical examples considered in this work, Saint Venant-Kirchhoff and compressible Neo-Hookean models are used. The associated initial boundary value problem for the elastodynamics is solved using the first-order quadrilateral/hexahedral elements while the generalised-alpha scheme \cite{KadapaCS2017} is used for integration in the time domain.

\subsection{Interface conditions}
The kinematic constraint and the traction equilibrium at the fluid-solid interface $\gamma^{f-s}$ are given by,
\begin{align}
\bm{v}^f &= \bm{v}^s, \\
\bm{\sigma}^f \cdot \bm{n}^f + \bm{\sigma}^s \cdot \bm{n}^s &= \mathbf{0}.
\end{align}

The coupling between the fluid and solid domains is resolved using a staggered scheme \cite{DettmerIJNME2013,KadapaOE2020}. The pseudocode for the staggered scheme is shown in Algorithm \ref{algo-stag-scheme}. The parameter $\beta$ is a relaxation parameter set by the user. The staggered scheme is first-order accurate for $\mathbf{F}^{s^P}_{n+1} = \mathbf{F}_n$, and second-order accurate for $\mathbf{F}^{s^P}_{n+1} = 2 \, \mathbf{F}_n - \mathbf{F}_{n-1}$. For the example of 3D plates in cross-flow, which involves a significant amount of added-mass, the first-order version is used. The reader is referred to Dettmer and Peri\'c \cite{DettmerIJNME2013}, Kadapa \cite{KadapaOE2020, KadapaArxiv2021} and Dettmer et al. \cite{DettmerIJNME2021} for the comprehensive details of the staggered schemes.

\begin{algorithm}
  \caption{Staggered scheme used in the present work}
  \label{algo-stag-scheme}
   \begin{algorithmic}[1]
    \State Predict force on the solid: $\mathbf{F}^{s^P}_{n+1} = \mathbf{F}_n$ or $\mathbf{F}^{s^P}_{n+1} = 2 \, \mathbf{F}_n - \mathbf{F}_{n-1}$
    \State Solve the solid problem force force $\mathbf{F}^{s^P}_{n+1}$
    \State Reposition immersed solid(s) and update the fluid mesh
    \State Solve the fluid problem to obtain force $\mathbf{F}^f_{n+1}$
    \State Average the force: $\mathbf{F}_{n+1} = - \beta \, \mathbf{F}^f_{n+1} + (1-\beta) \, \mathbf{F}^{s^P}_{n+1}$
    \State proceed to next time step
  \end{algorithmic}
\end{algorithm}

\section{Integration of cut-cells} \label{section-cutcells}
\begin{figure}[H]
\centering
\subfloat[]{\includegraphics[clip,scale=0.45]{cutfem-pentagon-domains.pdf}}
\subfloat[]{\includegraphics[clip,scale=0.45]{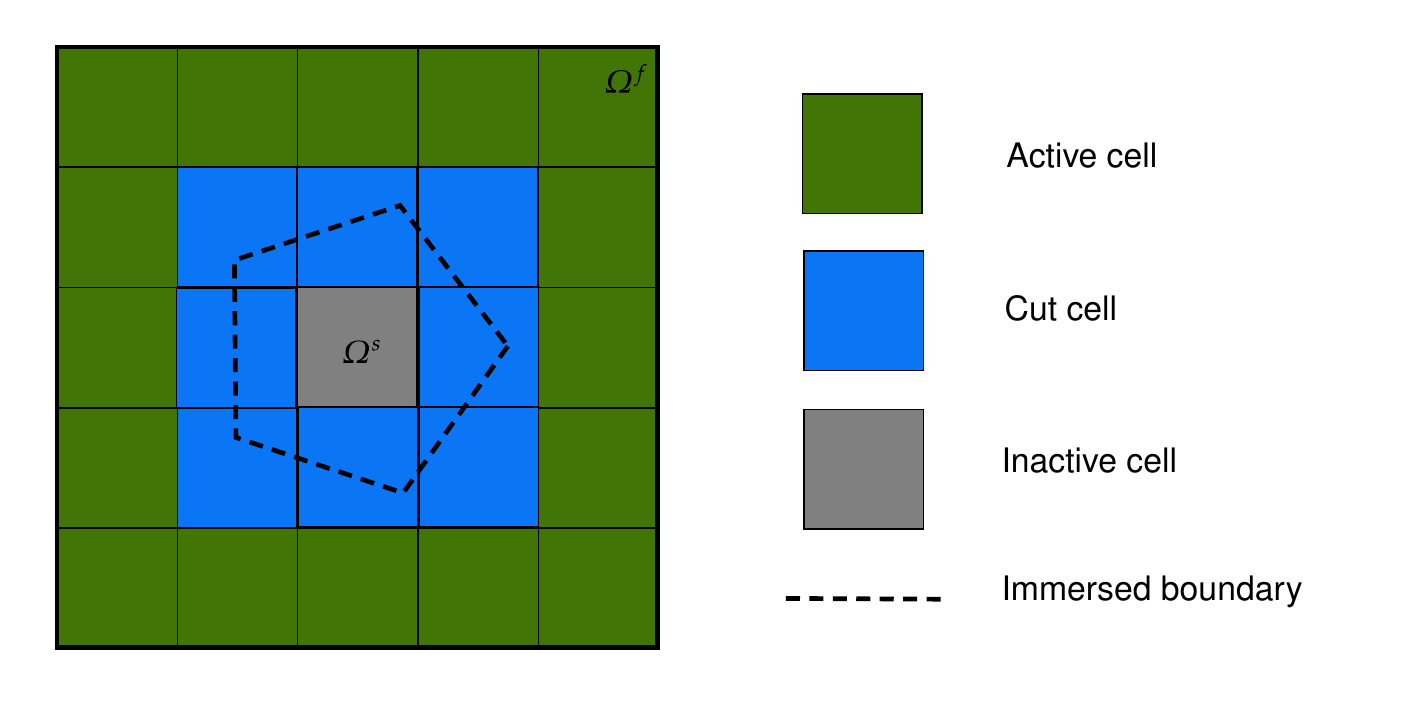}}
\caption{Cut discretisation: a.) geometry consisting of fluid and solid domains, and b.) discretisation with $5\times5$ elements along with the color coded for different elements.}
\label{cutfem-basic}
\end{figure}

The fundamental motivation behind numerical schemes for partial differential equations using cut-cell based methods is to minimise the cumbersome process of mesh generation for problems involving complex geometries, particularly in 3D. In these methods, numerical solutions are sought typically over a Cartesian grid that does not conform to the boundaries or interfaces, as illustrated in Fig. \ref{cutfem-basic} for a simple scenario consisting of a fluid domain and a solid domain. Some of the cells of the background grid are \textit{intersected/cut} by the interface between the two domains. Since the focus of the present work is fluid flow and FSI problems, without the loss of generality, it is assumed that numerical solutions on the Cartesian grids are sought in the fluid domain only. Moreover, in the finite element methodology used for the fluid problem, volume (or domain) integrals are evaluated as the summation of integrals over individual cells.

The cells of the background grid that lie completely inside the fluid domain are integrated using the standard Gauss quadrature rules for quadrilaterals. However, special numerical quadrature techniques need to be employed for the integration of cut cells due to the fact that the integration needs to be performed only on the portion of the cut cell that corresponds to the fluid domain. Several factors influence the selection of such a special numerical integration technique, with accuracy and cost of computation being the most important ones. Ideally, the numerical technique for cut-cell integration should (i) yield optimal convergence rates, (ii) produce accurate fluxes across the interface, (iii) be applicable to 3D problems, and (iv) be computationally cheap.

In the literature, several techniques have been proposed for the integration of cut cells. These techniques can be broadly classified into the following groups: a.) tessellation or subtriangulation, b.) adaptive quadrature, c.) conformal mapping, d.) equivalent polynomial approach, e.) moment fitting method and f.) methods based on the divergence theorem. Recently, \cite{StavrevCMAME2016} and \cite{KudelaCMAME2016} presented geometrically accurate techniques based on reparametrisation of cut cells. A detailed discussion of these techniques is beyond the scope of this article. The reader is referred to \cite{YograjTUMreport2015} and references therein for an elaborate discussion on different techniques used for the integration of cut cells. In this paper, we consider only sub-triangulation and adaptive integration methods because of their suitability in the context of the present work.

\subsection{Integration of cut-cells using subtriangulation}
\begin{figure}[H]
\centering
 \subfloat[$Q_1$ b-splines]{ \includegraphics[trim=0mm 0mm 0mm 0mm, clip, scale=0.5]{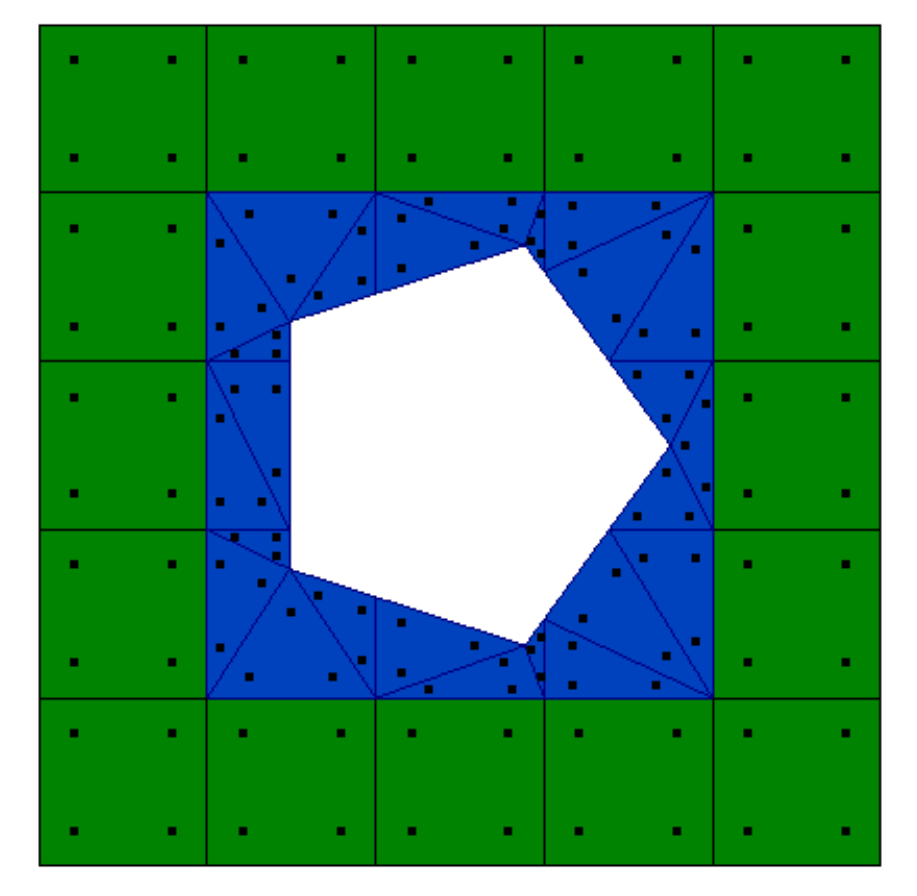} }
 \subfloat[$Q_2$ b-splines]{ \includegraphics[trim=0mm 0mm 0mm 0mm, clip, scale=0.5]{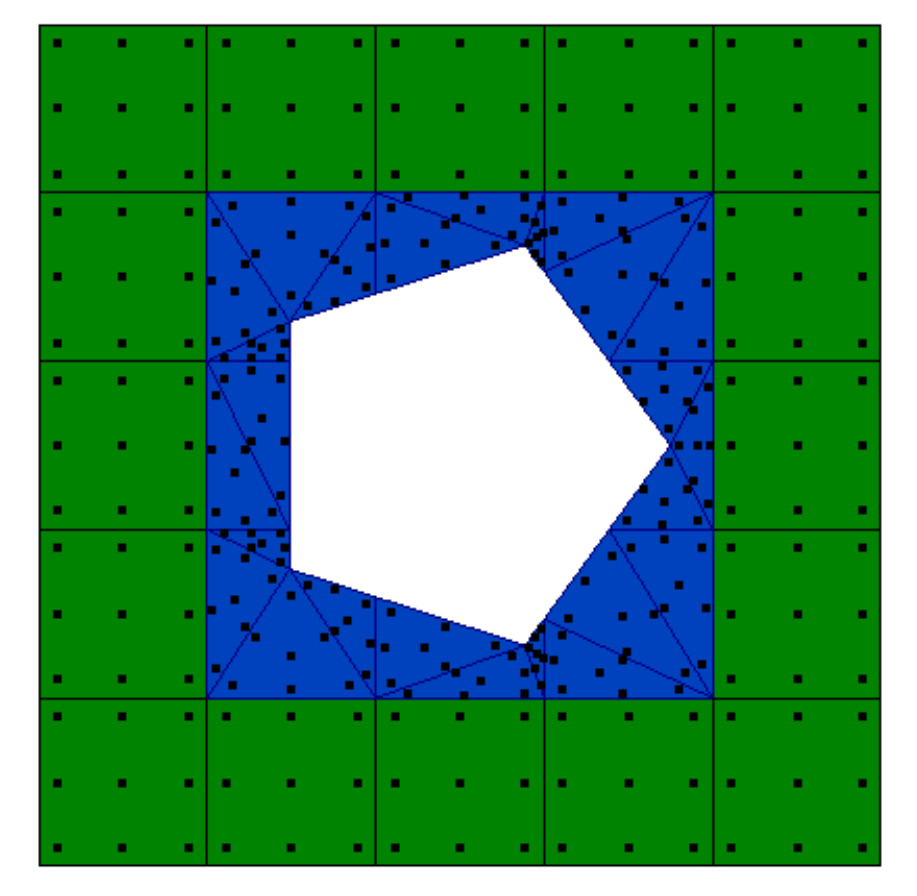} }
 \caption{Integration of cut cells using subtriangulation for linear ($Q_1$) and quadratic ($Q_2$) elements. Uncut cells are shown in green colour, cut cells in blue colour and Gauss points as black dots.}
 \label{fig-pentagon-st}
\end{figure}

In subtriangulation (ST), the active portion of a cut cell is integrated by splitting it into triangles and then applying the quadrature rules for individual triangles that correspond to the fluid domain. The concept of subtriangulation is illustrated schematically in Fig. \ref{fig-pentagon-st} for linear ($Q_1$) and quadratic ($Q_2$) elements.

While subtriangulation provides accurate integration of cut cells for interfaces discretised with straight lines using fewer quadrature points, its extension to higher-order discretisations of interfaces and for problems in three dimensions is significantly more challenging. For such cases, adaptive integration proves to be a viable alternative.

\subsection{Integration of cut-cells using adaptive integration}
\begin{figure}[H]
\centering
 \subfloat[Level-0]{ \includegraphics[trim=0mm 0mm 0mm 0mm, clip, scale=0.35]{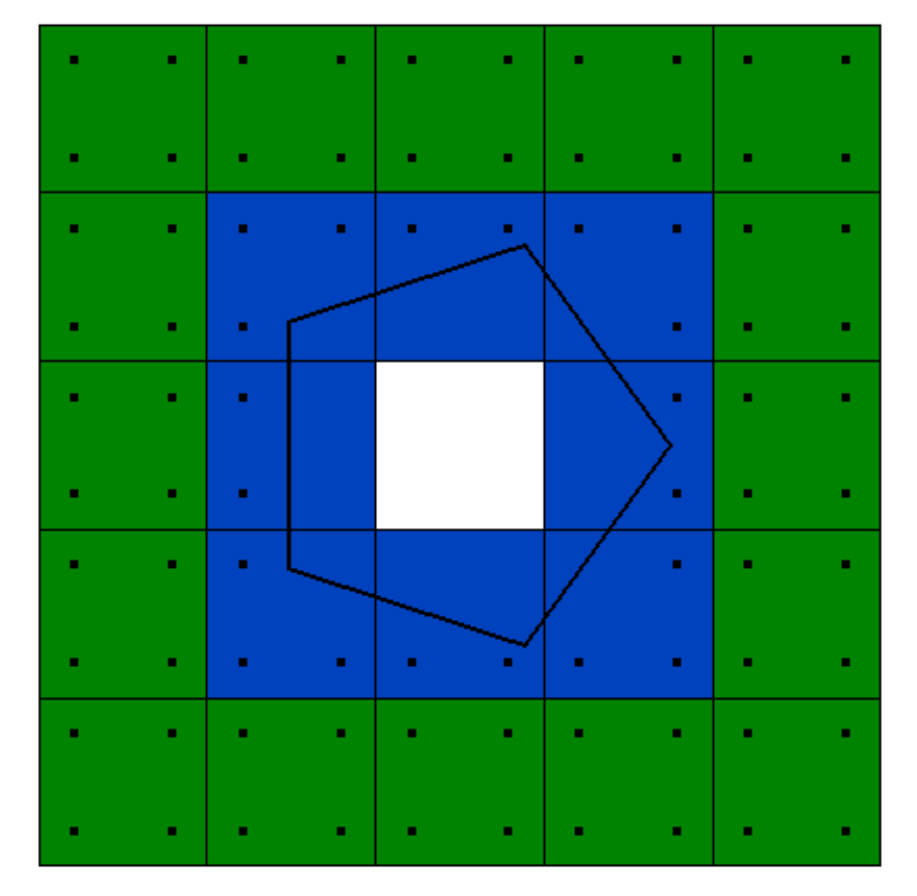} }
 \subfloat[Level-1]{ \includegraphics[trim=0mm 0mm 0mm 0mm, clip, scale=0.35]{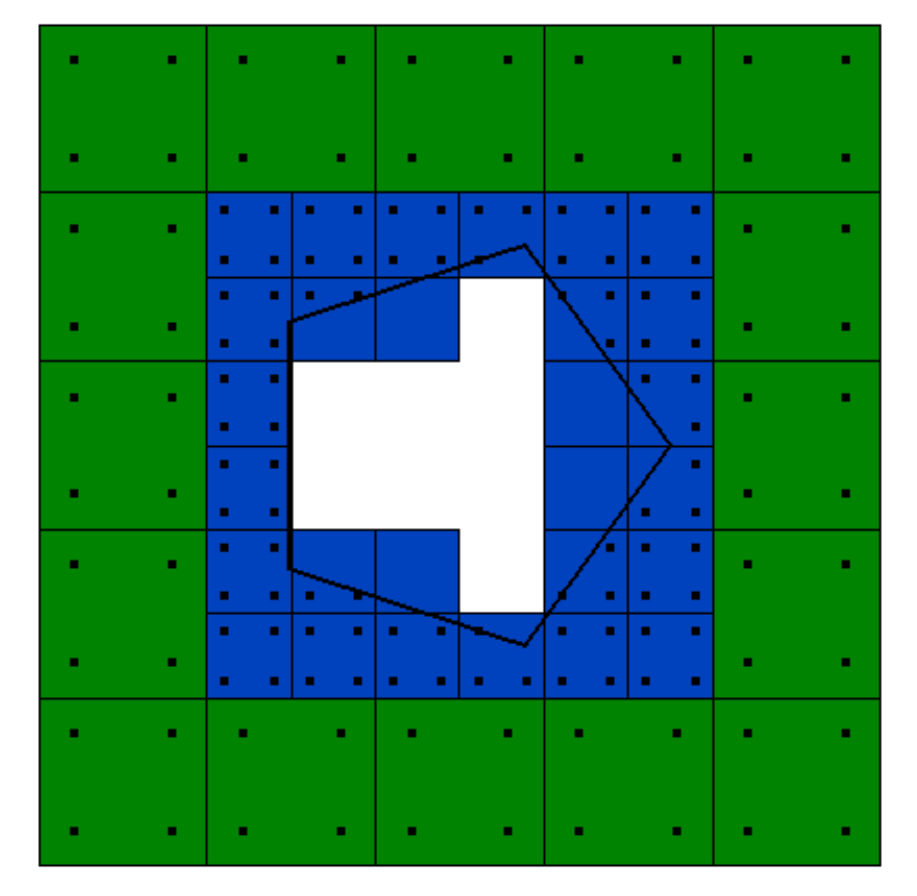} } \\
 \subfloat[Level-2]{ \includegraphics[trim=0mm 0mm 0mm 0mm, clip, scale=0.35]{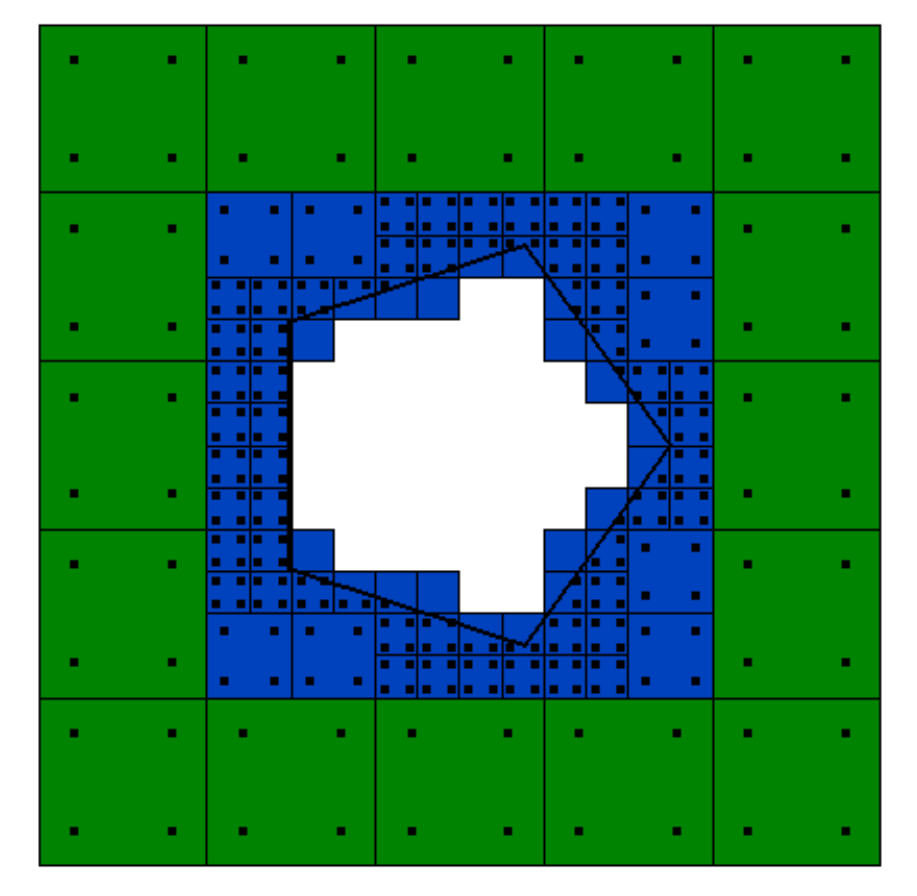} }
 \subfloat[Level-3]{ \includegraphics[trim=0mm 0mm 0mm 0mm, clip, scale=0.35]{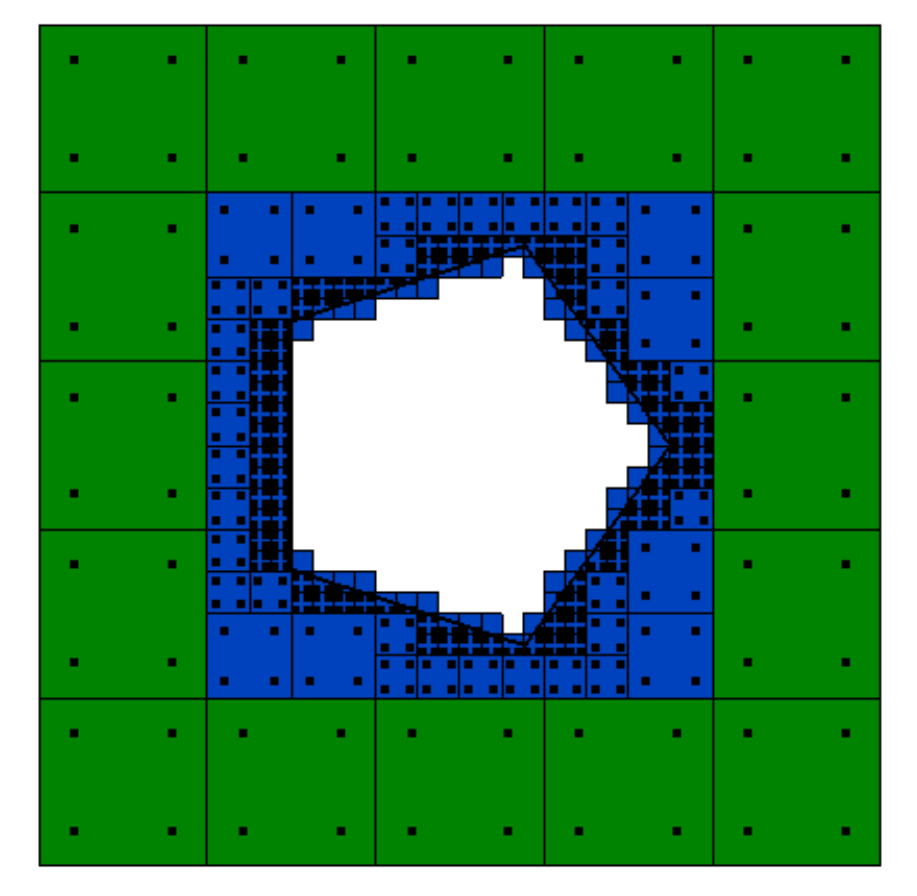} }
 \caption{Integration of cut cells: adaptive integration with \textit{quadtree} technique with a) one, b) two, c) three and c) four level(s) of recursive subdivision. Uncut cells are shown in green colour, cut cells in blue colour and Gauss points as black dots.}
 \label{fig-pentagon-ai}
\end{figure}

Integration of cut cells using adaptive integration relies on the recursive subdivision of a cut cell and then applying the quadrature rule for each relevant cell at finer levels, as illustrated schematically in Fig. \ref{fig-pentagon-ai}. This adaptive integration is usually performed using \textit{quadtree} subdivision in two dimensions and \textit{octree} subdivision in three dimensions although other variants of adaptive integration are also available based on binary subdivision and non-uniform refinement \cite{KadapaCOMPLAS2016, PetoAMSES2020}. According to the technique based on quadtree and octree, each cut cell is subdivided into four and eight smaller cells in two- and three- dimensions, respectively. The idea behind this technique lies in the fact that the quadrature points become increasingly clustered near the interface as the number of levels of the recursive subdivision is increased, as shown in Fig. \ref{fig-pentagon-ai}. Thus, the accuracy of integration of cut cells increases with an increasing number of levels.

The main advantages of the adaptive integration technique are: 
\begin{itemize}
\item Unlike subtriangulation, adaptive integration does not pose any restrictions on the type of representation of the interface. This technique requires only a point-location algorithm for the corresponding geometric representation.
\item It can be implemented quite robustly for both 2D and 3D problems using templates in C++.
\item The procedure can be parallelised efficiently for high-performance computing architectures.
\end{itemize}

Despite its simplicity and ease of implementation, the adaptive integration scheme proves to be computationally expensive when optimal convergence rates are to be recovered; inaccuracies in the integration of cut cells results in sub-optimal convergence rates. As demonstrated in Dettmer et al. \cite{DettmerCMAME2016} and also in Section \ref{subsec-Kovasznay}, an excessive number of the recursive levels are required to recover optimal convergence rates, and this number increases for higher-order discretisations of the background grid. The higher computational cost associated with the adaptive integration technique is due to the overhead costs incurred in the computation of the quadrature points up to the given level and subsequent evaluation of the element stiffness matrices and vectors for all of the integration points. Towards reducing the computational cost of adaptive integration technique, the so-called method of \emph{merging} is proposed in Dettmer et al. \cite{DettmerCMAME2016}. While the merging technique helps in bringing down the computational cost of element matrix and vector evaluations, it does not affect the order of accuracy. Note that, although other techniques such as binary subdivision, non-uniform subdivision and image compression \cite{PetoAMSES2020} can be used for achieving a reduction in computational cost, such techniques require sophisticated computer implementations.

\section{Numerical example} \label{section-examples}
We assess the performance of the adaptive integration technique using several numerical examples in two- and three- dimensions. We assess the accuracy using error norms, computational cost, force and displacement values. When using the sub-triangulation technique, three and seven quadrature points are used for each sub-triangle used, respectively, with $Q_1$ and $Q_2$ b-splines. For each boundary edge, 5 Gauss points are used. Spectral radii for the generalised-alpha time integration used for the solid and fluid problems are zero. The unsymmetric version of Nitsche's method is used in all the examples. Henceforth, adaptive integration and sub-triangulation are referred to as AI and ST, respectively.

Note that simulations of fluid-rigid solid interaction problems, for example, lock-in of a circular cylinder and galloping of a square body, showed a similar trend in the force and displacement observed with fluid-flexible solid interaction problems presented in this work. Therefore, for the sake of clarity, FSI problems with rigid solids are omitted from the paper.

\subsection{Kovasznay flow} \label{subsec-Kovasznay}
\begin{figure}[H]
\centering
 \subfloat[X-velocity]{\includegraphics[trim = 20mm 0mm 0mm 0mm, clip,scale=0.45]{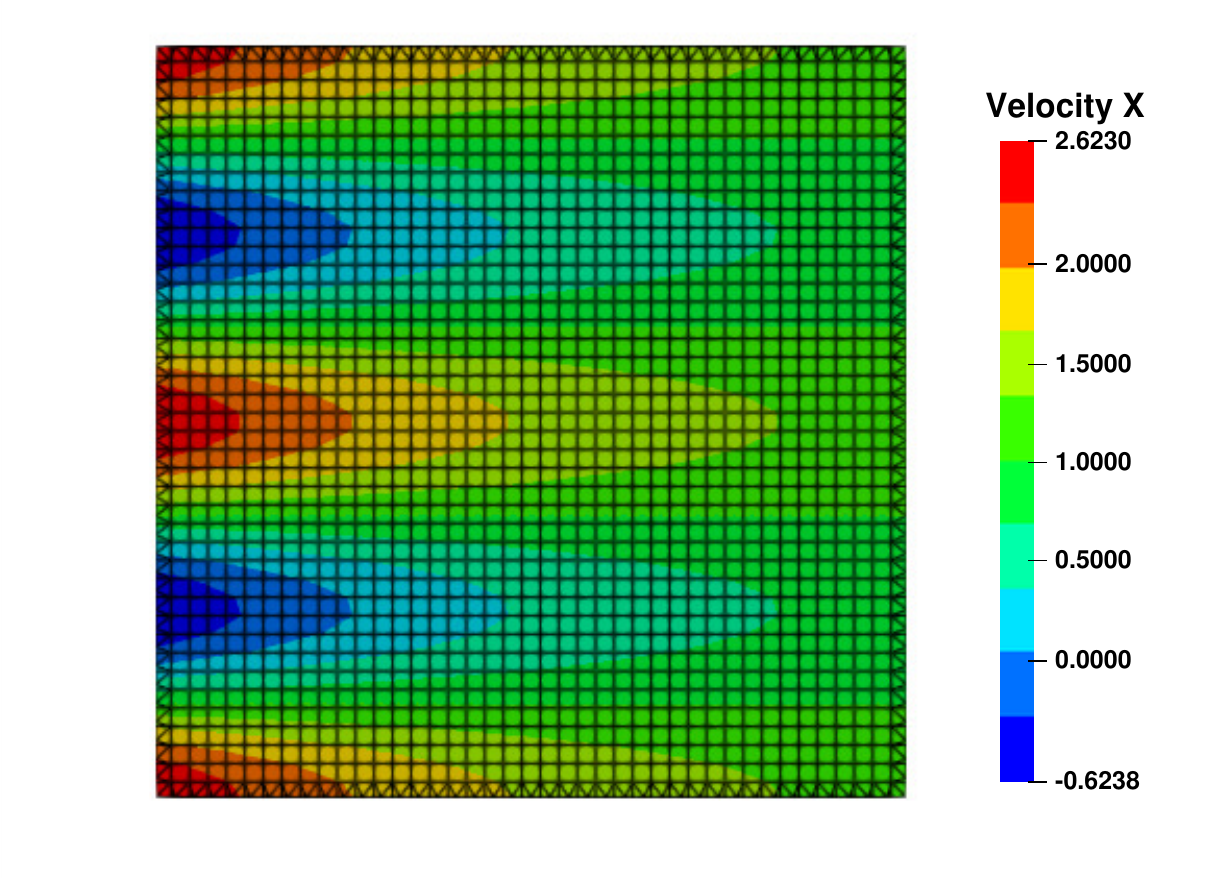}}
 \subfloat[Pressure]{\includegraphics[trim = 20mm 0mm 0mm 0mm, clip,scale=0.45]{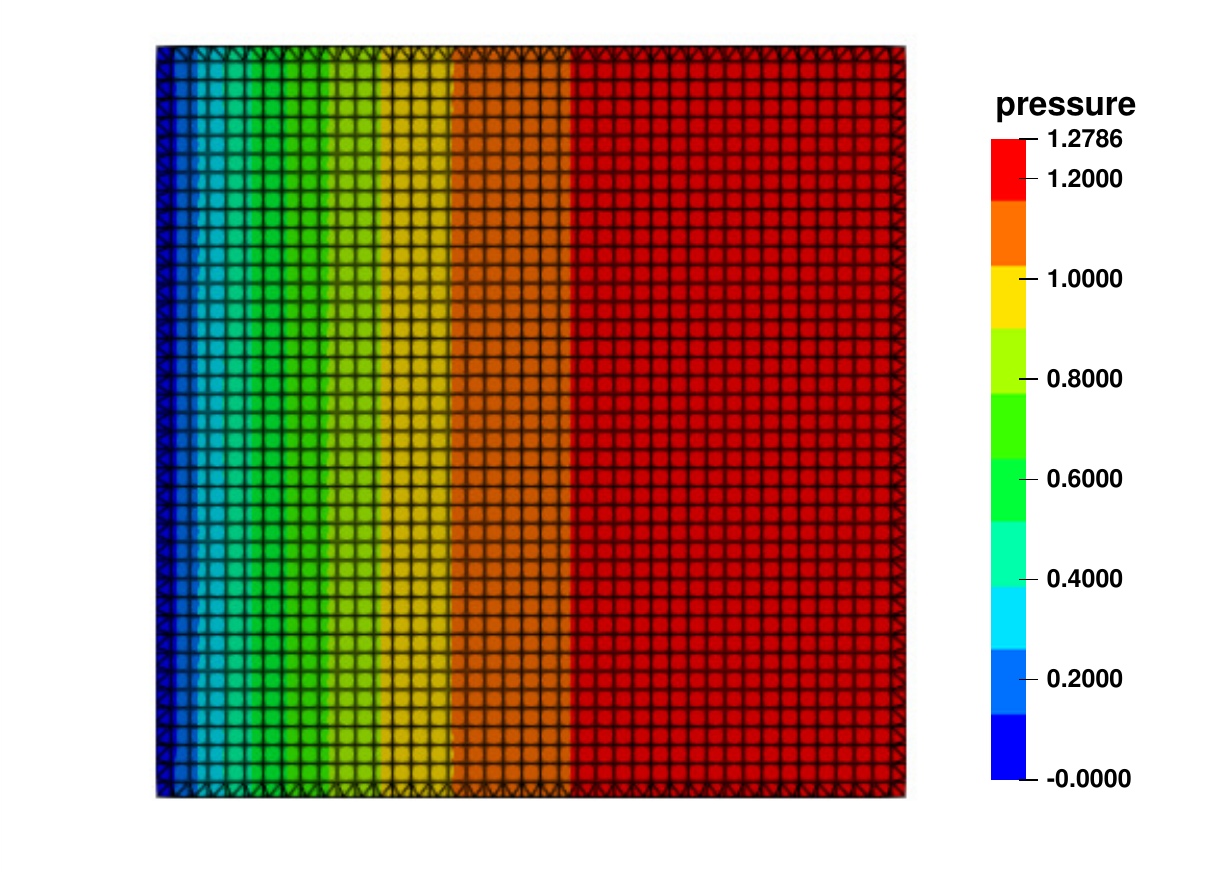}}
 \caption{Kovasznay flow: contour plots of X-velocity and pressure obtained with $Q_1$ elements.}
\label{fig-Kovasznay-contours}
\end{figure}

In this example, we assess the effect of adaptive integration on the convergence rates in error norms in fluid velocity and pressure using the problem of Kovasznay flow \cite{Kovasznay1948}. The domain of the problem is $\Omega^f=\left[-0.5,1.5 \right] \times \left[-0.5,1.5 \right]$. The analytical solution is given by the expressions,
\begin{align}
v_x^f(x,y) &= 1 - e^{\lambda \, x} \cos(2 \, \pi \, y), \\
v_y^f(x,y) &= \frac{\lambda}{2 \, \pi} \, e^{\lambda \, x} \sin(2 \, \pi \, y), \\
p(x,y) &= p_0 - \frac{1}{2} \, e^{2 \, \lambda \, x},
\end{align}
with
\begin{align}
\lambda = \frac{Re}{2} - \sqrt{\frac{Re^2}{4} + 4 \, \pi^2},
\end{align}
where $p_0$ is the reference pressure and $Re$ is the Reynolds number. The problem is modelled with 20, 40, 80 and 160 edges on each side of the domain, immersed on a background grid with $31 \times 31$, $61 \times 61$, $121 \times 121$ and $241 \times 241$ elements. The Reynolds number is assumed to be 40. The horizontal velocity and pressure obtained with a level-2 hierarchical mesh using $Q_1$ elements are shown in Fig. \ref{fig-Kovasznay-contours}.

\begin{figure}[H]
 \begin{center}
  \subfloat[AI - Level 2]{\includegraphics[trim = 20mm 0mm 30mm 0mm, clip,scale=0.5]{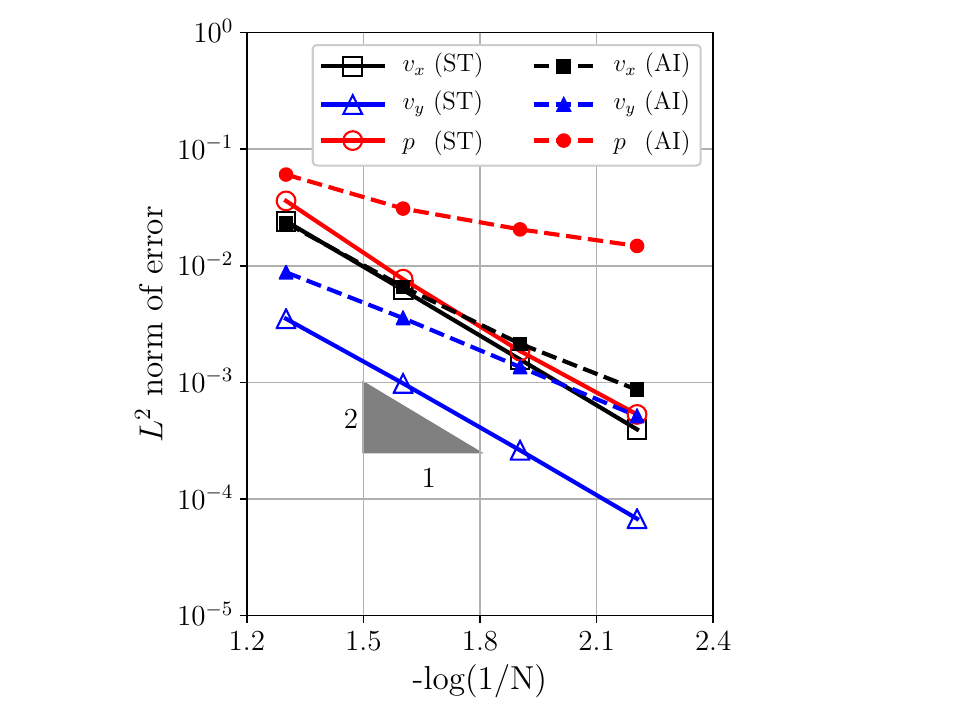}}
  \subfloat[AI - Level 3]{\includegraphics[trim = 20mm 0mm 30mm 0mm, clip,scale=0.5]{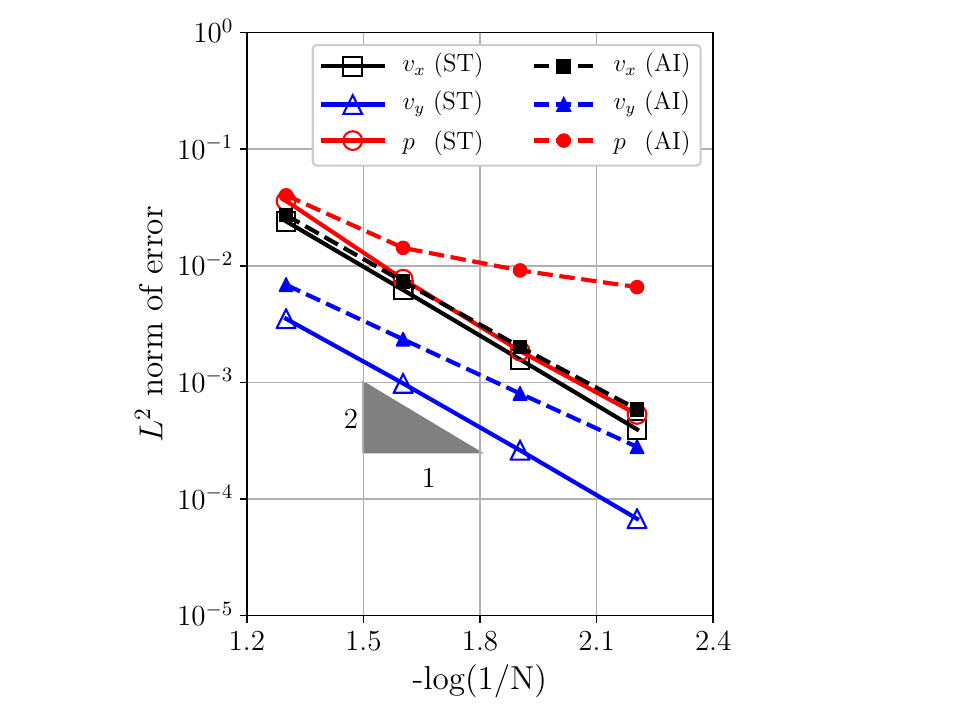}}
  \subfloat[AI - Level 4]{\includegraphics[trim = 20mm 0mm 30mm 0mm, clip,scale=0.5]{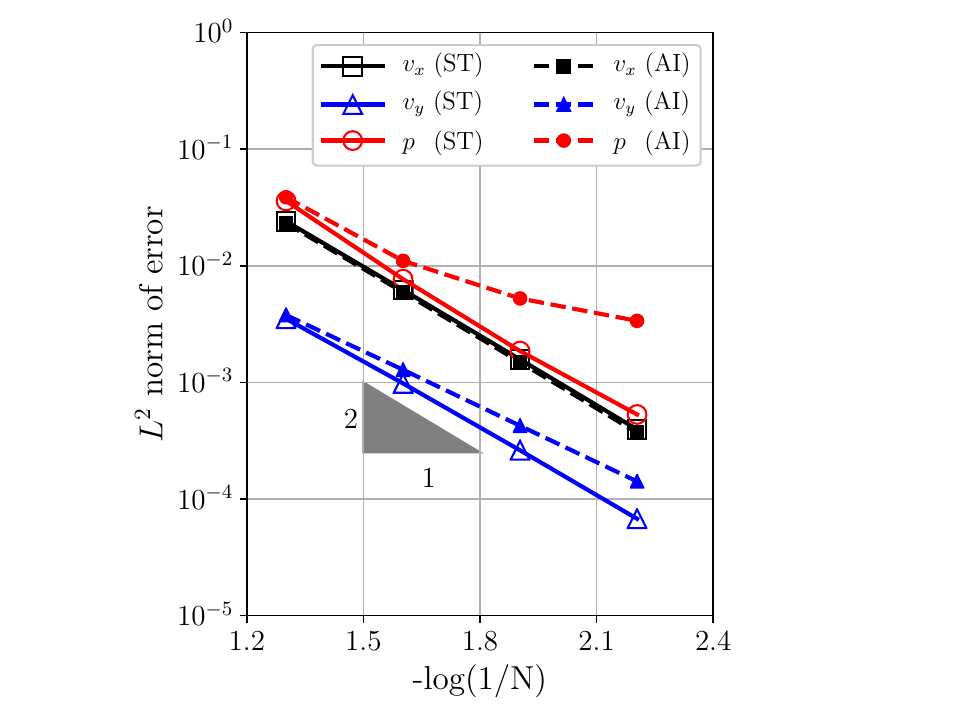}} \\
  \subfloat[AI - Level 6]{\includegraphics[trim = 20mm 0mm 30mm 0mm, clip,scale=0.5]{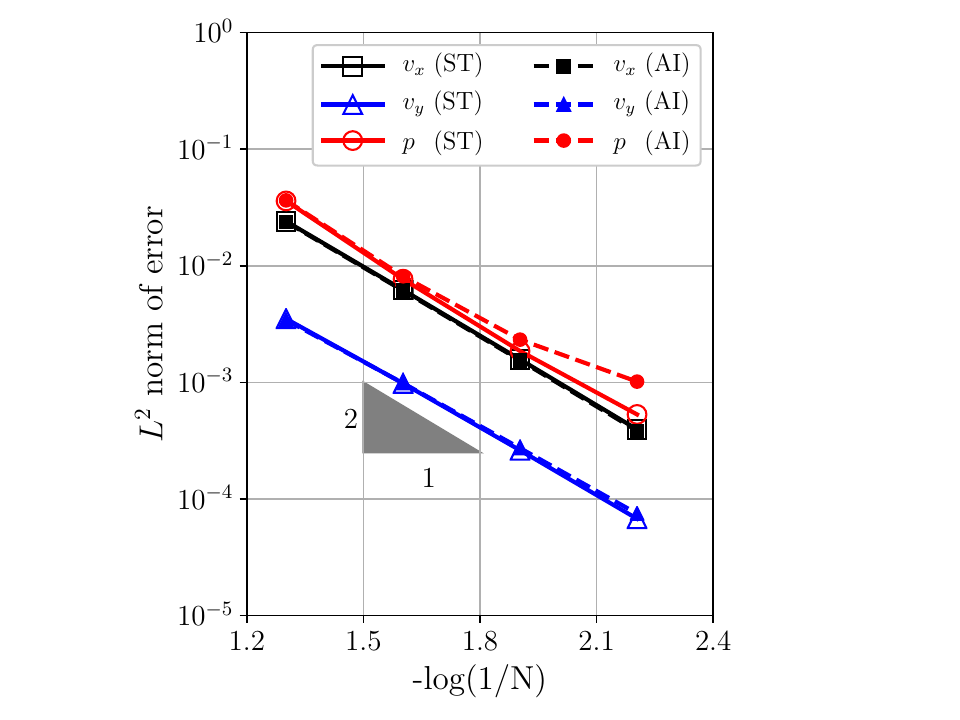}}
  \subfloat[AI - Level 8]{\includegraphics[trim = 20mm 0mm 30mm 0mm, clip,scale=0.5]{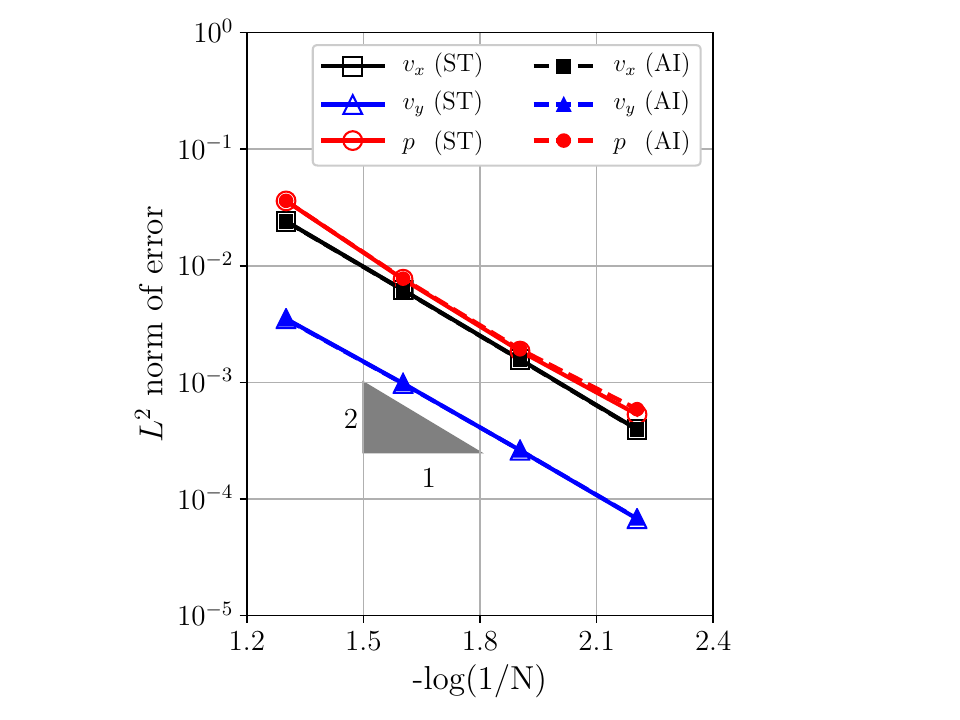}}
  \subfloat[AI - Level 10]{\includegraphics[trim = 20mm 0mm 30mm 0mm, clip,scale=0.5]{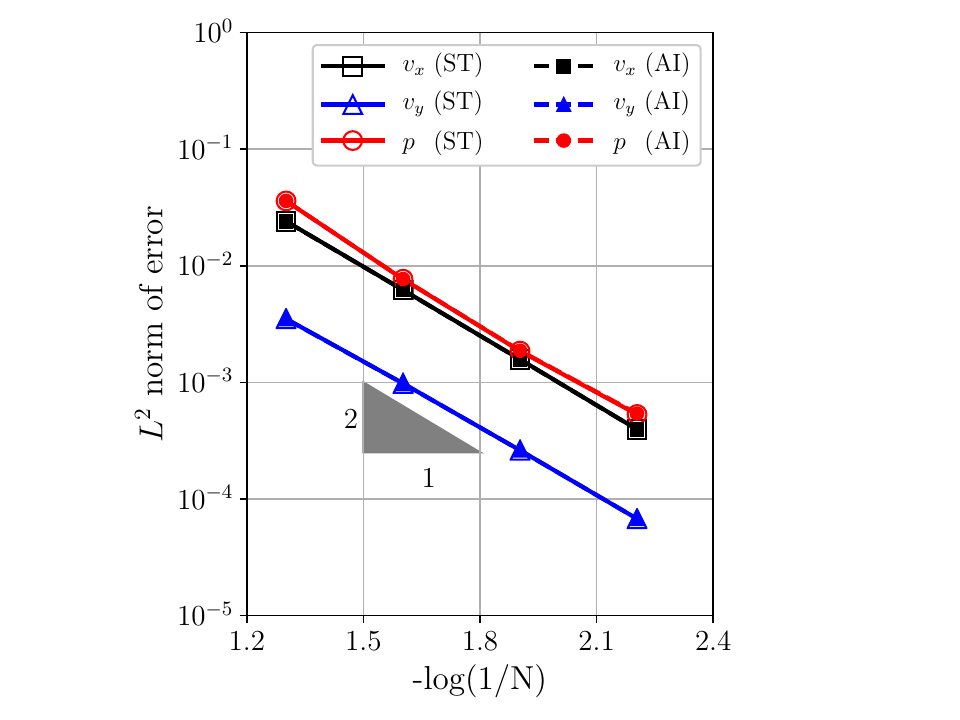}}
 \caption{Kovasznay flow: error norms in velocity and pressure for $Q_1$ b-splines. $N$ is number of edges along one side.}
 \label{fig-Kovasznay-q1}
 \end{center}
\end{figure}

\begin{figure}[H]
 \begin{center}
  \subfloat[AI - Level 2]{\includegraphics[trim = 20mm 0mm 30mm 0mm, clip,scale=0.5]{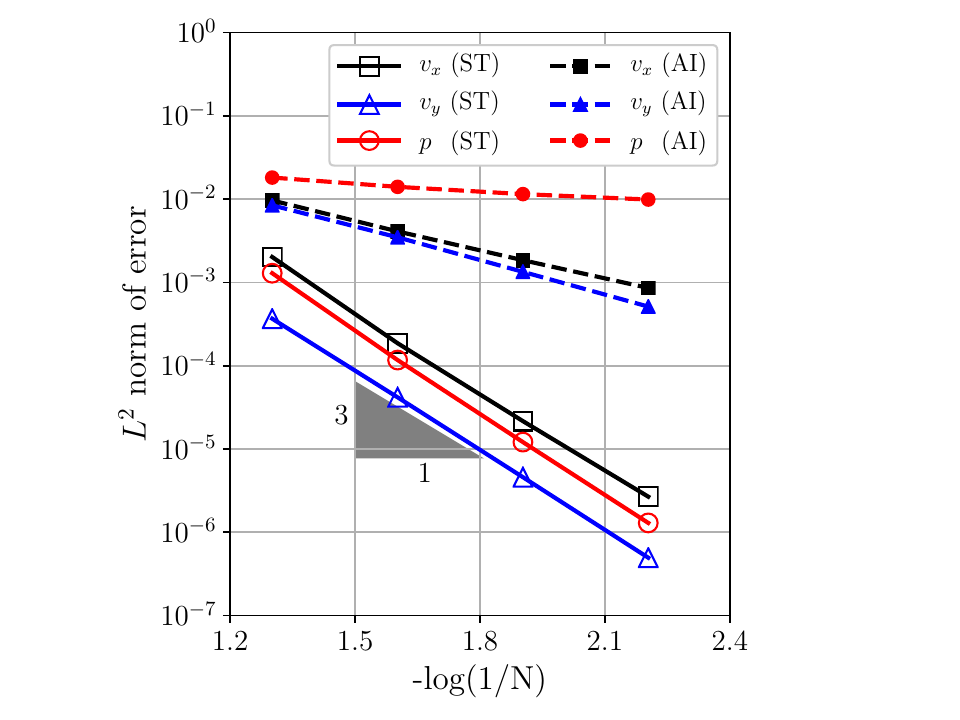}}
  \subfloat[AI - Level 3]{\includegraphics[trim = 20mm 0mm 30mm 0mm, clip,scale=0.5]{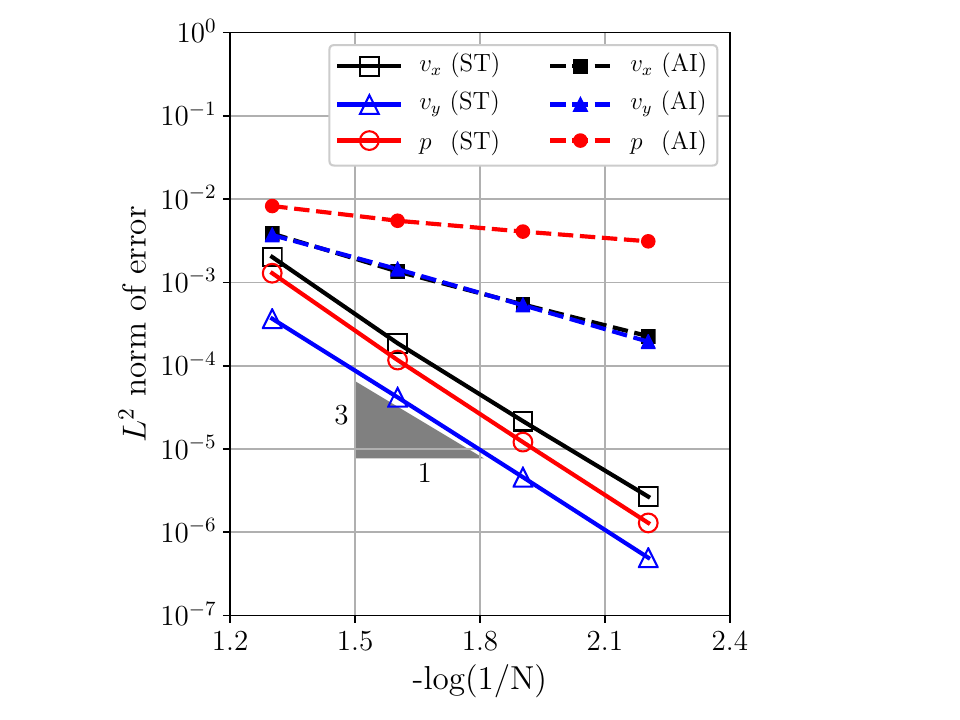}}
  \subfloat[AI - Level 4]{\includegraphics[trim = 20mm 0mm 30mm 0mm, clip,scale=0.5]{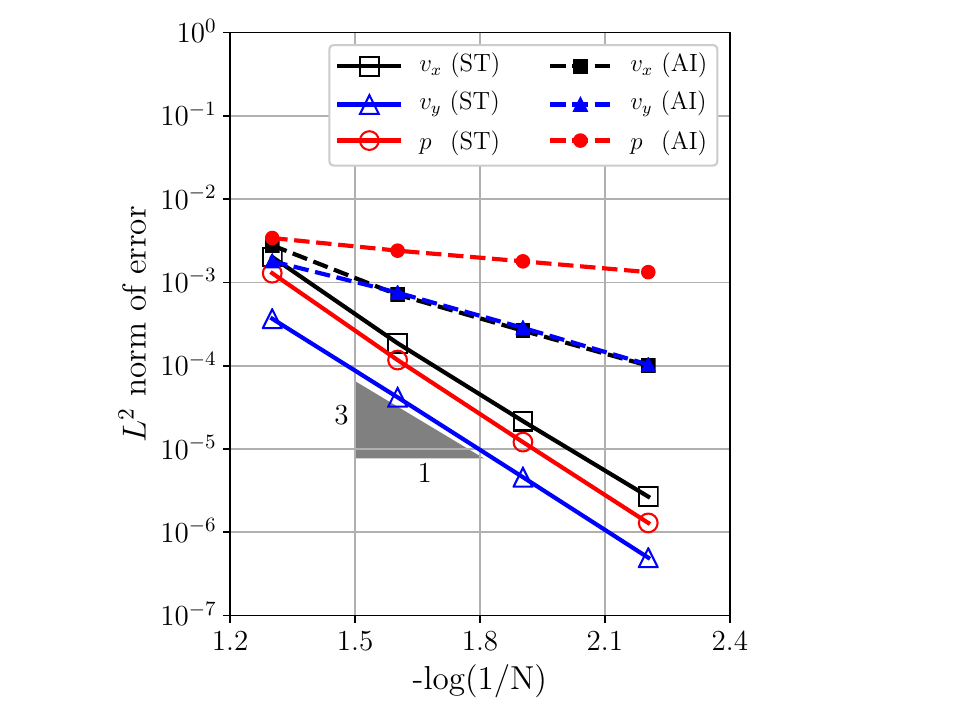}} \\
  \subfloat[AI - Level 6]{\includegraphics[trim = 20mm 0mm 30mm 0mm, clip,scale=0.5]{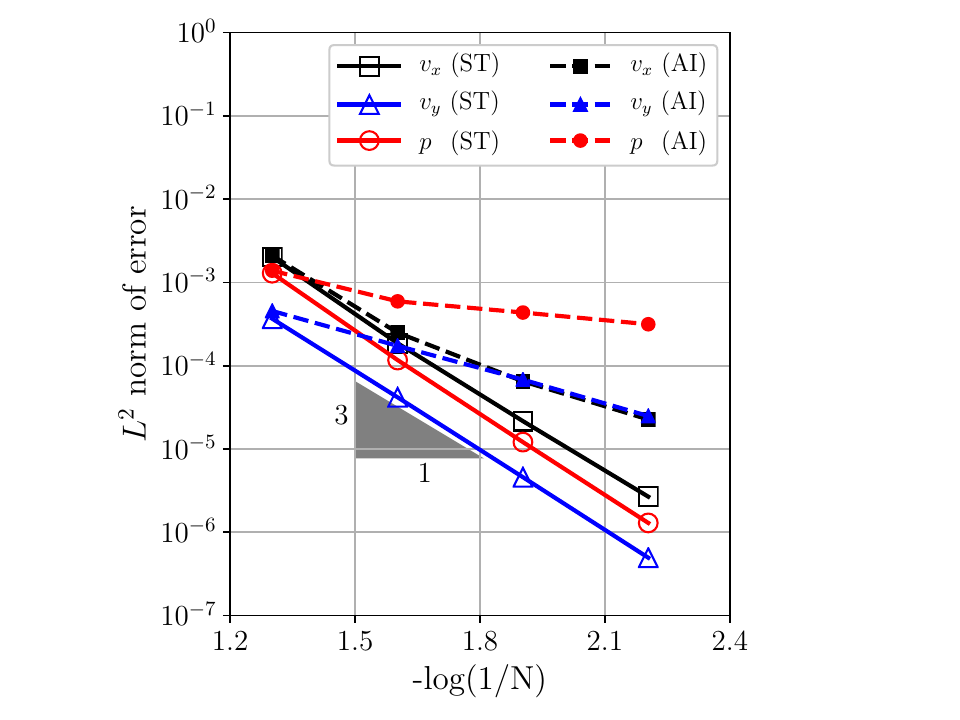}}
  \subfloat[AI - Level 8]{\includegraphics[trim = 20mm 0mm 30mm 0mm, clip,scale=0.5]{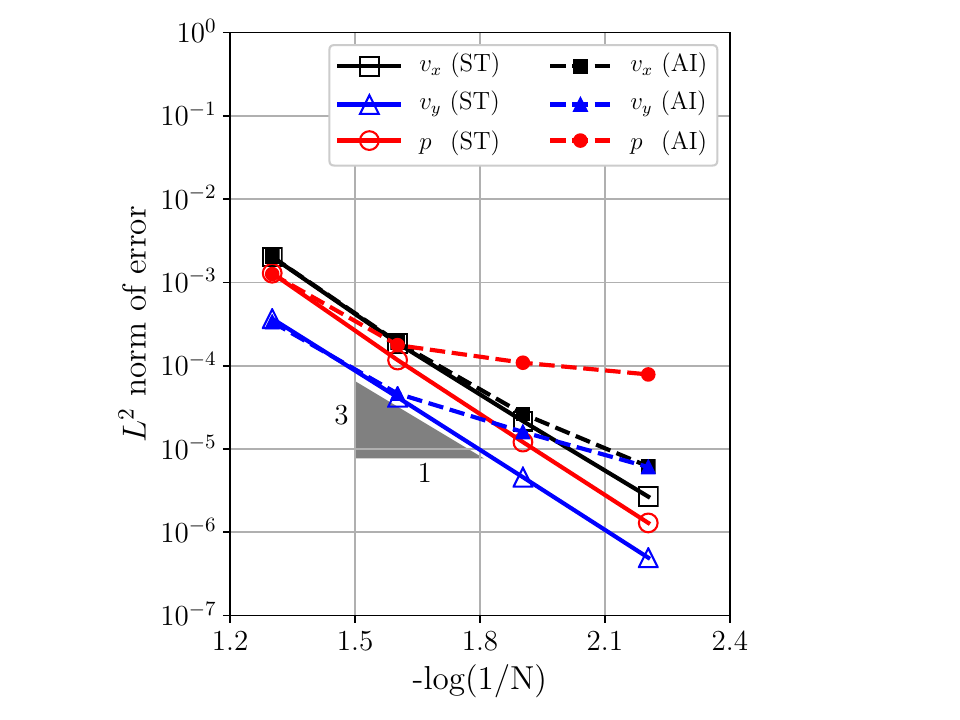}}
  \subfloat[AI - Level 10]{\includegraphics[trim = 20mm 0mm 30mm 0mm, clip,scale=0.5]{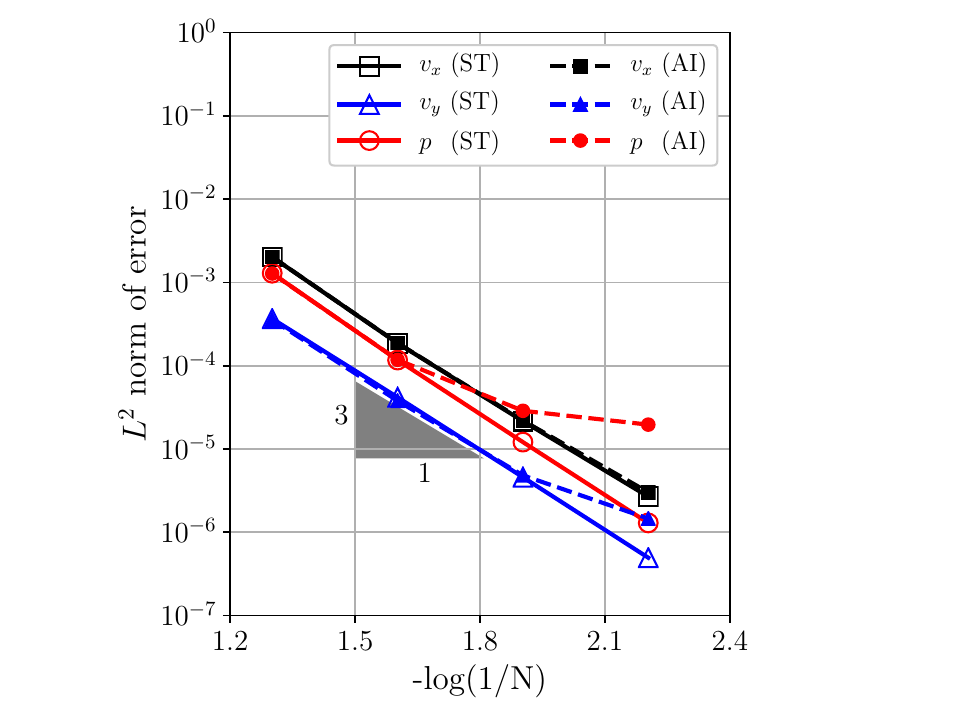}}
 \caption{Kovasznay flow: error norms in velocity and pressure for $Q_2$ b-splines.}
 \label{fig-Kovasznay-q2}
 \end{center}
\end{figure}

Error norms in velocity and pressure computed using subtriangulation and adaptive integration are shown in Figs. \ref{fig-Kovasznay-q1} and \ref{fig-Kovasznay-q2}, respectively, for $Q_1$ and $Q_2$ b-splines. As shown, subtriangulation yields optimal convergence rates right away for both cases, while the optimal rates are recovered by increasing the number of levels of refinement when using the adaptive integration technique. Moreover, it is apparent from the convergence graphs that the number of levels of adaptive integration required for recovering optimal convergence rates is higher for $Q_2$ elements than that for $Q_1$ elements.

\begin{figure}[H]
 \begin{center}
  \subfloat[]{\includegraphics[trim = 0mm 0mm 0mm 0mm, clip,scale=0.5]{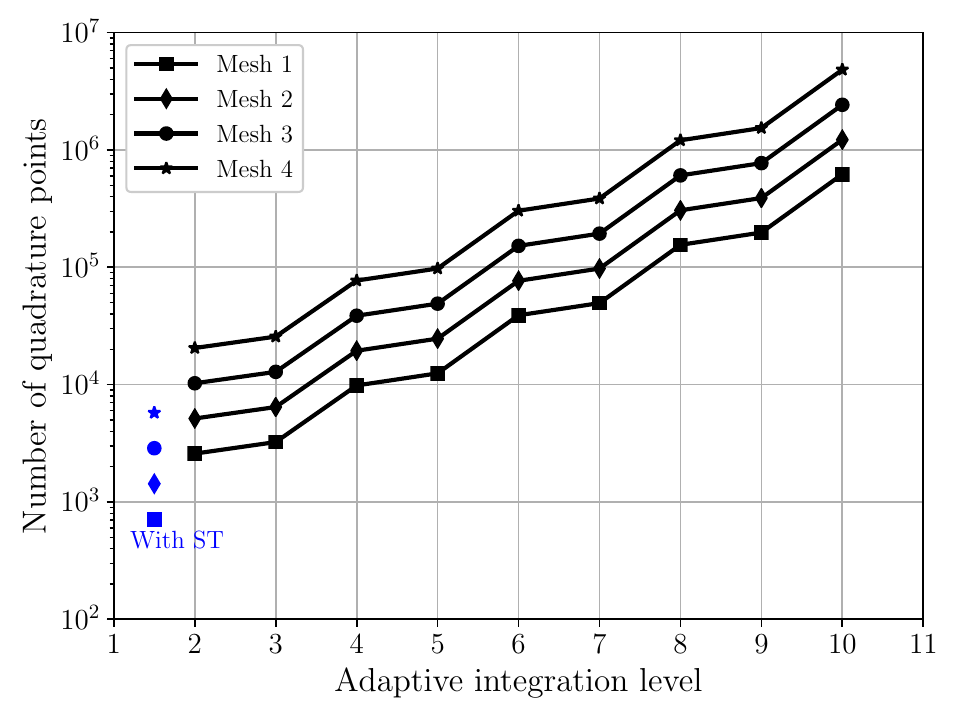}}
  \subfloat[]{\includegraphics[trim = 0mm 0mm 0mm 0mm, clip,scale=0.5]{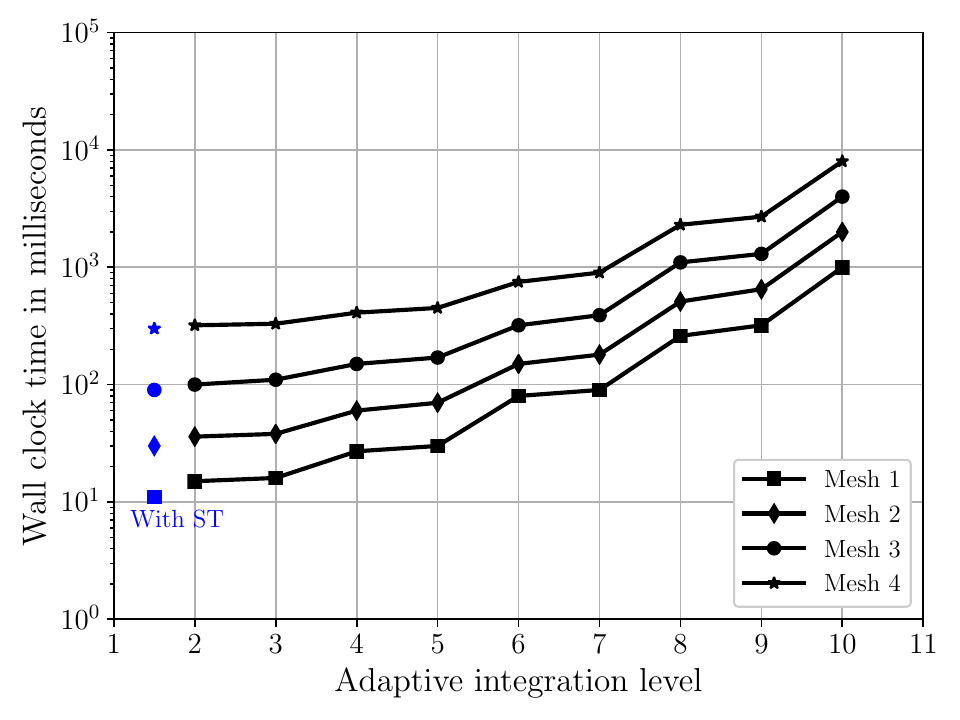}}
 \caption{Kovasznay flow: number of quadrature points and wall clock time for computing the global stiffness matrix for $Q_1$ b-splines.}
 \label{fig-Kovasznay-time-q1}
 \end{center}
\end{figure}

\begin{figure}[H]
 \begin{center}
  \subfloat[]{\includegraphics[trim = 0mm 0mm 0mm 0mm, clip,scale=0.5]{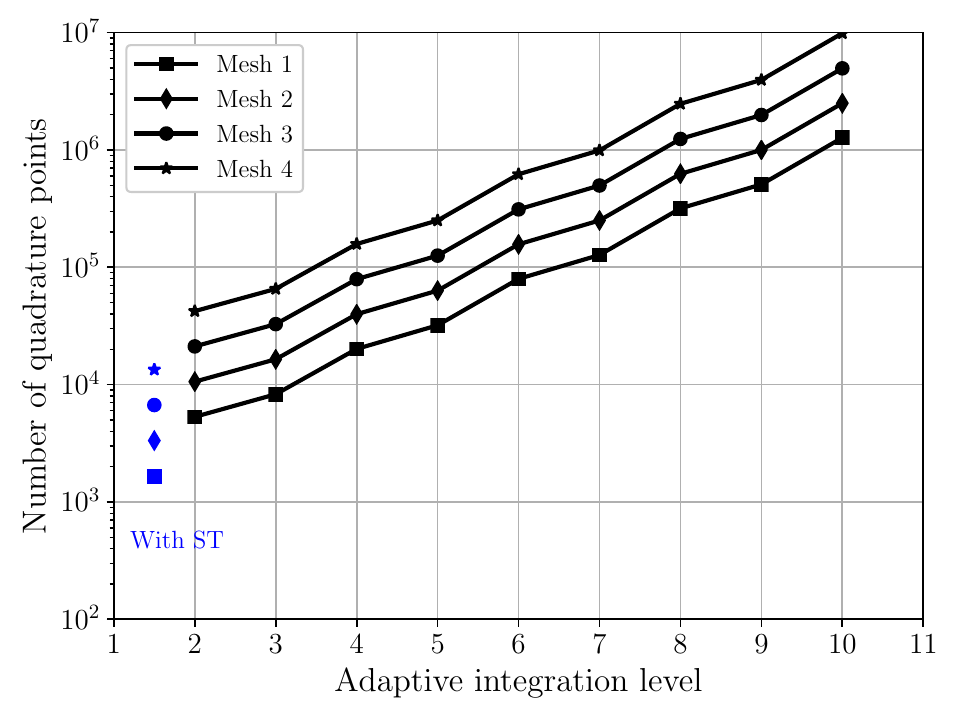}}
  \subfloat[]{\includegraphics[trim = 0mm 0mm 0mm 0mm, clip,scale=0.5]{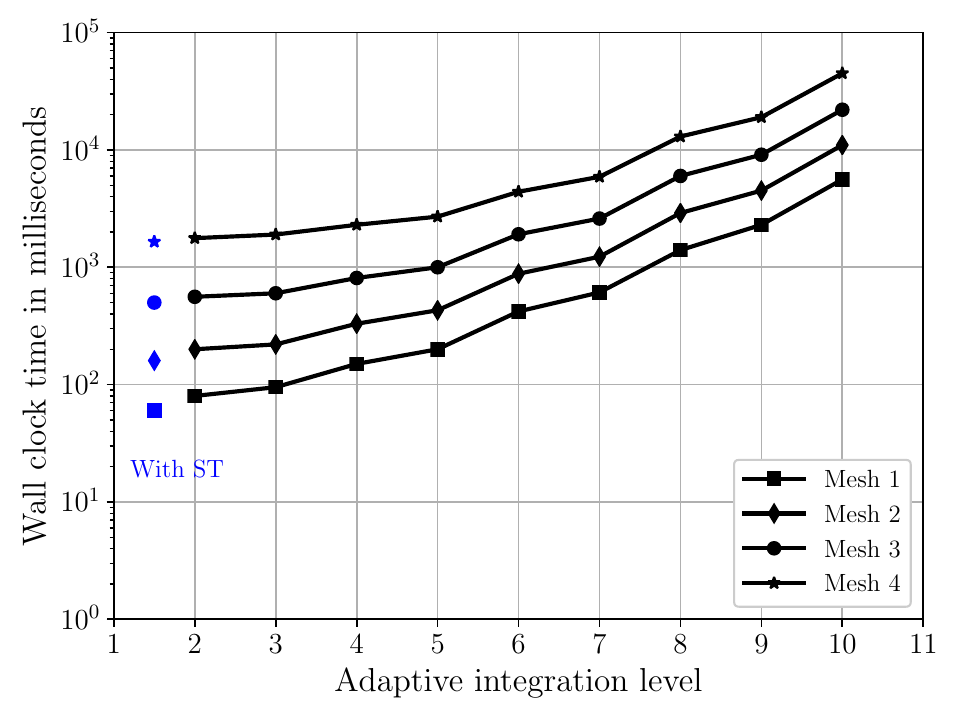}}
 \caption{Kovasznay flow: number of quadrature points and wall clock time for computing the global stiffness matrix for $Q_2$ b-splines.}
 \label{fig-Kovasznay-time-q2}
 \end{center}
\end{figure}

Although the optimal convergence rates can be recovered by increasing the number of levels of refinement in adaptive integration, this comes with added computational overheads due to an increase in the number of quadrature points with increasing levels of refinement. The number of quadrature points and the corresponding wall clock time (measured on a single Intel i7-8750H CPU) increases exponentially as the level of refinement in the adaptive integration technique is increased, as shown in Figs. \ref{fig-Kovasznay-time-q1} and \ref{fig-Kovasznay-time-q2}, respectively, for $Q_1$ and $Q_2$ elements. Note that the wall clock time for the matrix solver remains almost the same for all the levels of adaptive integration; therefore, it is not included in the discussion. The wall clock time required for computing and assembling the global stiffness matrix per iteration increases by more than an order of magnitude for level 10 when compared with that of subtriangulation. This significant increase in the computational cost for recovering the optimal convergence rates is a drawback of the standard quadtree-based adaptive integration technique. However, as demonstrated in the examples that follow, it is not necessary to use such finer levels of adaptive integration for FSI problems. Global quantities such as forces and displacements of acceptable accuracy for FSI problems can be obtained using fewer levels of adaptive integration; thus, making adaptive integration a computationally viable alternative for subtriangulation and one that can be readily extended to 3D problems.

\subsection{Steady flow past a fixed square body at $Re=20$}
\begin{figure}[H]
\centering
  \subfloat[]{ \label{fixed-square-geom} \includegraphics[trim = 0mm 0mm 0mm 0mm, clip, scale=0.9]{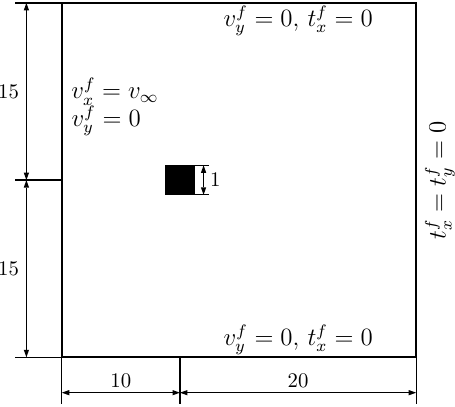}}
  \subfloat[]{ \label{fixed-square-mesh} \includegraphics[trim = 0mm 0mm 0mm 0mm, clip, scale=0.4]{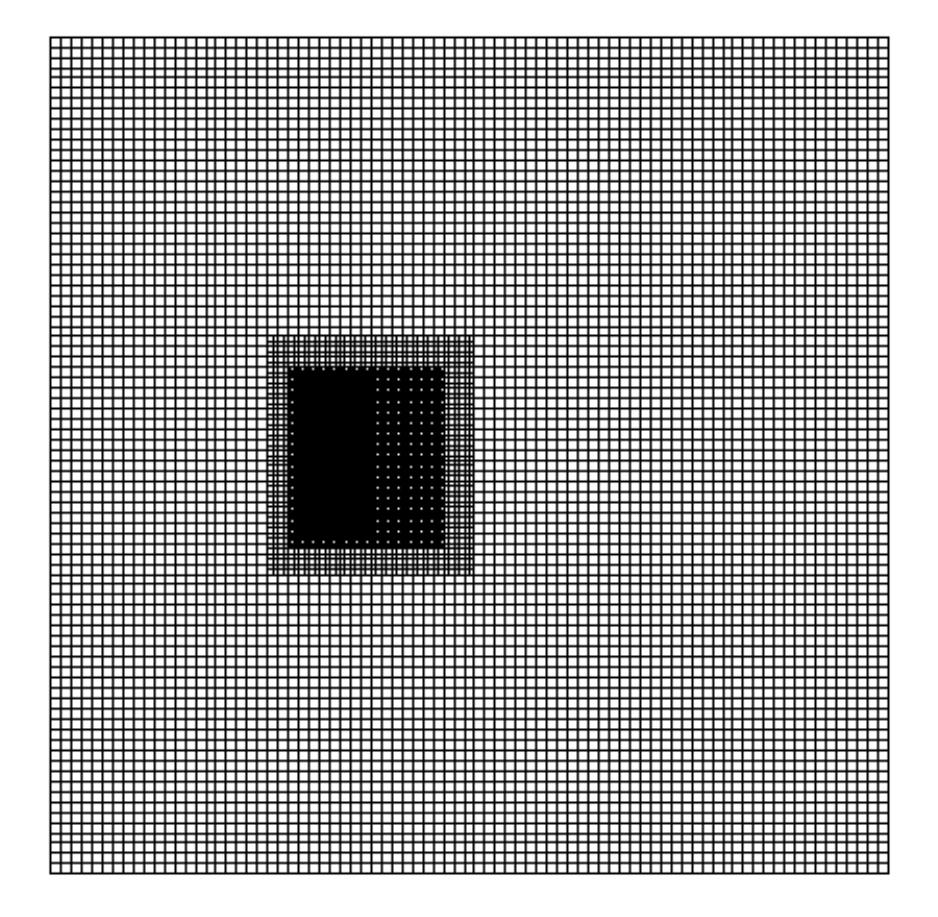}}
  \caption{Flow past a fixed square: a.) geometry and boundary conditions and b.) hierarchical b-spline mesh.}
 \label{fixed-square-geom-main}
\end{figure}

\begin{figure}[H]
 \begin{center}
  \includegraphics[trim = 0mm 0mm 0mm 0mm, clip, scale=0.6]{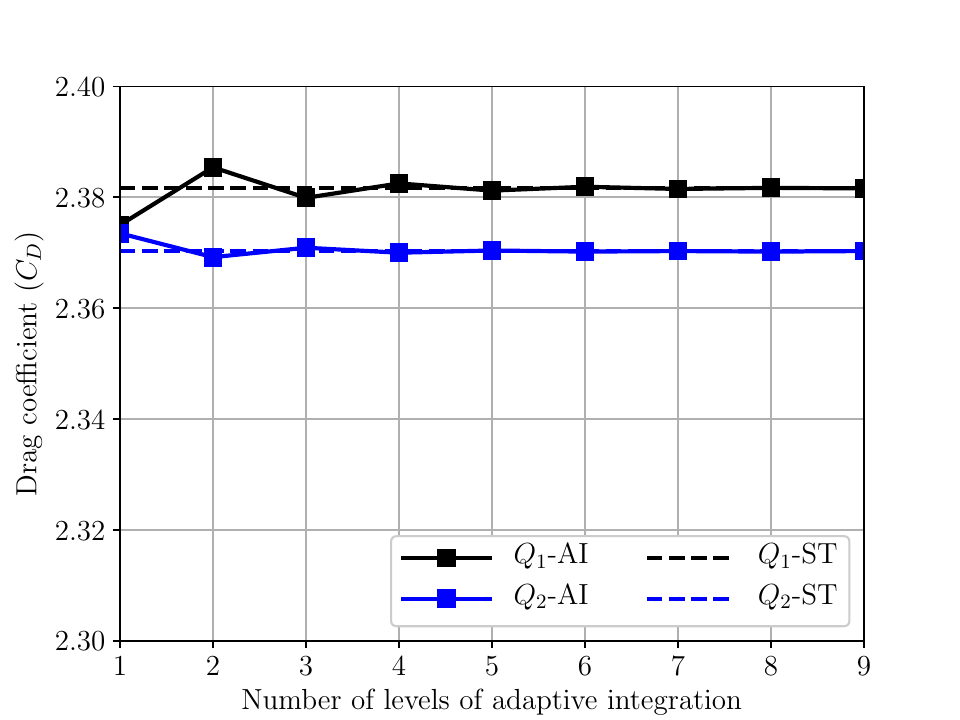}
  \caption{Flow past a fixed square: variation of drag coefficient with different levels of adaptive integration with $Q_1$, and $Q_2$ b-splines.}
  \label{fixed-square-re20-drag}
 \end{center}
\end{figure}

This example is concerned with the steady flow over a fixed rigid square body. This particular example is chosen because the geometry of the square is represented exactly irrespective of the description used. The geometry and boundary conditions of the problem are as shown in Fig. \ref{fixed-square-geom}.  Properties of the fluid are: density, $\rho^f = 1.0$ and viscosity, $\mu^f = 0.05$. An uniform velocity of $v_{\infty}=1.0$ is imposed at the inlet in X-direction so that the Reynolds number is, $Re = \rho D v_{\infty}/\mu = 20$. Simulations are carried out on a level-3 hierarchical mesh shown in Fig. \ref{fixed-square-mesh} with $Q_1$ and $Q_2$ b-splines. The effect of adaptive integration on the numerical results is assessed by computing the drag coefficient ($C_D$) for each simulation and comparing it with that obtained using subtriangulation. Computed values of drag coefficients are plotted in Fig. \ref{fixed-square-re20-drag}. As expected, as the number of levels of adaptive integration is increased, the values of $C_D$ obtained with adaptive integration converge towards those obtained with sub-triangulation. Comparison of $C_D$ values obtained in the present work match well with the reference values from the literature, as presented in Table. \ref{table-square2d}.

\graphicspath{{./figures/}}

\renewcommand{\arraystretch}{1.2}
\begin{table}[H]
\centering
\begin{tabular}{|p{5cm}|p{2cm}|p{2cm}|p{2cm}|p{2cm}|}
\hline
                 &  $Re=20$   & \multicolumn{3}{|c|}{  $Re=100$ }    \\
\hline
Data             &  $C_{D}$   &  $C_{D_{avg}}$ &  $C_{L_{rms}}$ &  $St$ \\
\hline
Sen et al. \cite{SenIJNMF2011}             &  2.2140  &  1.5300  &  0.1850  &  0.1350  \\
Sharma and Eswaran \cite{BreuerIJHFF2000}  &  2.3500  &  1.5000  &  0.1900  &  0.1480  \\
Breuer et al. \cite{SharmaNHptA2004}       &  2.4000  &  1.4000  &     -    &  0.1400  \\
Zhao et al. \cite{ZhaoPoF2013}             &     -    &  1.4520  &  0.1908  &  0.1447  \\
Present - $Q_1$ b-splines                  &  2.3817  &  1.4776  &  0.1897  &  0.1470  \\
Present - $Q_2$ b-splines                  &  2.3703  &  1.4302  &  0.1810  &  0.1520  \\
\hline
\end{tabular}
\caption{Flow past a fixed square: comparison of drag coefficient ($C_D$), lift coefficient ($C_L$) and Strouhal number ($St$).}
 \label{table-square2d}
\end{table}
\renewcommand{\arraystretch}{1.0}

\subsection{Unsteady flow past a fixed square body at $Re=100$}
We now study unsteady flow past a fixed square body for a Reynolds number of 100. The setup of the problem is the same as the one from the previous example. The viscosity of the fluid is adjusted to $\mu^f = 0.01$ so that $Re=100$. The same mesh used in the previous example is considered. Using a constant time step, $\Delta t = 0.1$, simulations are carried out with $Q_1$ and $Q_2$ b-splines and with subtriangulation and different levels of adaptive integration for each order b-splines. The effect of adaptive integration on the numerical results is assessed by computing the drag coefficient, lift coefficient ($C_L$) and Strouhal number ($St$) for each simulation. The results obtained with the subtriangulation are used for comparison.

\begin{figure}[H]
 \begin{center}
  \includegraphics[trim = 0mm 0mm 0mm 0mm, clip, scale=0.35]{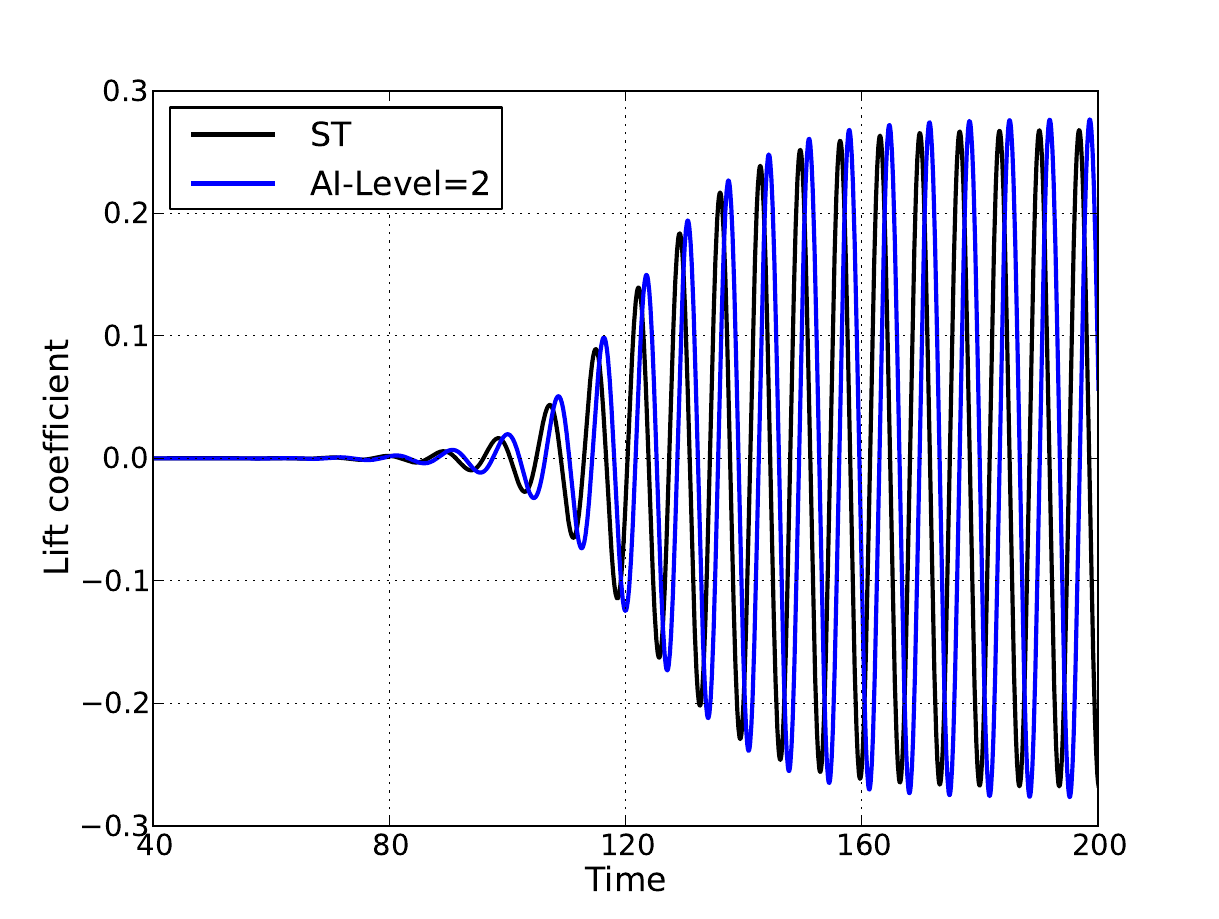}
  \includegraphics[trim = 0mm 0mm 0mm 0mm, clip, scale=0.35]{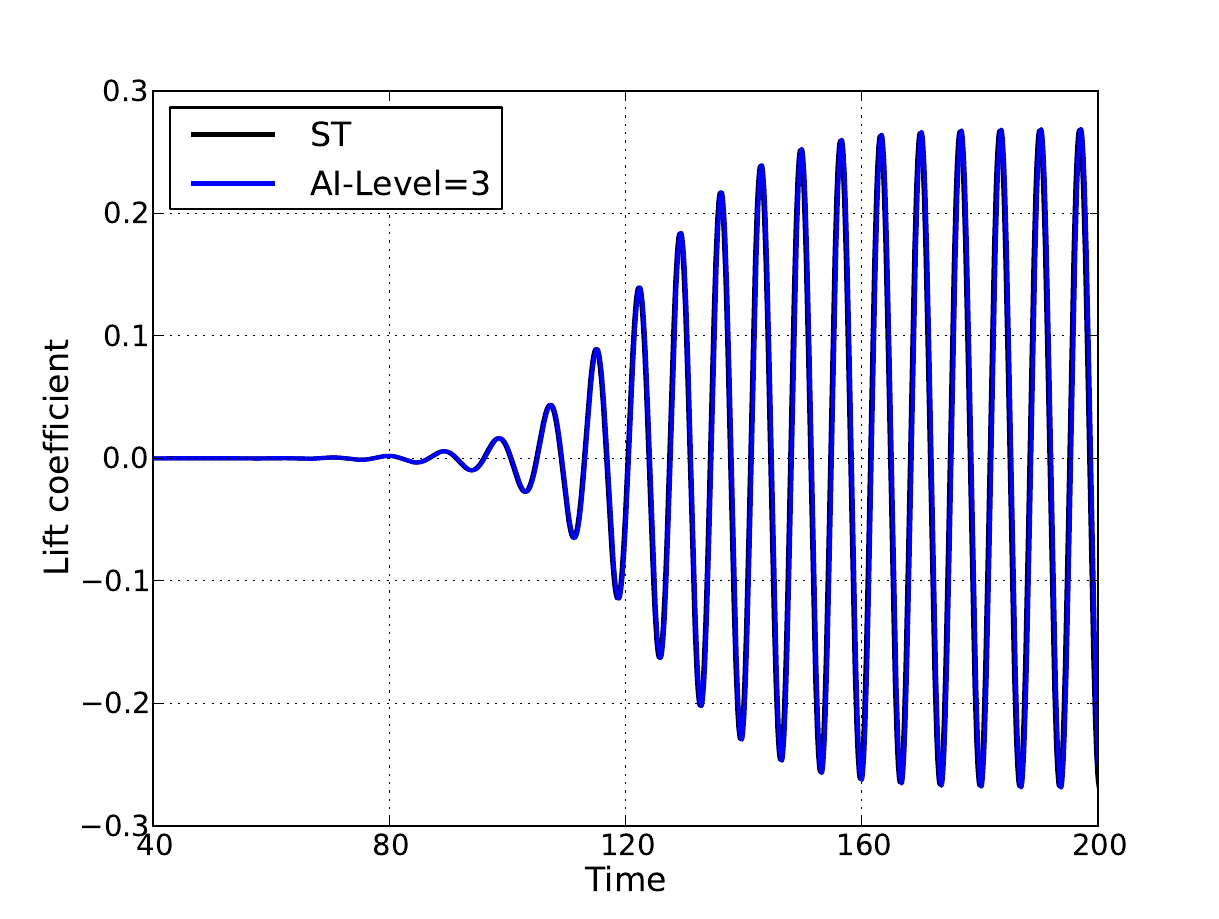}
  \includegraphics[trim = 0mm 0mm 0mm 0mm, clip, scale=0.35]{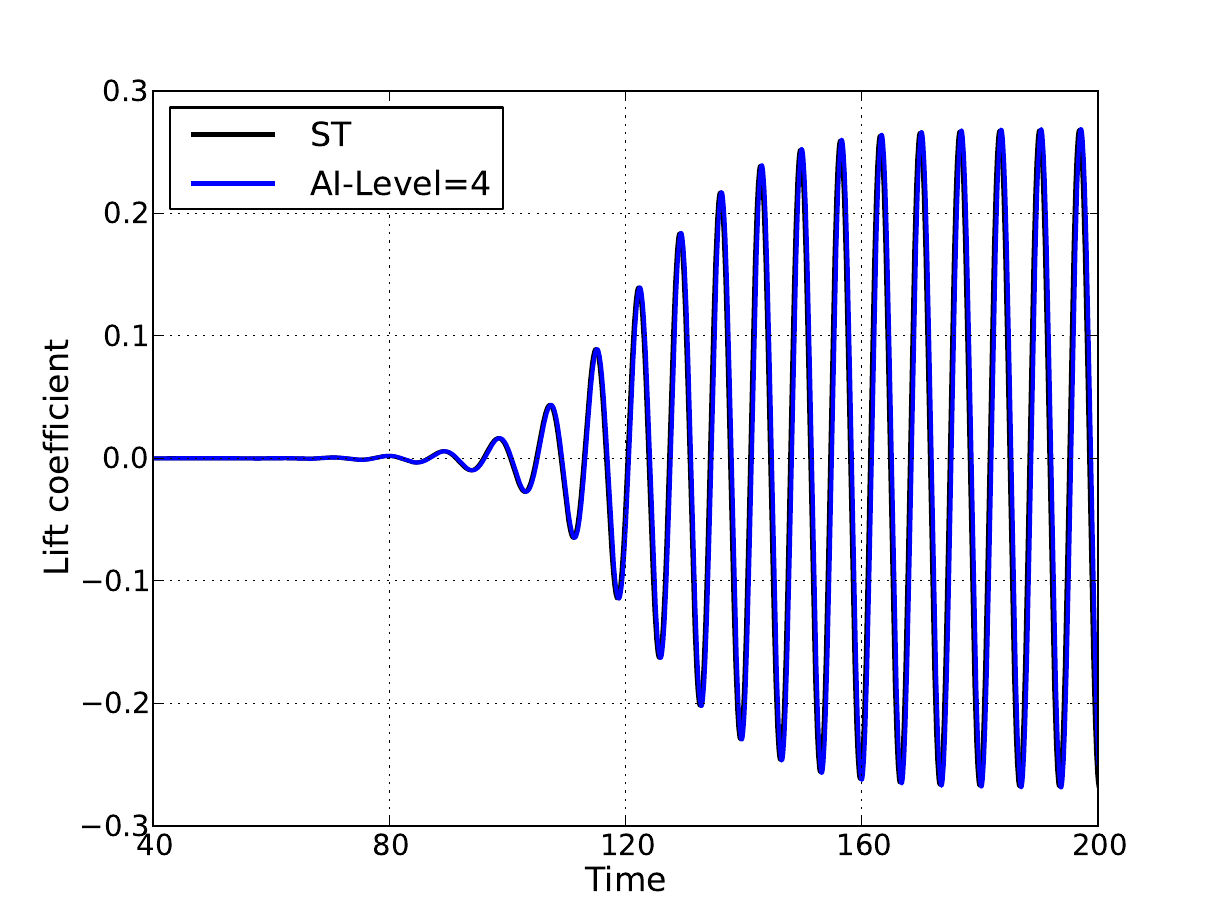}
  \includegraphics[trim = 0mm 0mm 0mm 0mm, clip, scale=0.35]{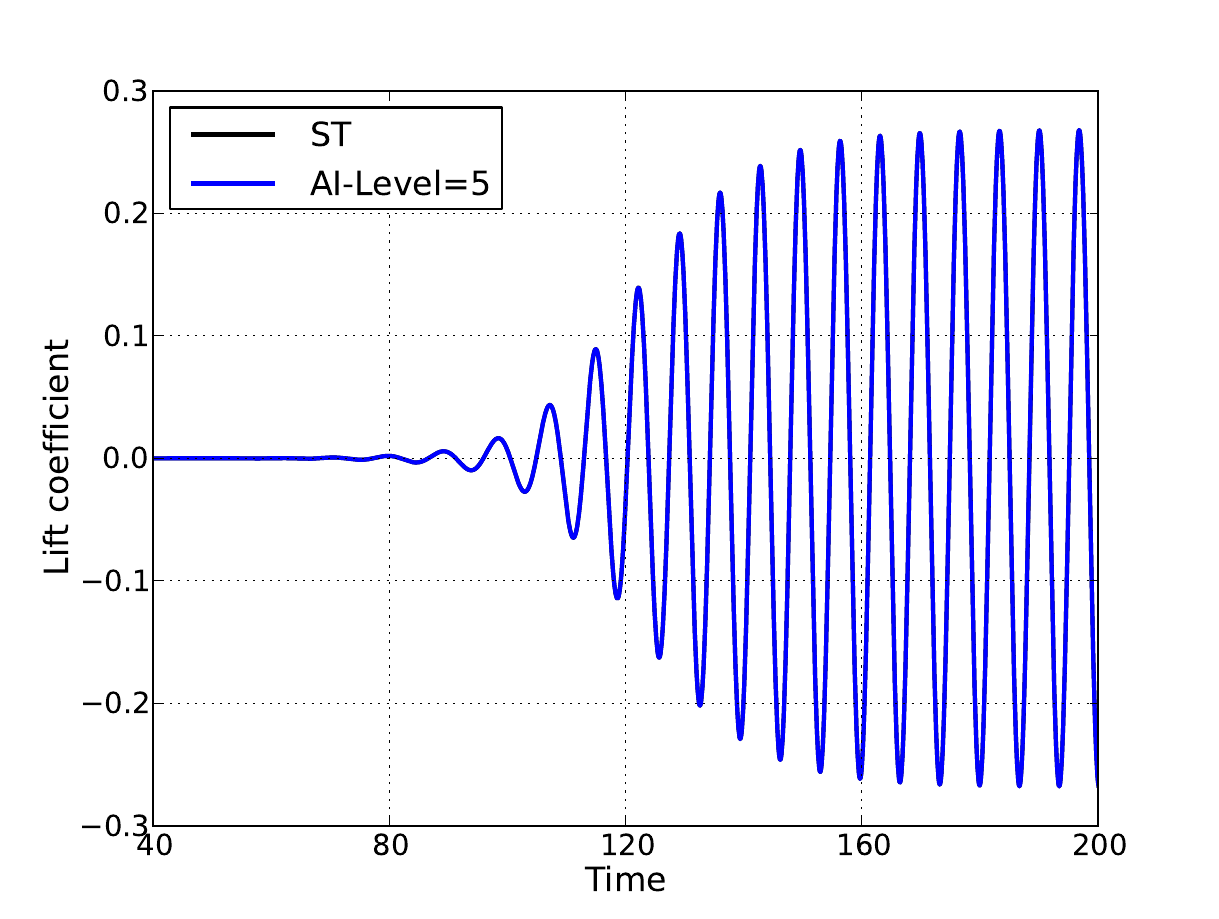}
  \caption{Flow past a fixed square at $Re=100$: evolution of lift coefficient with different levels of adaptive integration for $Q_1$ b-splines.}
  \label{fixed-square-re100-q1}
 \end{center}
\end{figure}

\begin{figure}[H]
 \begin{center}
  \includegraphics[trim = 0mm 0mm 0mm 0mm, clip, scale=0.35]{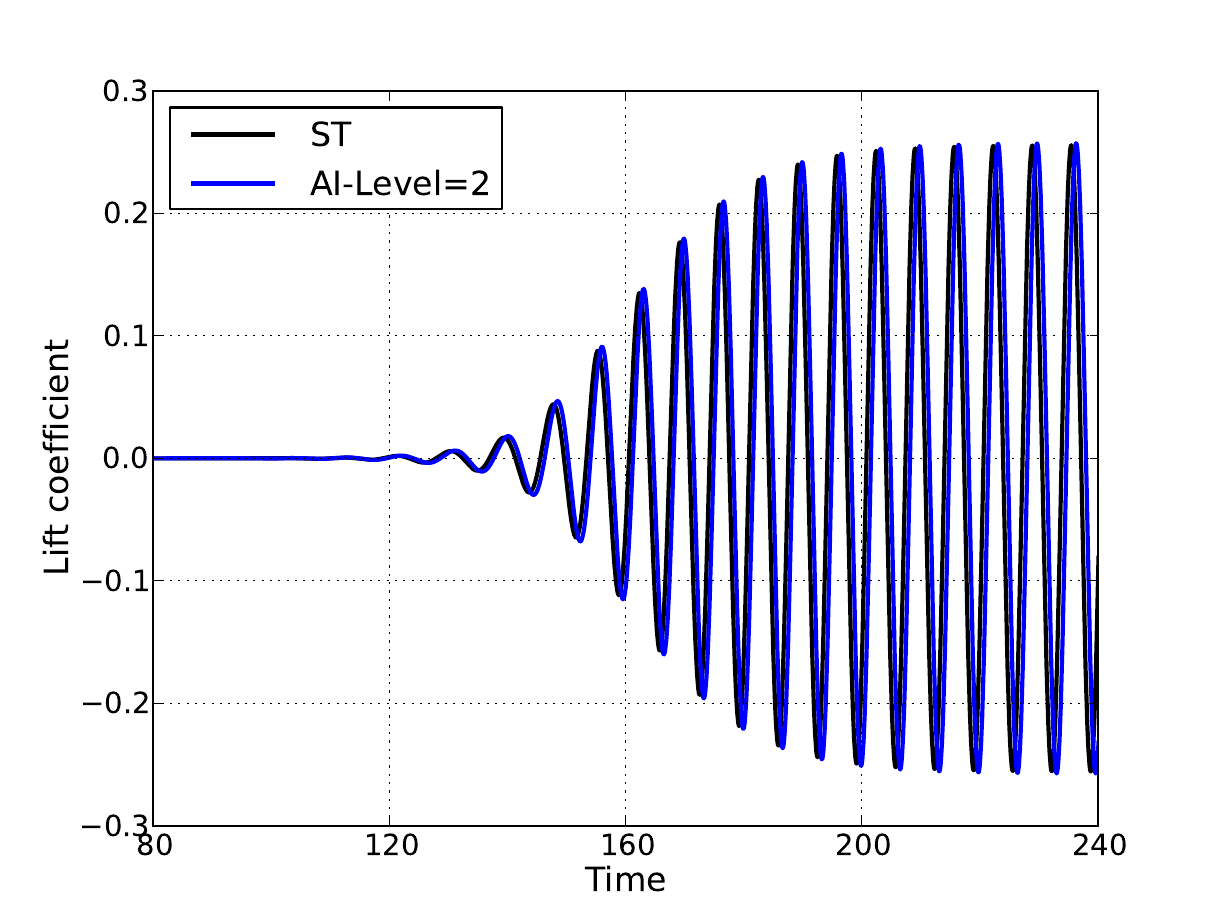}
  \includegraphics[trim = 0mm 0mm 0mm 0mm, clip, scale=0.35]{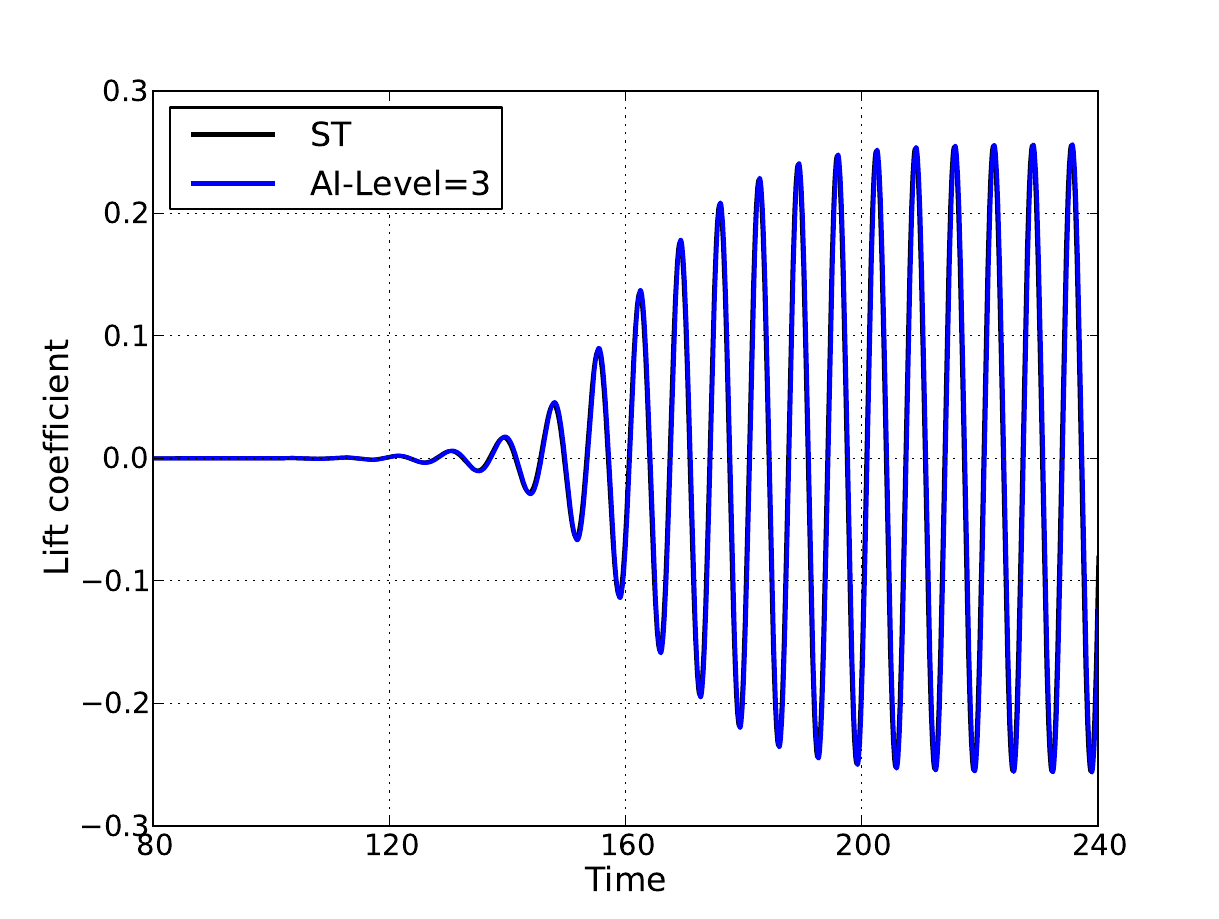}
  \includegraphics[trim = 0mm 0mm 0mm 0mm, clip, scale=0.35]{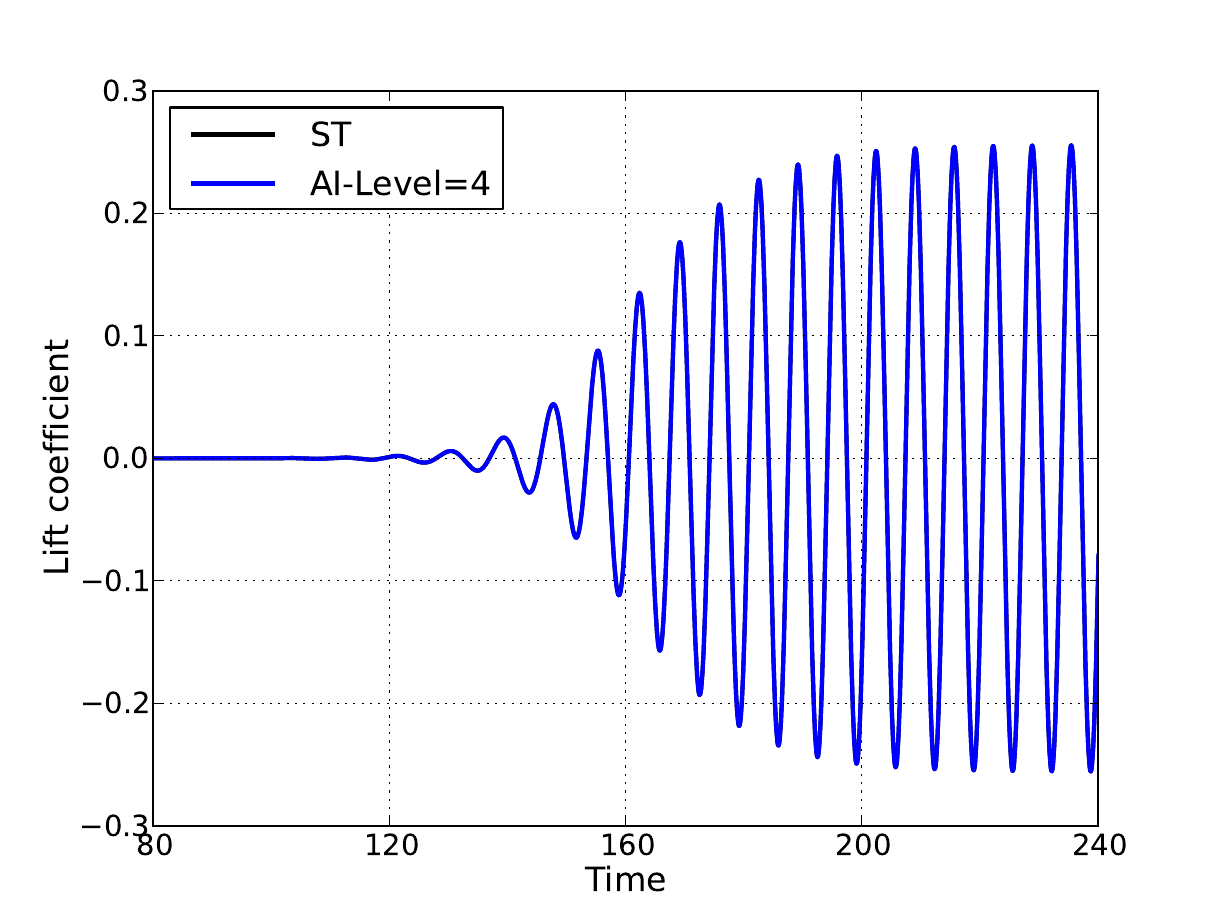}
  \includegraphics[trim = 0mm 0mm 0mm 0mm, clip, scale=0.35]{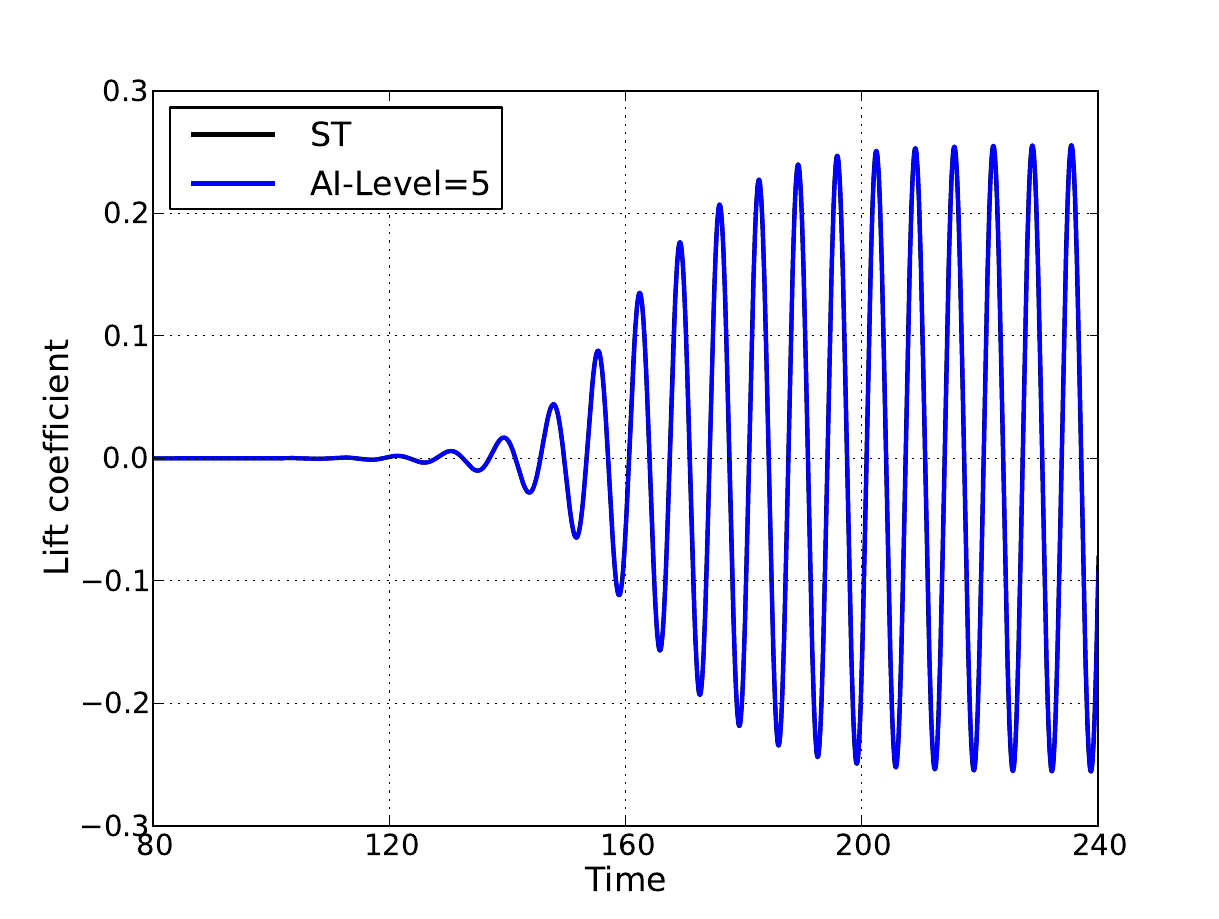}
 \caption{Flow past a fixed square at $Re=100$: evolution of lift coefficient with different levels of adaptive integration for $Q_2$ b-splines.}
 \label{fixed-square-re100-q2}
 \end{center}
\end{figure}

Figure \ref{fixed-square-re100-q1} shows the evolution of lift coefficient obtained with different levels of adaptive integration for $Q_1$ b-splines, and Fig. \ref{fixed-square-re100-q2} shows the corresponding graphs for $Q_2$ b-splines. As shown, the graphs of lift coefficient obtained with three and higher levels of adaptive integration match well with those obtained with subtriangulation. The mean value of $C_D$, root mean square value of $C_L$ and $St$ are in good agreement with the values from the literature. It is also worth pointing out that graphs of forces are free from spurious oscillations.

The results obtained for this and previous examples show that it is not necessary to use excessive levels of adaptive integration in order to obtain force coefficients and vortex shedding frequency that match well with the ones obtained with exact integration using subtriangulation. For both the orders of b-spline discretisations considered, three levels of adaptive integration are sufficient to compute force coefficient and frequencies of acceptable accuracy.

%
%
%
%
%
%
\subsection{Unsteady flow past a fixed cylinder in 3D}
\begin{figure}[H]
\centering
 \includegraphics[trim=0mm 0mm 0mm 0mm, clip, scale=1.2]{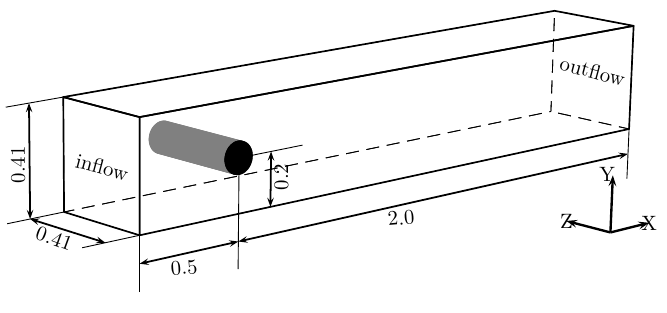}
 \caption{Flow past a fixed cylinder in 3D: geometry and boundary conditions.}
 \label{fig-cylinder3d-geom}
\end{figure}

We now assess the performance of adaptive integration for computing the forces for 3D problems using the example of unsteady flow past a fixed circular cylinder proposed in Schafer et al. \cite{SchaferCylinder3D}. The setup of the problem is depicted in Fig. \ref{fig-cylinder3d-geom}. The density and viscosity of the fluid, respectively, are $\rho=1.0$ and $\mu=0.001$. The velocity profile at the inlet is given as,
\begin{equation}
v_x^f(0,y,z,t) = \frac{16}{H^4} \, U_m \, y \, z \, (H-y) \, (H-z) \, \sin(\pi \, t/8); \quad v_y^f=v_z^f=0,
\end{equation}
with $U_m=2.25$. The time internal of interest is $0 \leq t \leq 8$ during which $0 \leq Re \leq 100$.

\begin{figure}[H]
\centering
 \subfloat[]{\includegraphics[trim=0mm 0mm 0mm 0mm, clip, scale=0.4]{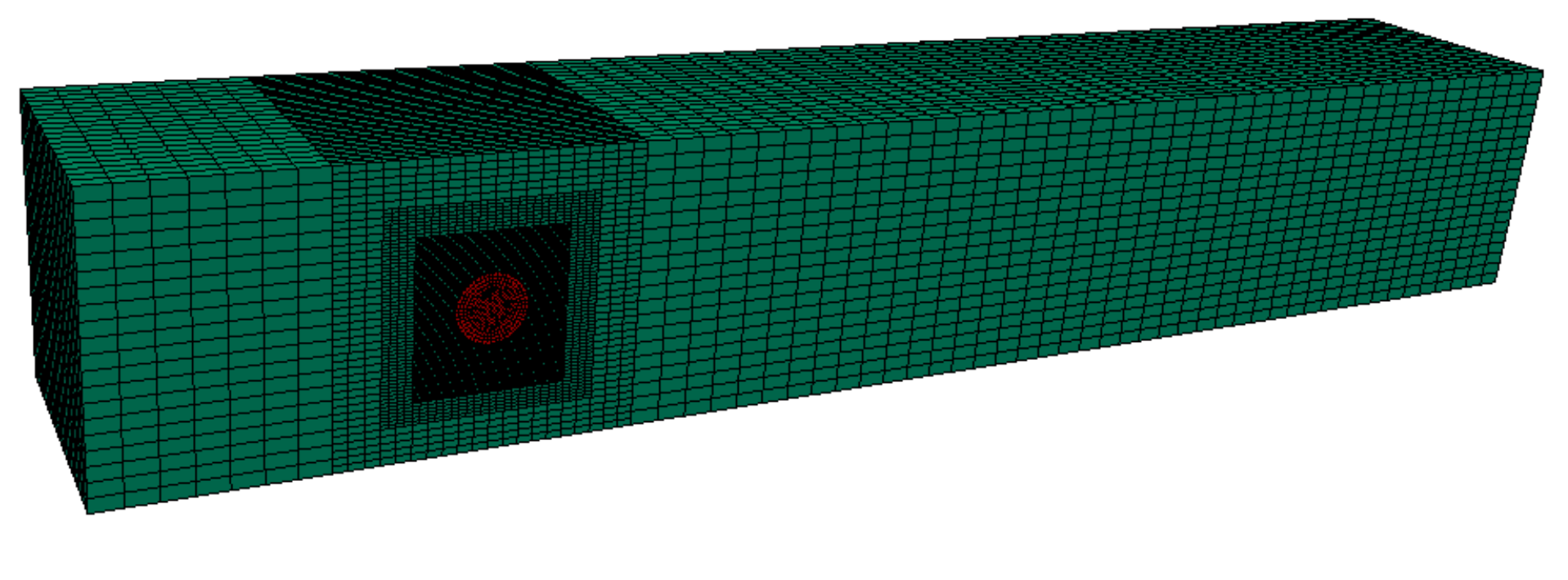}} \\
 \subfloat[]{\includegraphics[trim=0mm 0mm 0mm 0mm, clip, scale=0.2]{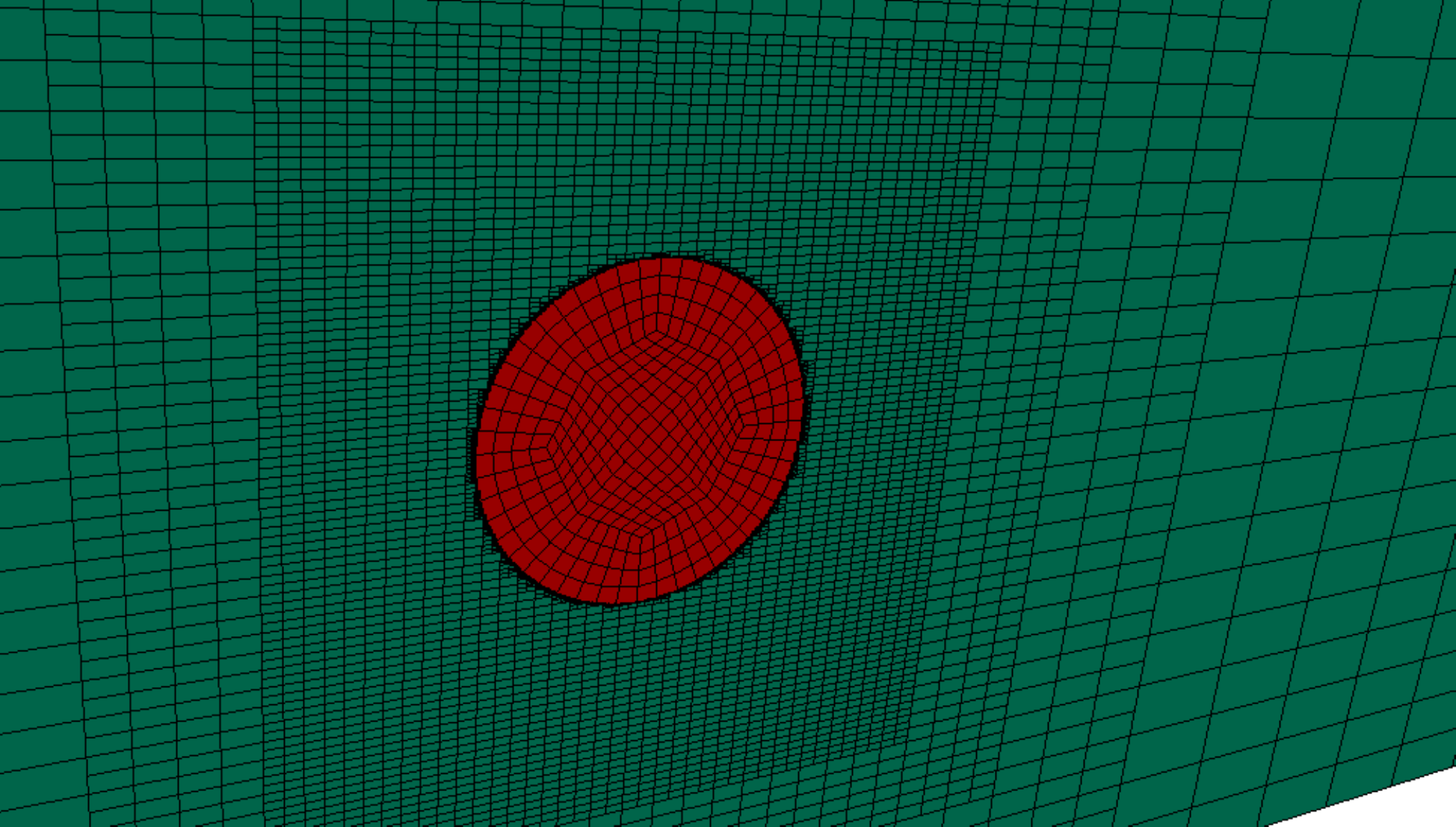}}
 \subfloat[]{\includegraphics[trim=0mm 0mm 0mm 0mm, clip, scale=0.4]{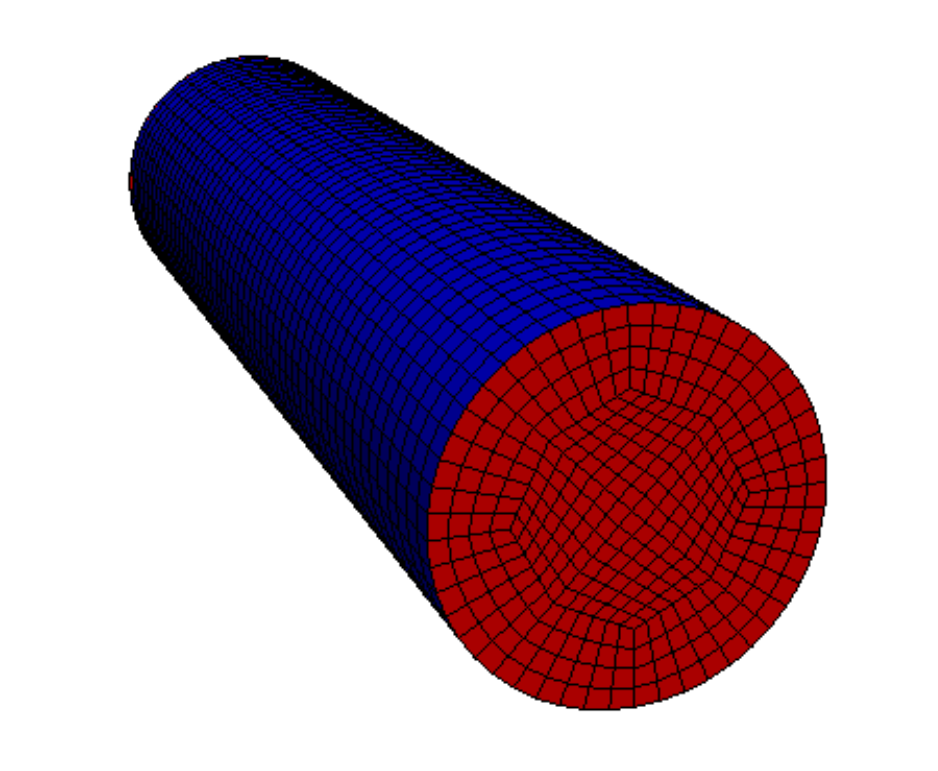}}
 \caption{Flow past a fixed cylinder in 3D: (a) fluid mesh with three levels of hierarchical refinement, (b) zoomed view of the fluid mesh in the vicinity of the cylinder and (c) discretisation used for the surface of the cylinder.}
 \label{fig-cylinder3d-mesh}
\end{figure}

A hierarchically refined mesh with levels of local refinement in the vicinity of the cylinder as shown in Fig.  \ref{fig-cylinder3d-mesh} is used with $Q_1$ b-splines, in which case the total number of degrees of freedom is 2272908. The surface of the cylinder is discretised with 3072 linear quadrilateral elements. The interface integrals are evaluated using 16 quadrature points per quadrilateral. Using a constant time step, $\Delta t = 0.08$ s, simulations are performed for different levels of adaptive refinement. Note that, due to the issues associated with the implementation of subtetrahedralisation, only adaptive integration is used in 3D. It is deemed appropriate based on the numerical experiments conducted and the results obtained.

\begin{figure}[H]
\centering
 \includegraphics[trim=0mm 0mm 0mm 0mm, clip, scale=0.8]{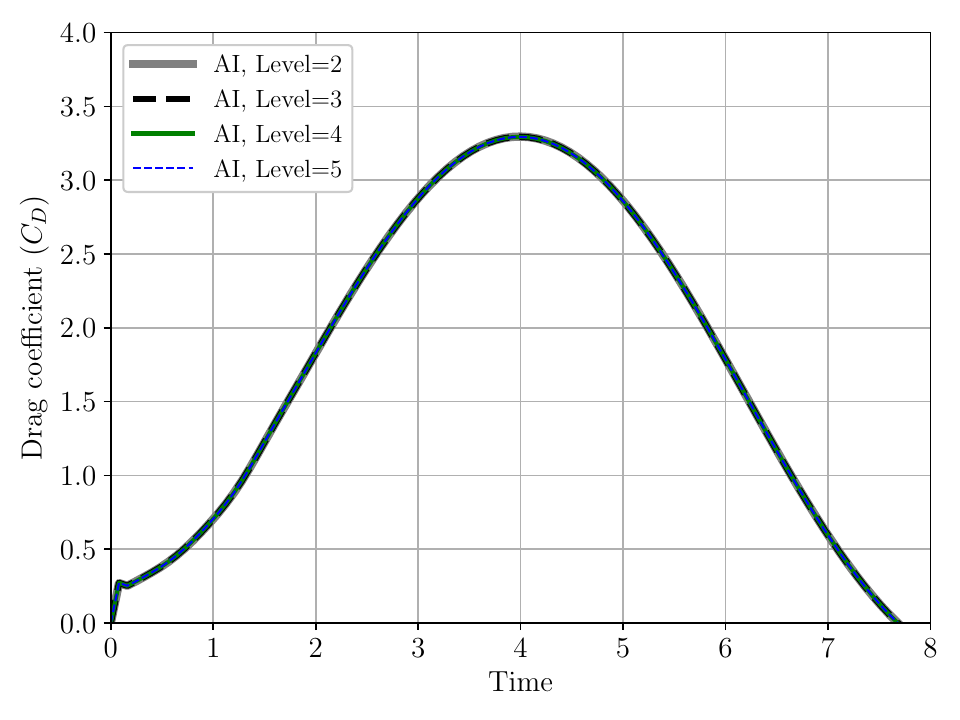}
 \caption{Flow past a fixed cylinder in 3D: drag coefficient obtained with different levels of adaptive integration.}
 \label{fig-cylinder3d-graph}
\end{figure}

As shown in Fig. \ref{fig-cylinder3d-graph}, there is no noticeable difference in the drag coefficient ($C_D$) values obtained with different levels of adaptive integration. It is also worth noting that the drag coefficient is free from spurious oscillations. The maximum values of $C_D$ obtained in the present work agree well with the reference values, as shown in Table. \ref{table-cylinder3d}.


\renewcommand{\arraystretch}{1.2}
\begin{table}[H]
\centering
\begin{tabular}{|p{6cm}|p{3cm}|}
\hline
Data             &  $C_{D_{max}}$ \\
\hline
Sch\"afer and Turek \cite{SchaferCylinder3D} &   3.2 - 3.3  \\
\hline
Present - AI, Level=2    & 3.2954 \\
Present - AI, Level=3    & 3.2937 \\
Present - AI, Level=4    & 3.2934 \\
Present - AI, Level=5    & 3.2935 \\
\hline
\end{tabular}
\caption{Flow past a fixed cylinder in 3D: maximum values of drag coefficient obtained with different levels of adaptive integration.}
 \label{table-cylinder3d}
\end{table}
\renewcommand{\arraystretch}{1.0}

\subsection{Turek-Hron FSI benchmark problem in 2D}
\begin{figure}[H]
 \begin{center}
  \includegraphics[trim = 0mm 0mm 0mm 0mm, clip,scale=0.9]{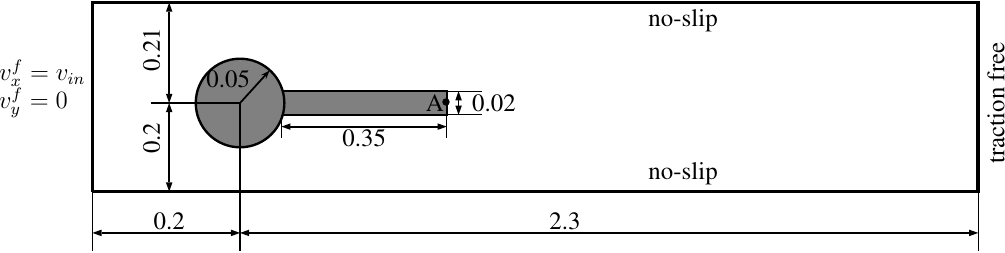}
  \caption{FSI2 benchmark in 2D: geometry and boundary conditions of the problem.}
  \label{fig-Turek-geom}
 \end{center}
\end{figure}

In this example, we assess the accuracy of adaptive integration for FSI problems using the benchmark problem for fluid-flexible solid interaction proposed by Turek and Hron \cite{TurekFSIflex2006}. The geometry and boundary conditions of the problem are as shown in Fig.\ref{fig-Turek-geom}. The density of the fluid is ${\rho_f} = {10^3}$ kg/m$^3$ and its viscosity is ${\mu_f} = 1 $ Pa s. The material properties of the solid are: density, ${\rho_s} = {10^4}$ kg/m$^3$, Young’s modulus, $E_s = 1.4 \times {10^6}$ N/m$^2$, and Poisson’s ratio, and $\nu  = 0.4$. The Saint Venant-Kirchhoff constitutive model is used to model the finite strain deformation behaviour of the solid. The horizontal velocity at the inlet is of parabolic type, given as, $v_{in}=\frac{6}{0.1681} y [0.41-y]$. The inlet velocity is increased sinusoidally to $v_{in}$ during the first second and then kept constant at $v_{in}$ for the rest of the simulation.

\begin{figure}[H]
 \begin{center}
  \includegraphics[trim = 0mm 0mm 0mm 0mm, clip,scale=0.4]{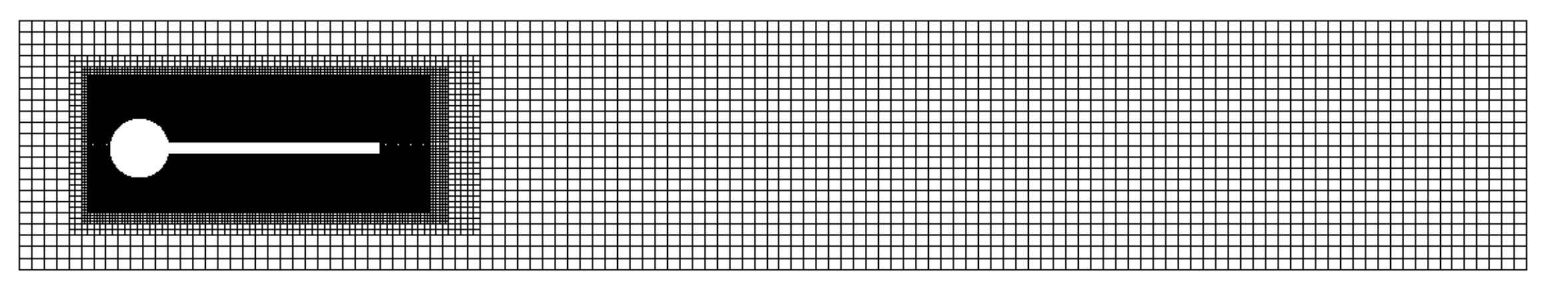}
  \caption{FSI2 benchmark in 2D: hierarchically refined mesh used for the fluid domain.}
  \label{fig-Turek-mesh}
 \end{center}
\end{figure}

Starting with a mesh of $121 \times 22$ elements, the background mesh for the fluid domain is refined locally up to three levels in the vicinity of the solid body, as shown in Fig. \ref{fig-Turek-mesh}. The beam part of the solid is discretised with $200 \times 10$ bilinear quadrilateral elements. The relaxation parameter for the staggered scheme is $\beta=0.05$, and the uniform time step size considered for all the simulations is $\Delta t=0.002$ s. Simulations are conducted for level-2 and level-3 meshes using $Q_1$ and $Q_2$ b-splines with subtriangulation and different levels of adaptive integration. The accuracy of the results obtained with adaptive integration is assessed by comparing the displacement of point A (see Fig. \ref{fig-Turek-geom}) and total force on the solid in the Y-direction against the ones obtained with sub-triangulation as well as the results from the literature. The force response is smoothed using a three-point moving average.

\begin{figure}[H]
 \begin{center}
  \includegraphics[trim = 0mm 20mm 0mm 30mm, clip, scale=0.5]{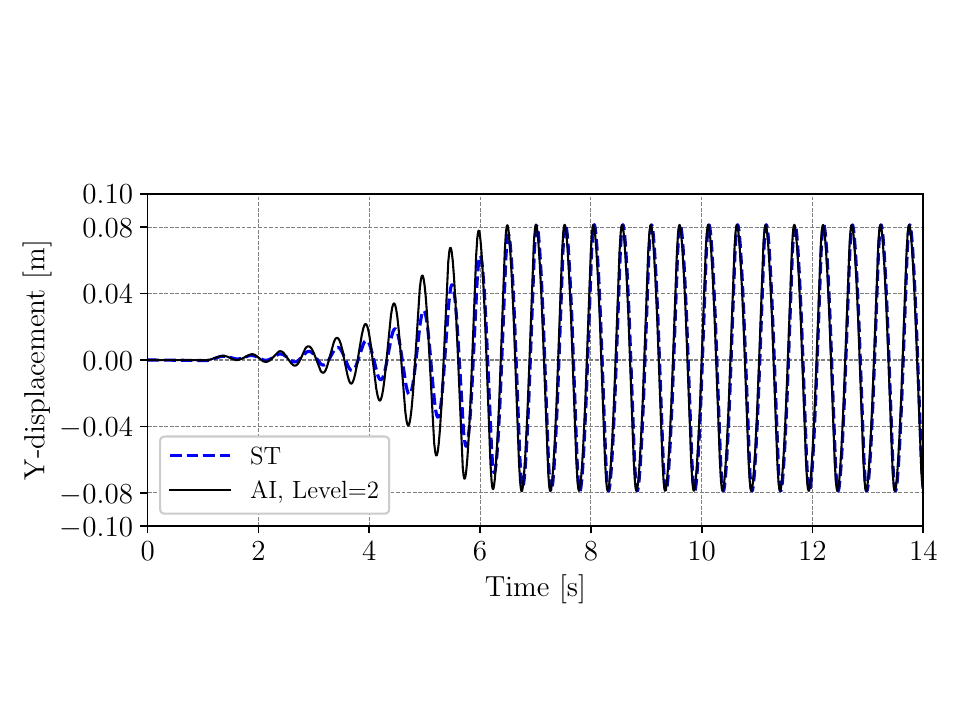}
  \includegraphics[trim = 0mm 20mm 0mm 30mm, clip, scale=0.5]{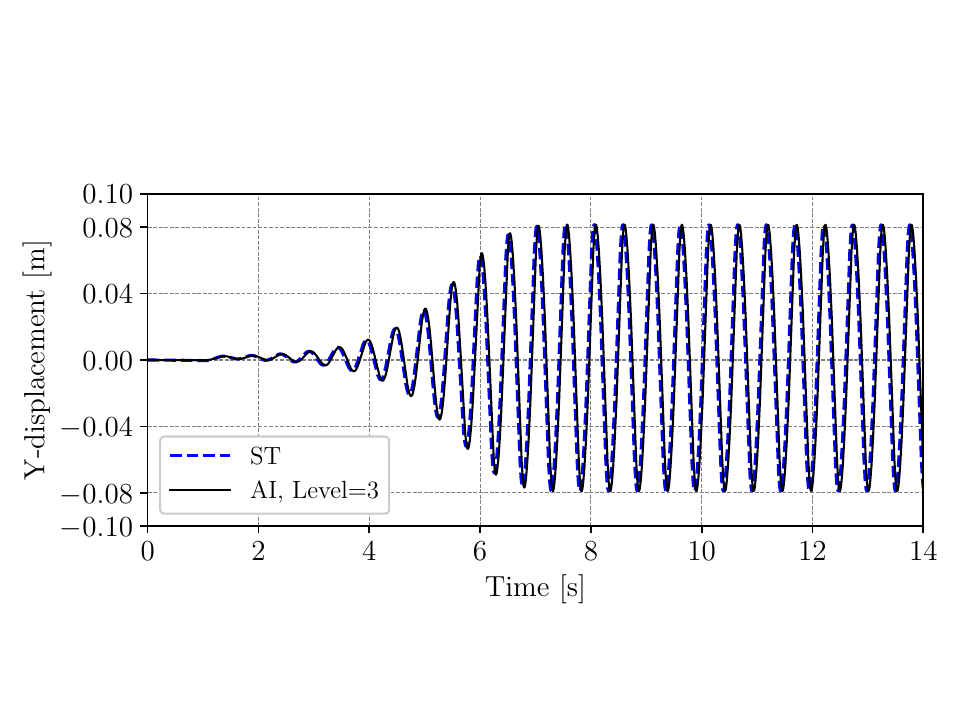}
  \includegraphics[trim = 0mm 20mm 0mm 30mm, clip, scale=0.5]{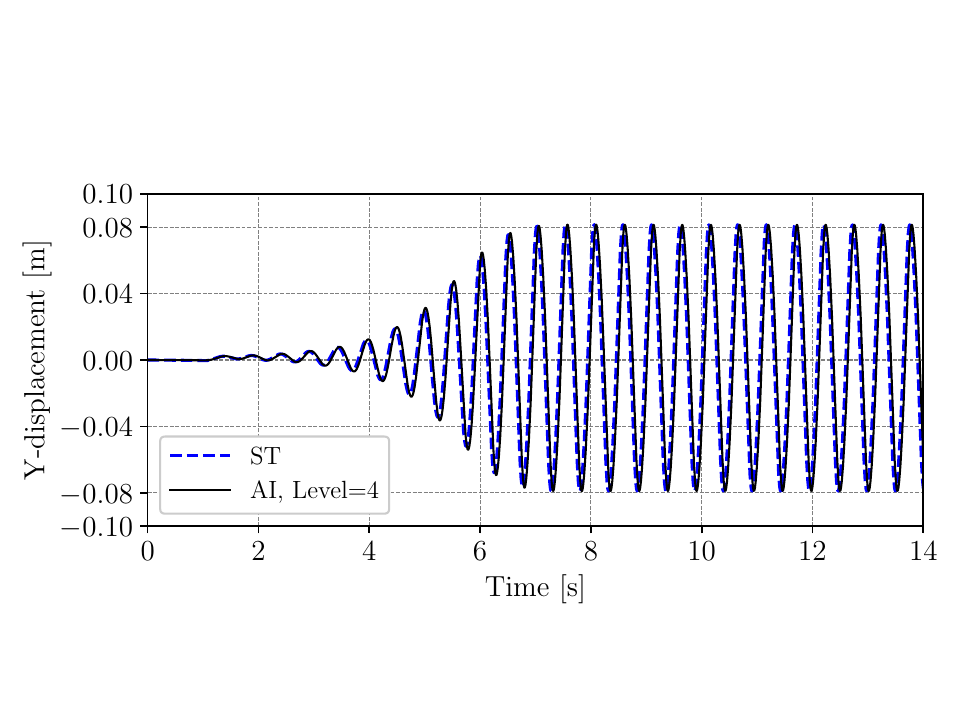}
  \includegraphics[trim = 0mm 20mm 0mm 30mm, clip, scale=0.5]{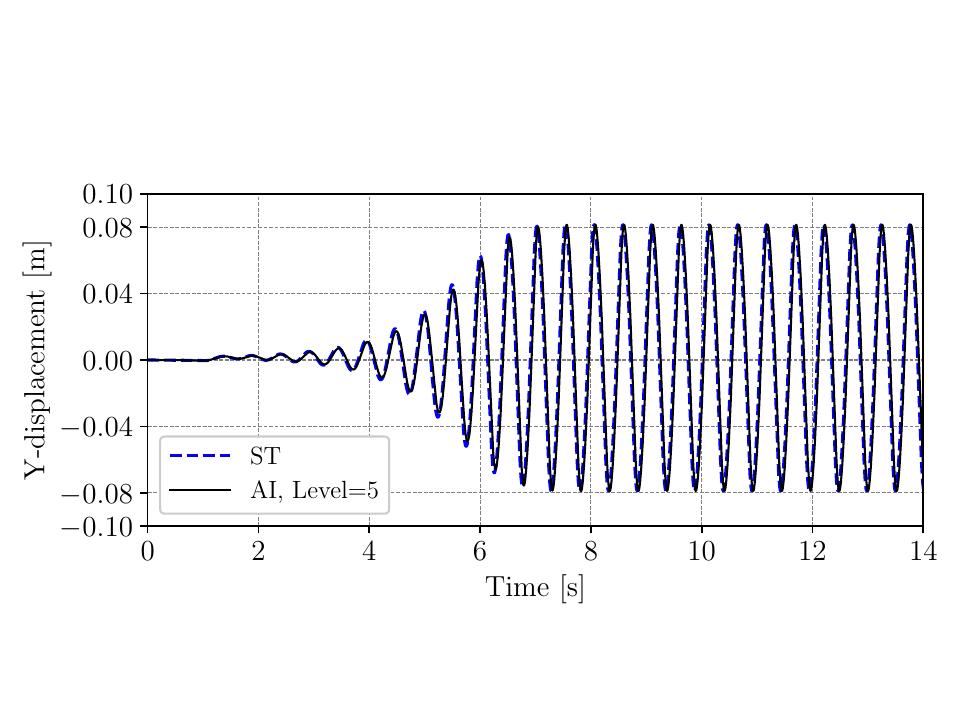}
  \caption{FSI2 benchmark in 2D: evolution of vertical displacement of point A for different levels of adaptive integration obtained with $Q_1$ b-splines and level-2 mesh.}
 \label{fig-Turek-disp-l2-q1-disp}
 \end{center}
\end{figure}

\begin{figure}[H]
 \begin{center}
  \includegraphics[trim = 0mm 10mm 0mm 25mm, clip, scale=0.5]{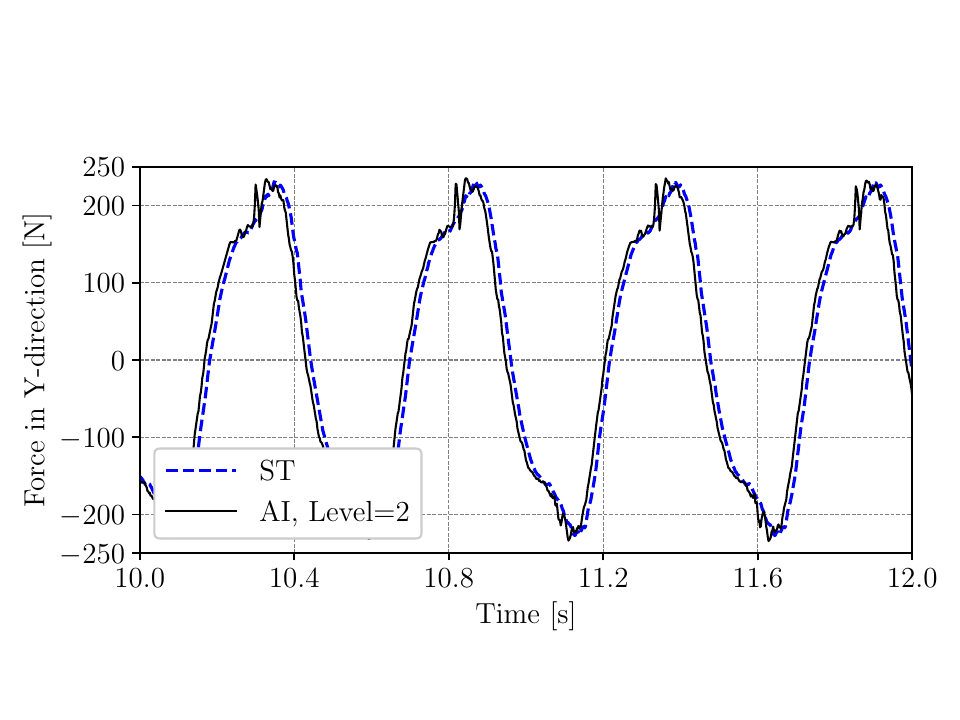}
  \includegraphics[trim = 0mm 10mm 0mm 25mm, clip, scale=0.5]{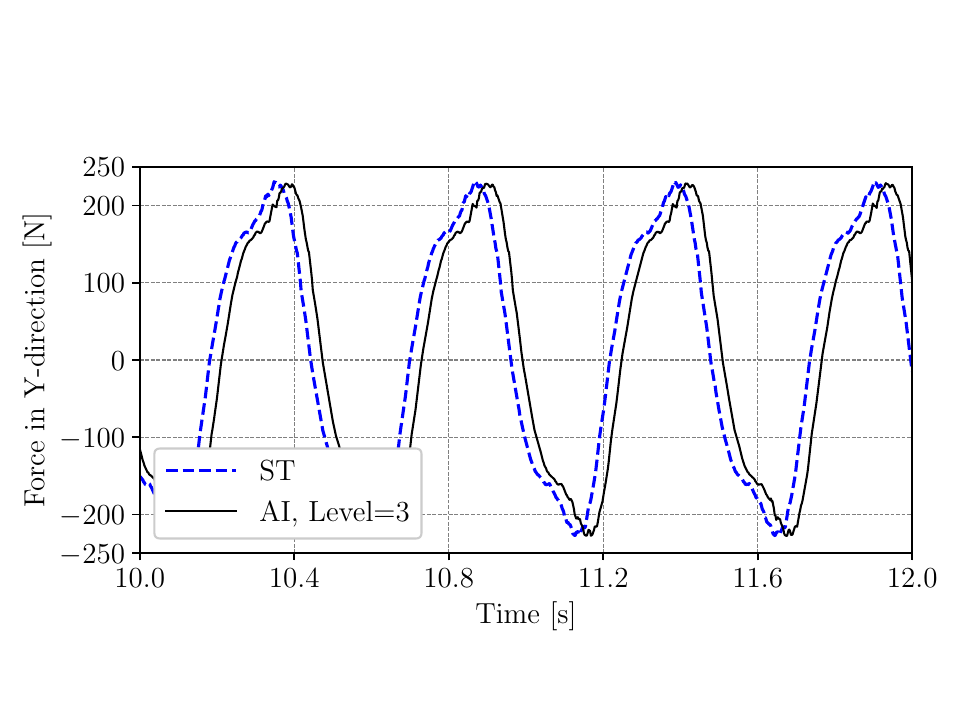}
  \includegraphics[trim = 0mm 10mm 0mm 25mm, clip, scale=0.5]{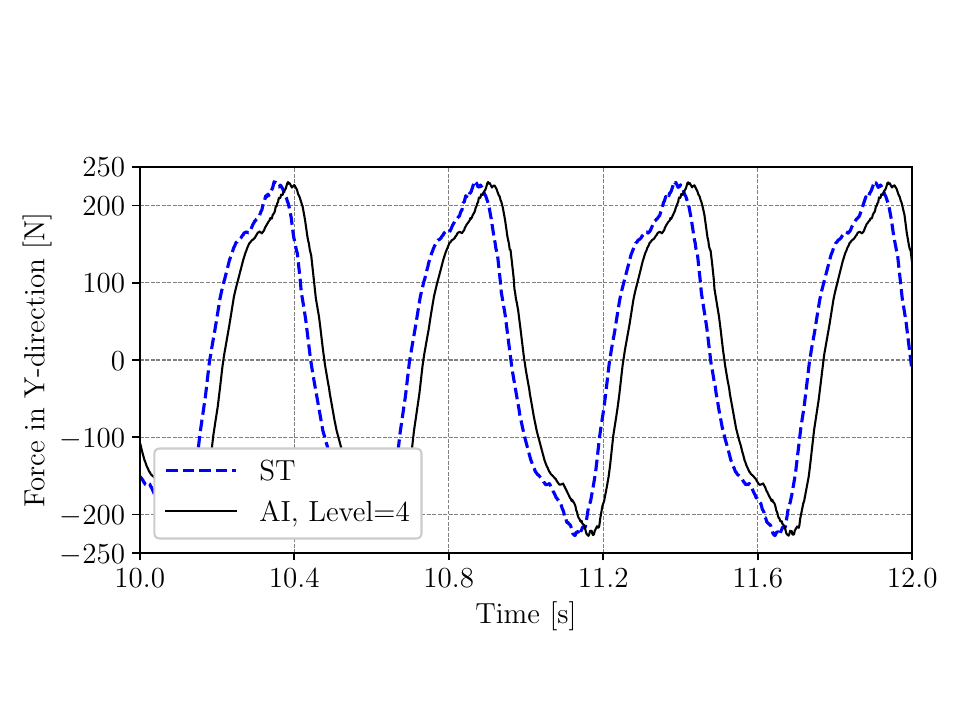}
  \includegraphics[trim = 0mm 10mm 0mm 25mm, clip, scale=0.5]{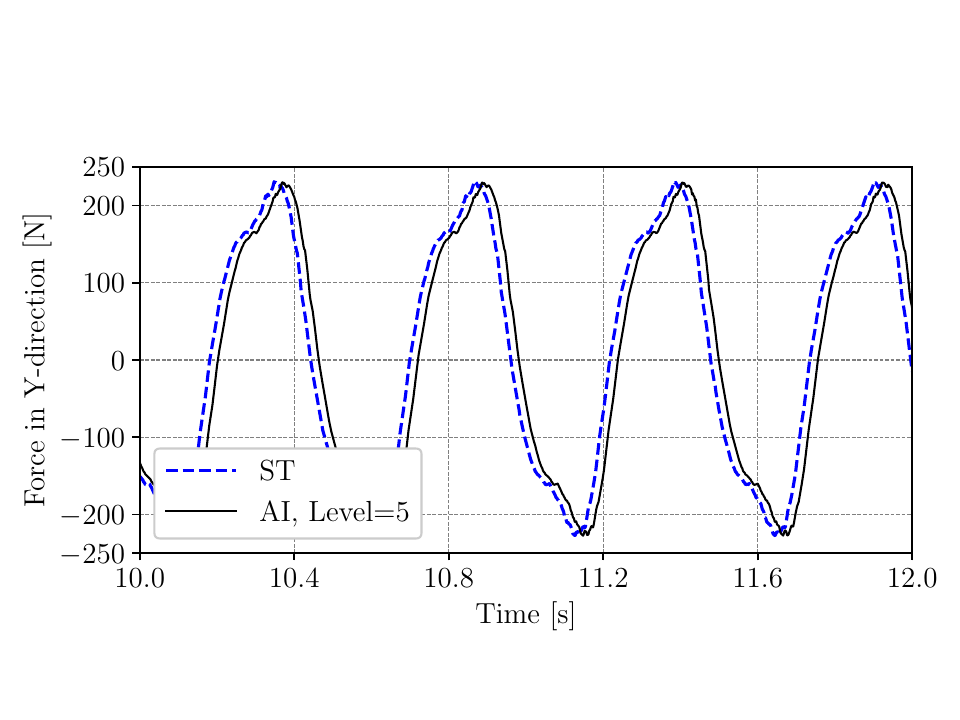}
  \caption{FSI2 benchmark in 2D: evolution of total vertical force on the solid for different levels of adaptive integration obtained with $Q_1$ b-splines and level-2 mesh.}
 \label{fig-Turek-disp-l2-q1-force}
 \end{center}
\end{figure}

\begin{figure}[H]
 \begin{center}
  \includegraphics[trim = 0mm 20mm 0mm 30mm, clip, scale=0.5]{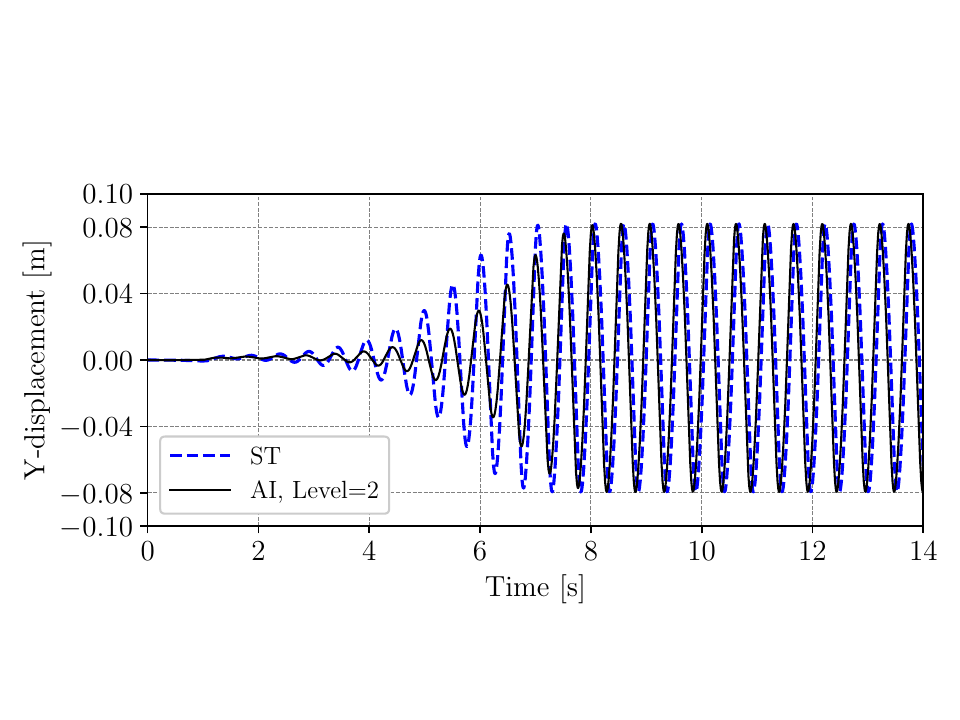}
  \includegraphics[trim = 0mm 20mm 0mm 30mm, clip, scale=0.5]{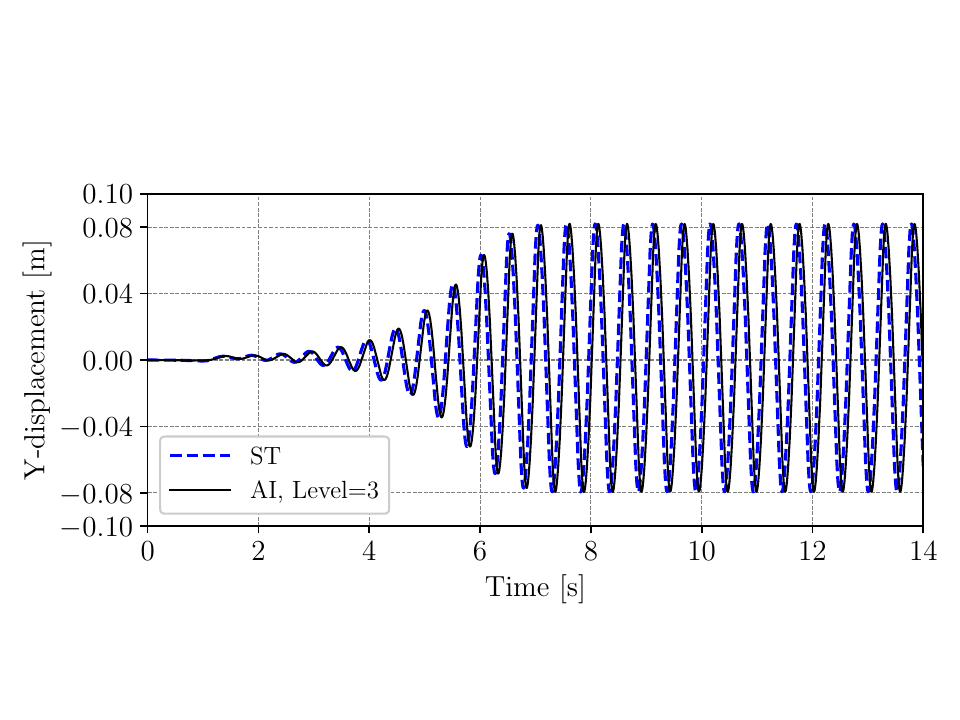}
  \includegraphics[trim = 0mm 20mm 0mm 30mm, clip, scale=0.5]{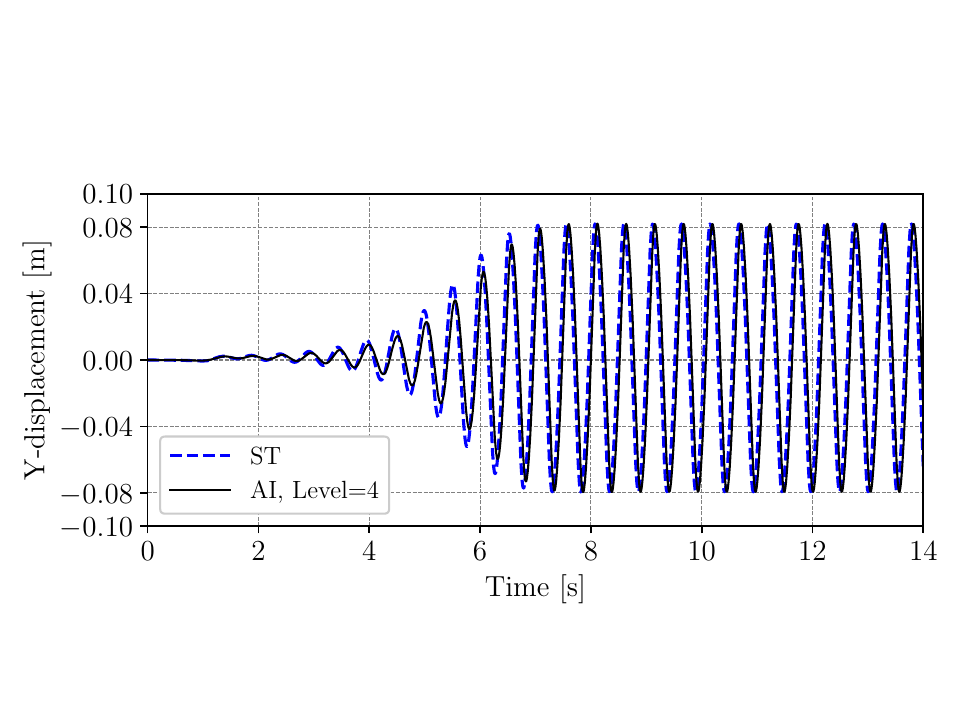}
  \includegraphics[trim = 0mm 20mm 0mm 30mm, clip, scale=0.5]{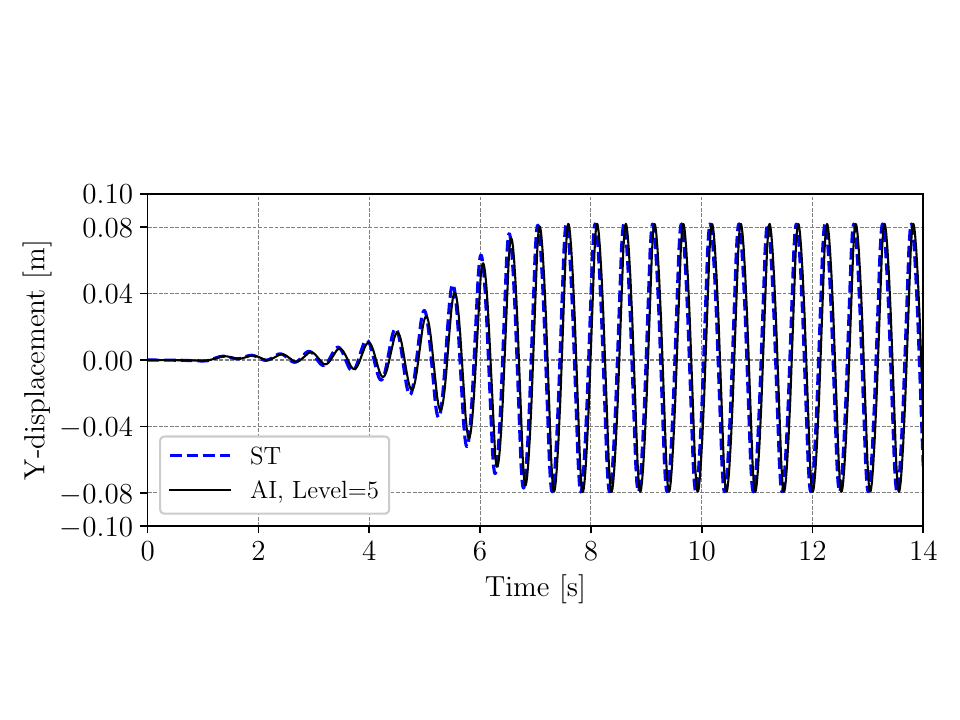}
  \caption{FSI2 benchmark in 2D: evolution of vertical displacement of point A for different levels of adaptive integration obtained with $Q_1$ b-splines and level-3 mesh.}
 \label{fig-Turek-disp-l3-q1-disp}
 \end{center}
\end{figure}

\begin{figure}[H]
 \begin{center}
  \includegraphics[trim = 0mm 10mm 0mm 25mm, clip, scale=0.5]{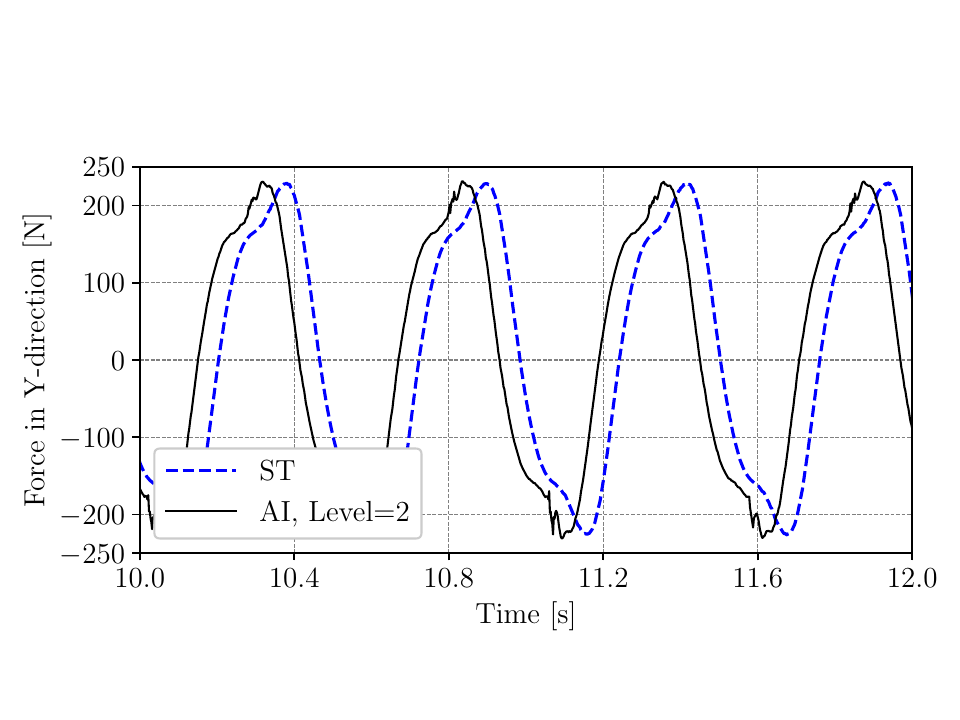}
  \includegraphics[trim = 0mm 10mm 0mm 25mm, clip, scale=0.5]{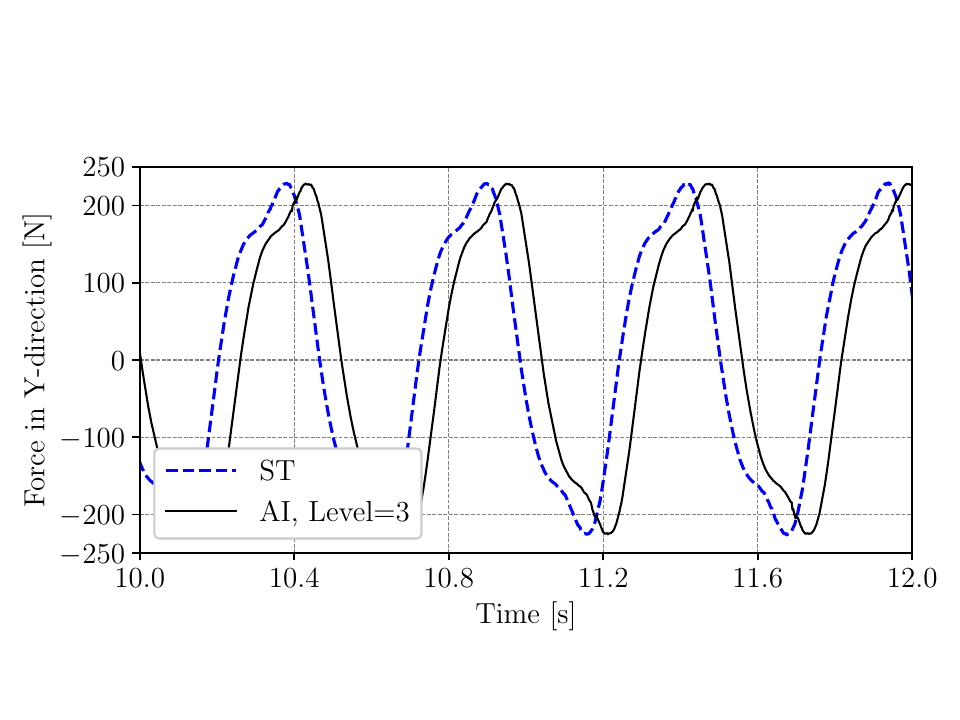}
  \includegraphics[trim = 0mm 10mm 0mm 25mm, clip, scale=0.5]{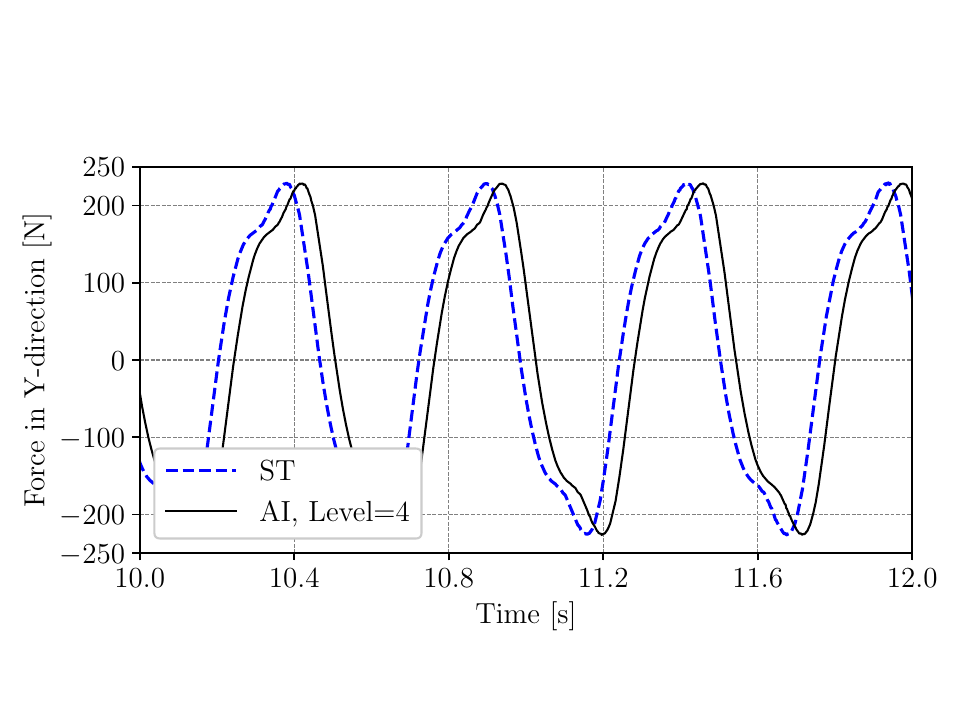}
  \includegraphics[trim = 0mm 10mm 0mm 25mm, clip, scale=0.5]{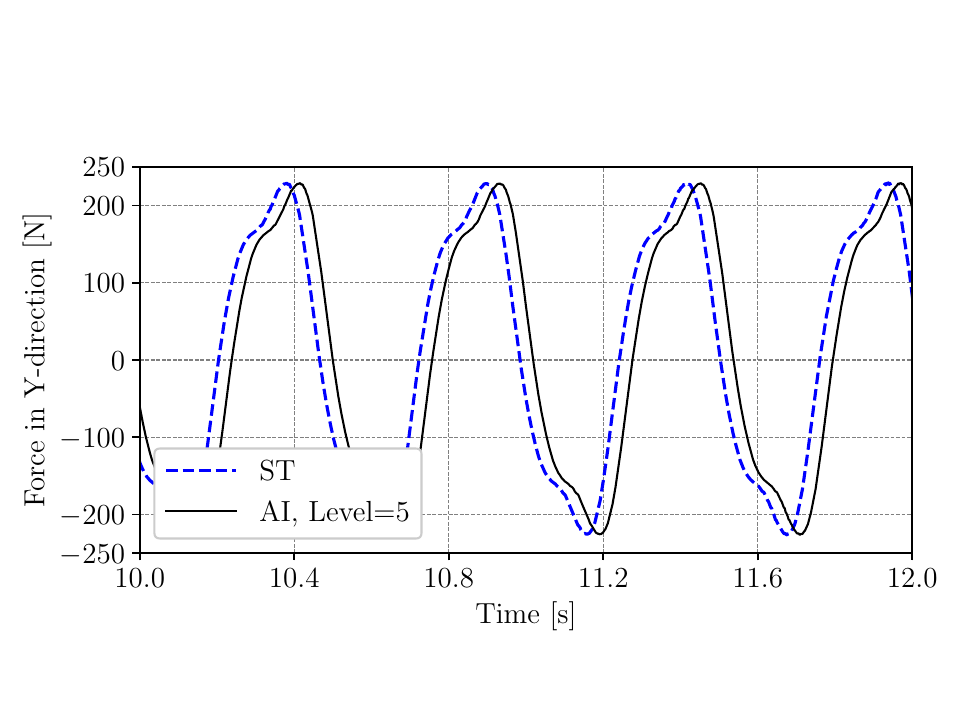}
  \caption{FSI2 benchmark in 2D: evolution of total vertical force on the solid for different levels of adaptive integration obtained with $Q_1$ b-splines and level-3 mesh.}
 \label{fig-Turek-disp-l3-q1-force}
 \end{center}
\end{figure}

\begin{figure}[H]
 \begin{center}
  \includegraphics[trim = 0mm 20mm 0mm 30mm, clip, scale=0.5]{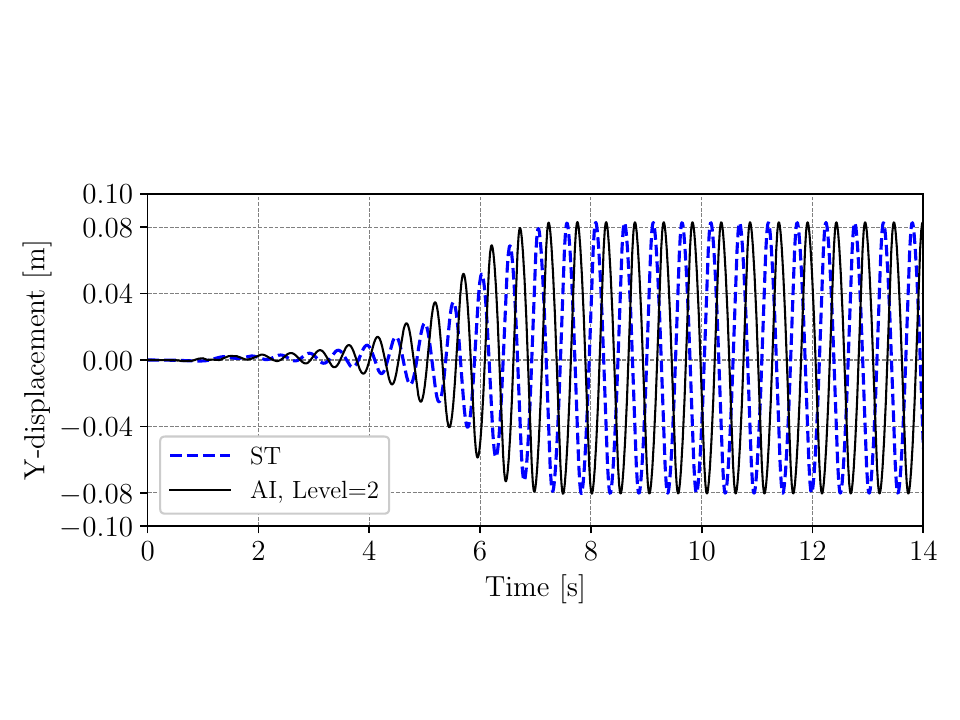}
  \includegraphics[trim = 0mm 20mm 0mm 30mm, clip, scale=0.5]{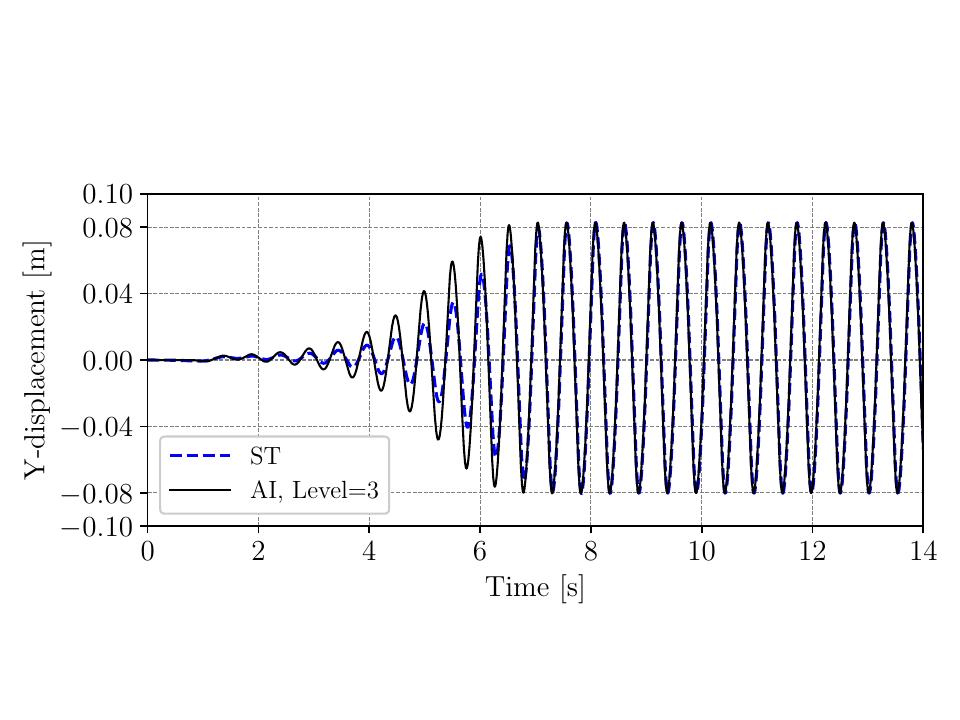}
  \includegraphics[trim = 0mm 20mm 0mm 30mm, clip, scale=0.5]{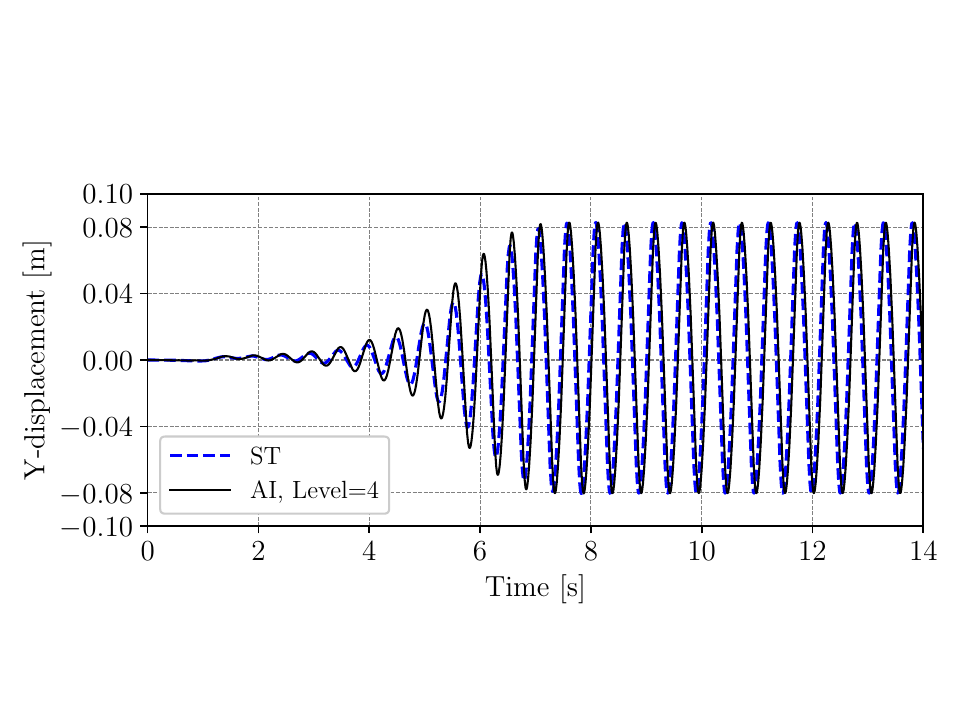}
  \includegraphics[trim = 0mm 20mm 0mm 30mm, clip, scale=0.5]{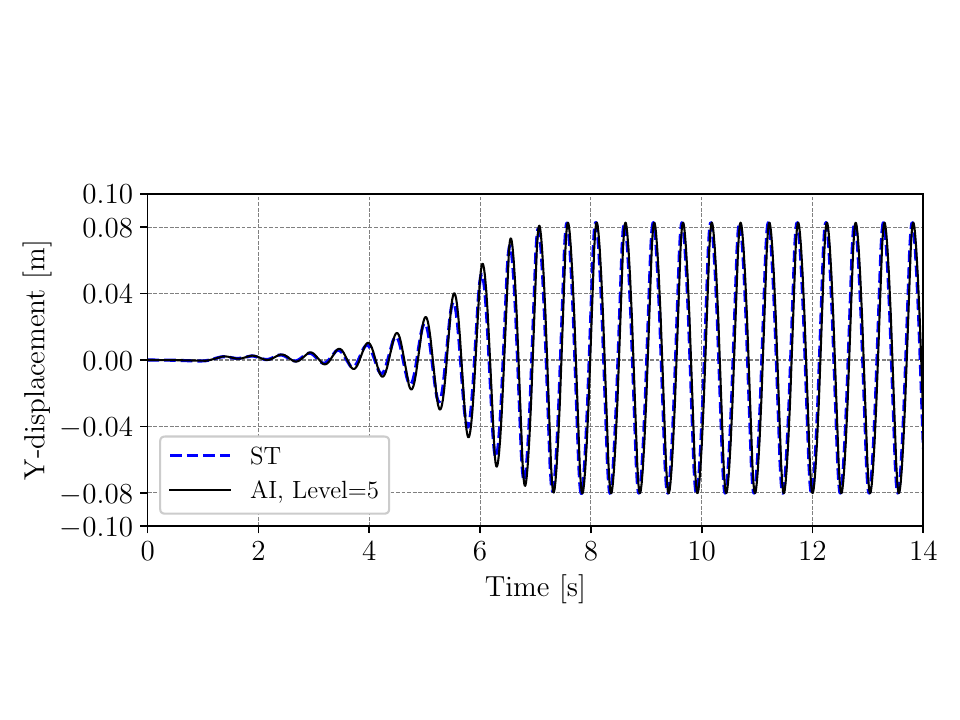}
  \caption{FSI2 benchmark in 2D: evolution of vertical displacement of point A for different levels of adaptive integration obtained with $Q_2$ b-splines and level-2 mesh.}
 \label{fig-Turek-disp-l2-q2-disp}
 \end{center}
\end{figure}

\begin{figure}[H]
 \begin{center}
  \includegraphics[trim = 0mm 10mm 0mm 25mm, clip, scale=0.5]{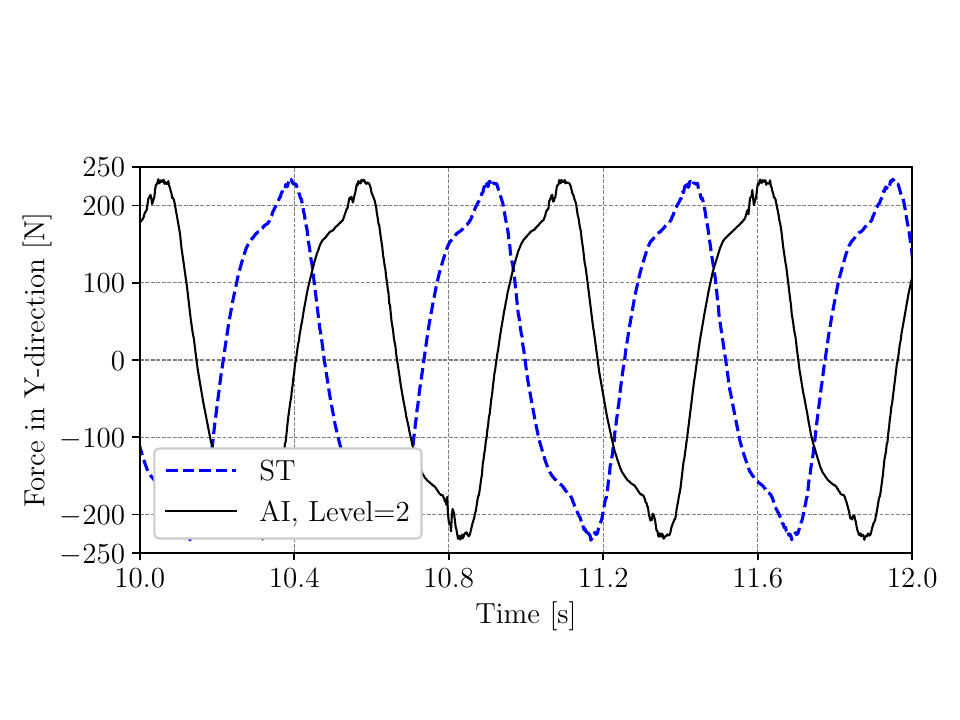}
  \includegraphics[trim = 0mm 10mm 0mm 25mm, clip, scale=0.5]{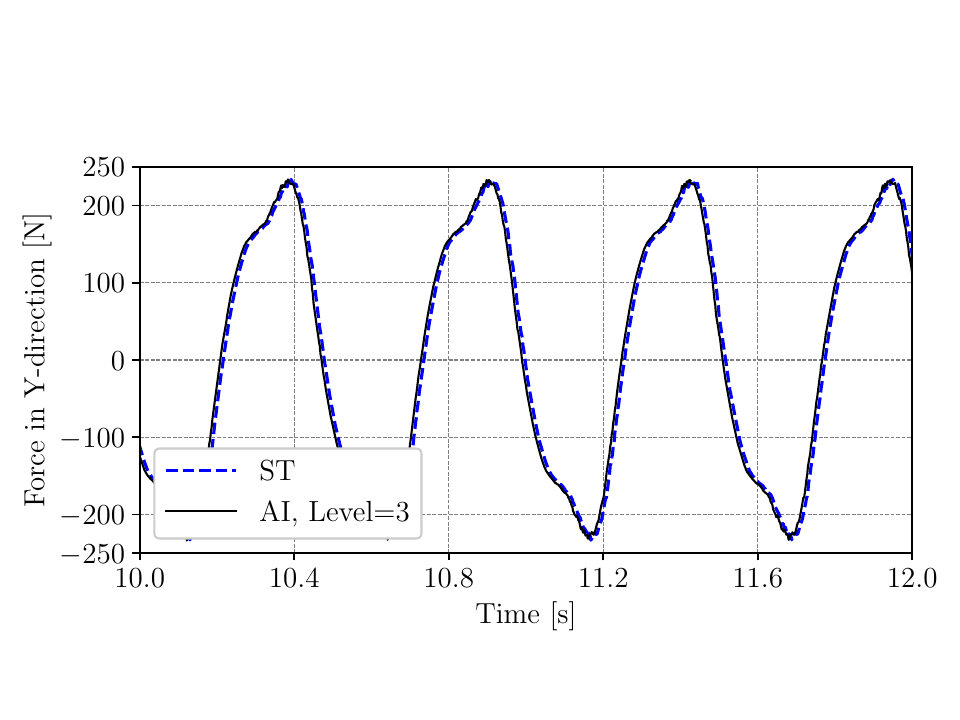}
  \includegraphics[trim = 0mm 10mm 0mm 25mm, clip, scale=0.5]{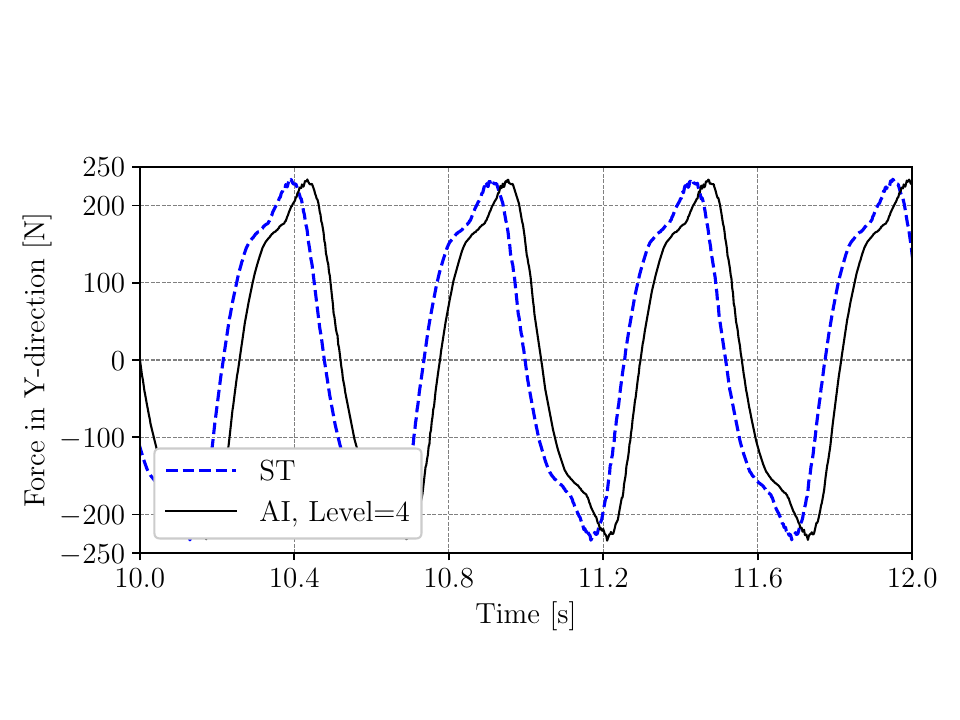}
  \includegraphics[trim = 0mm 10mm 0mm 25mm, clip, scale=0.5]{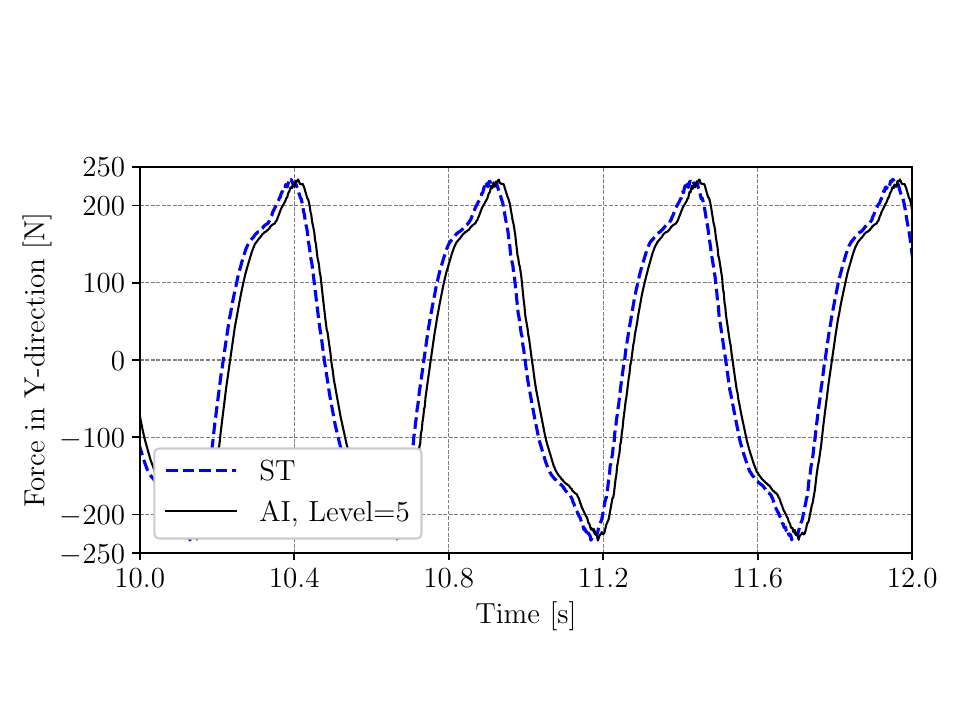}
  \caption{FSI2 benchmark in 2D: evolution of total vertical force on the solid for different levels of adaptive integration obtained with $Q_2$ b-splines and level-2 mesh.}
 \label{fig-Turek-disp-l2-q2-force}
 \end{center}
\end{figure}

\begin{figure}[H]
\begin{center}
	\includegraphics[trim = 0mm 10mm 0mm 25mm, clip, scale=0.5]{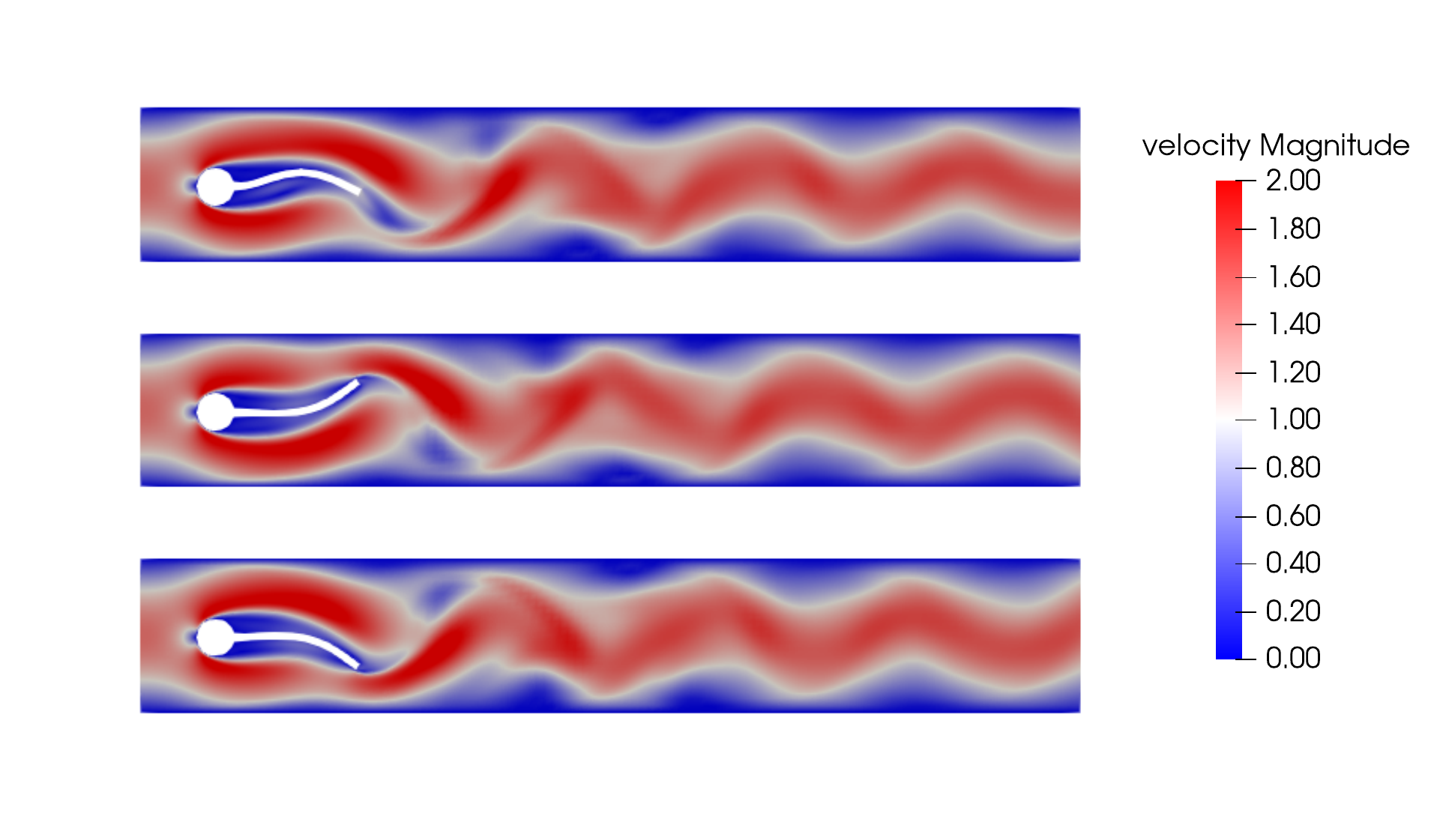}
	\caption{FSI2 benchmark in 2D: contour plots of velocity magnitude at $t=8$ s (top), $t=8.14$ s (middle) and $t=8.4$ s (bottom) obtained with $Q_1$ b-splines and level-3 mesh using level-3 adaptive integration.}
	\label{fig-Turek-contours-velo}
\end{center}
\end{figure}

\begin{figure}[H]
\begin{center}
	\includegraphics[trim = 0mm 10mm 0mm 25mm, clip, scale=0.5]{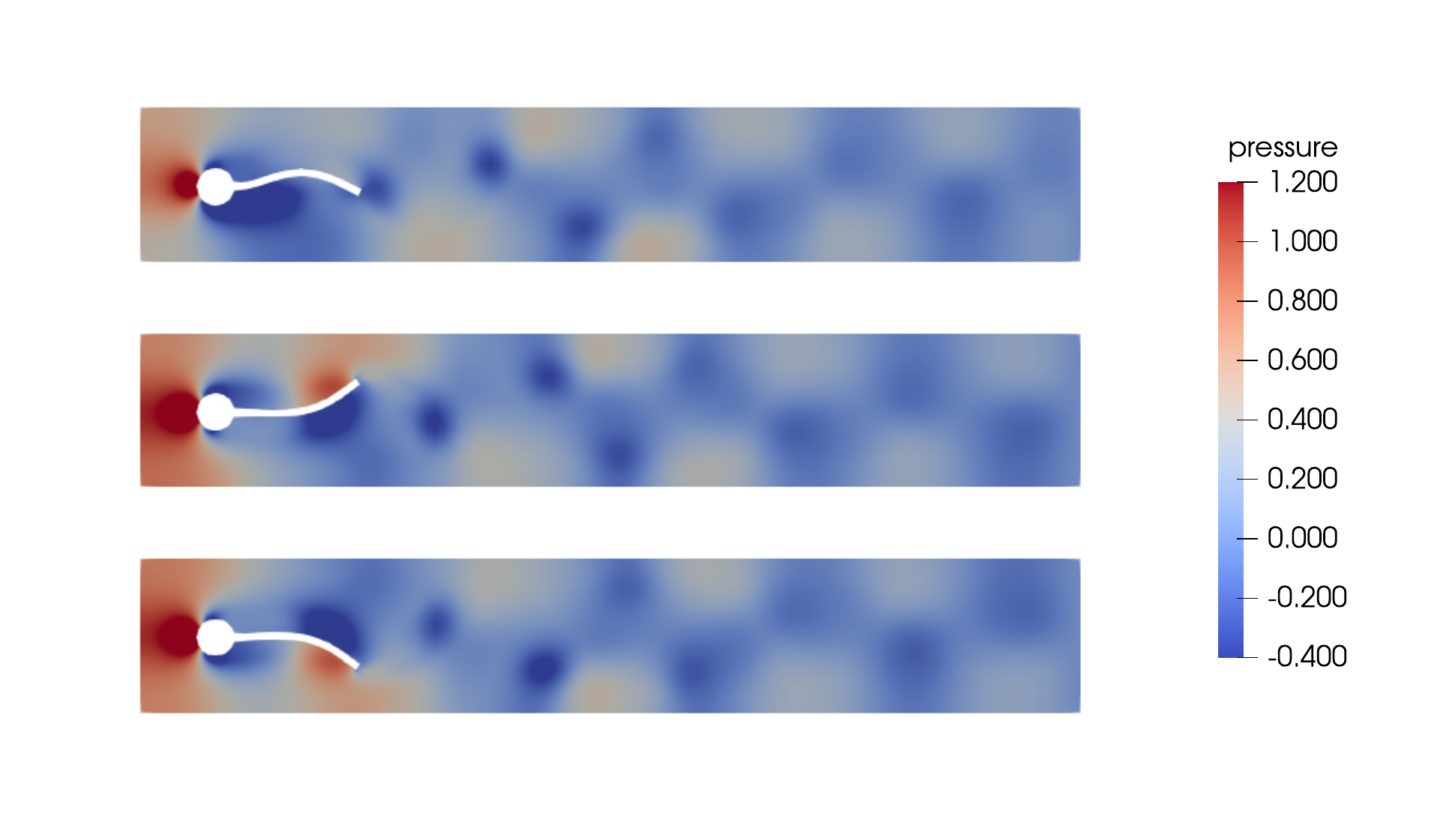}
	\caption{FSI2 benchmark in 2D: contour plots of pressure at $t=8$ s (top), $t=8.14$ s (middle) and $t=8.4$ s (bottom) obtained with $Q_1$ b-splines and level-3 mesh using level-3 adaptive integration.}
	\label{fig-Turek-contours-pres}
\end{center}
\end{figure}

The evolution of the Y-displacement of point A and Y-component of force is shown, respectively, in Figs. \ref{fig-Turek-disp-l2-q1-disp} and \ref{fig-Turek-disp-l2-q1-force} for the level-2 mesh with $Q_1$ b-splines for four different levels of adaptive integration. The corresponding graphs for the level-3 mesh with $Q_1$ b-splines are shown in Figs. \ref{fig-Turek-disp-l3-q1-disp} and \ref{fig-Turek-disp-l3-q1-force}, and the ones for level-2 mesh and $Q_2$ b-splines are shown in Figs. \ref{fig-Turek-disp-l2-q2-disp} and \ref{fig-Turek-disp-l2-q2-force}. The amplitude and frequency are also compared in Table. \ref{table-Turekbeam}. As shown, the results obtained with three and higher levels of adaptive integration match well with those of subtriangulation. From the graphs for forces, we can observe the disappearance of spurious oscillations in the forces with three and higher levels of adaptive integration. Contour plots of velocity magnitude and pressure at three different time instants obtained with the level-3 mesh and $Q_1$ b-splines using three levels of adaptive integration are presented in Figs. \ref{fig-Turek-contours-velo} and \ref{fig-Turek-contours-pres}. The results obtained for this example indicate that three levels of adaptive integration are sufficient for computing accurate numerical results for laminar fluid-flexible structure interaction problems. As observed in the example of Kovasznay flow, the computational cost of using three levels of adaptive integration is comparable to that of subtriangulation.


\renewcommand{\arraystretch}{1.2}
\begin{table}[H]
\centering
\begin{tabular}{|l|c|c|c|c|}
\hline
 Data          &  max($d^s_y \times 10^3$) & $f_o$  & $F^s_y$ \\
\hline
Turek and Hron \cite{TurekFSIflex2006} - Level-4, $\Delta t=0.002$  &   $1.25 \pm 80.70$   &  2.00  & $0.97 \pm 233.2$  \\
\hline
Present - Level-2, $Q_1$, ST             &   $1.26 \pm 80.19$   &  1.95 & $1.12 \pm 234.4$  \\
Present - Level-2, $Q_1$, AI, Level=2    &   $1.25 \pm 80.21$   &  1.95 & $1.18 \pm 252.0$ \\
Present - Level-2, $Q_1$, AI, Level=3    &   $1.26 \pm 80.24$   &  1.95 & $1.34 \pm 233.8$ \\
Present - Level-2, $Q_1$, AI, Level=4    &   $1.28 \pm 80.19$   &  1.95 & $1.34 \pm 234.4$ \\
Present - Level-2, $Q_1$, AI, Level=5    &   $1.26 \pm 80.19$   &  1.95 & $1.21 \pm 234.2$ \\
\hline
Present - Level-3, $Q_1$, ST             &   $1.26 \pm 80.71$   &  1.95 & $1.48 \pm 229.6$ \\
Present - Level-3, $Q_1$, AI, Level=2    &   $1.27 \pm 80.74$   &  1.95 & $0.95 \pm 238.9$ \\
Present - Level-3, $Q_1$, AI, Level=3    &   $1.27 \pm 80.71$   &  1.95 & $2.07 \pm 228.3$ \\
Present - Level-3, $Q_1$, AI, Level=4    &   $1.25 \pm 80.78$   &  1.95 & $1.89 \pm 228.6$ \\
Present - Level-3, $Q_1$, AI, Level=5    &   $1.25 \pm 80.73$   &  1.95 & $1.83 \pm 228.9$ \\
\hline
Present - Level-2, $Q_2$, ST             &   $1.24 \pm 81.54$   &  1.96 & $1.75 \pm 234.8$ \\
Present - Level-2, $Q_2$, AI, Level=2    &   $1.24 \pm 81.77$   &  1.97 & $3.33 \pm 245.2$ \\
Present - Level-2, $Q_2$, AI, Level=3    &   $1.23 \pm 81.52$   &  1.95 & $1.81 \pm 235.3$ \\
Present - Level-2, $Q_2$, AI, Level=4    &   $1.24 \pm 81.49$   &  1.95 & $2.19 \pm 235.1$ \\
Present - Level-2, $Q_2$, AI, Level=5    &   $1.22 \pm 81.52$   &  1.95 & $1.38 \pm 235.0$ \\
\hline
\end{tabular}
\caption{FSI2 benchmark in 2D: summary of vertical displacement of point A ($d^s_y$), total lift force ($F^s_y$) and frequency of oscillations ($f_o$).}
 \label{table-Turekbeam}
\end{table}
\renewcommand{\arraystretch}{1.0}

\subsection{Flow past flexible plates in 3D}
\begin{figure}[H]
 \begin{center}
  \includegraphics[trim = 0mm 0mm 0mm 0mm, clip,scale=0.8]{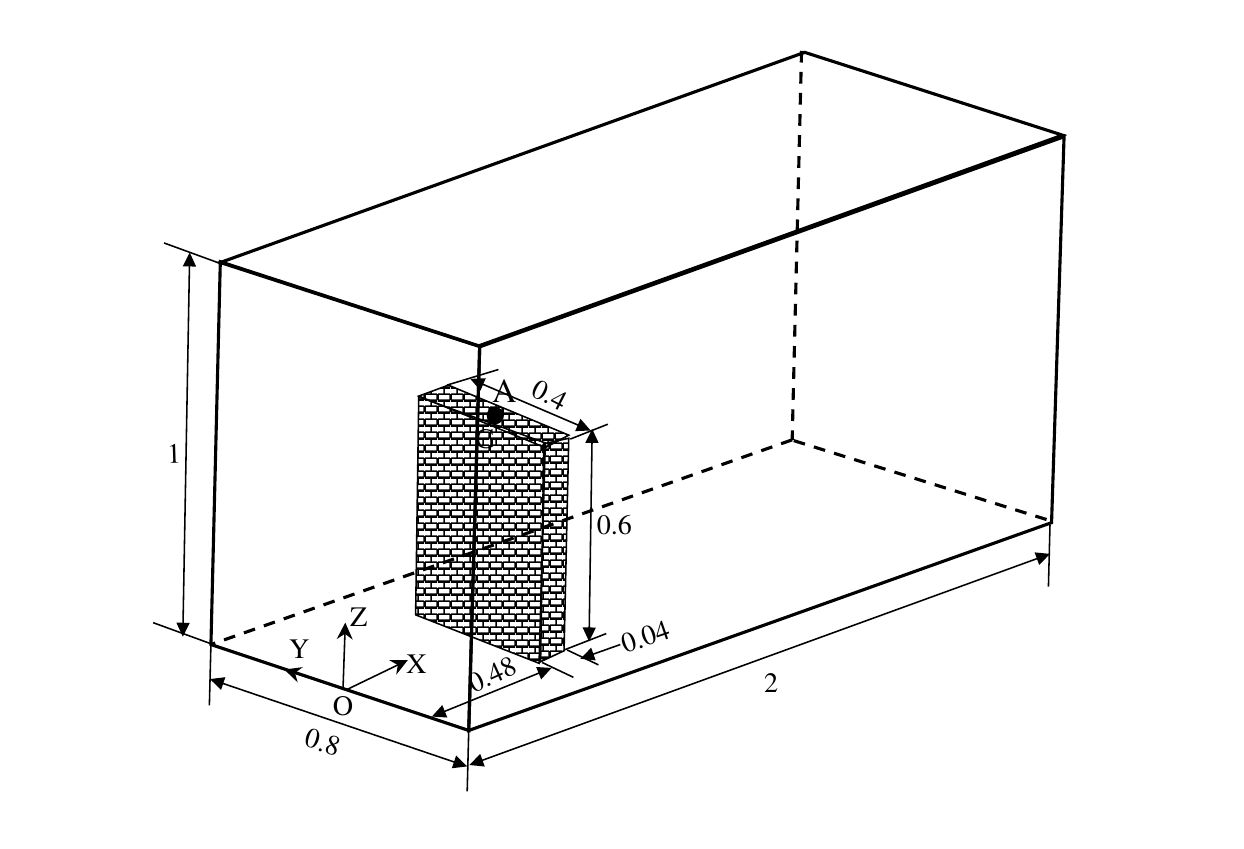}
  \caption{Flexible plate in cross-flow: evolution of horizontal displacement at point XX for different levels adaptive integrations obtained with $Q_1$ b-splines.}
  \label{fig-thinplate-geom}
 \end{center}
\end{figure}

\begin{figure}[H]
 \begin{center}
  \includegraphics[trim = 0mm 0mm 0mm 0mm, clip,scale=0.8]{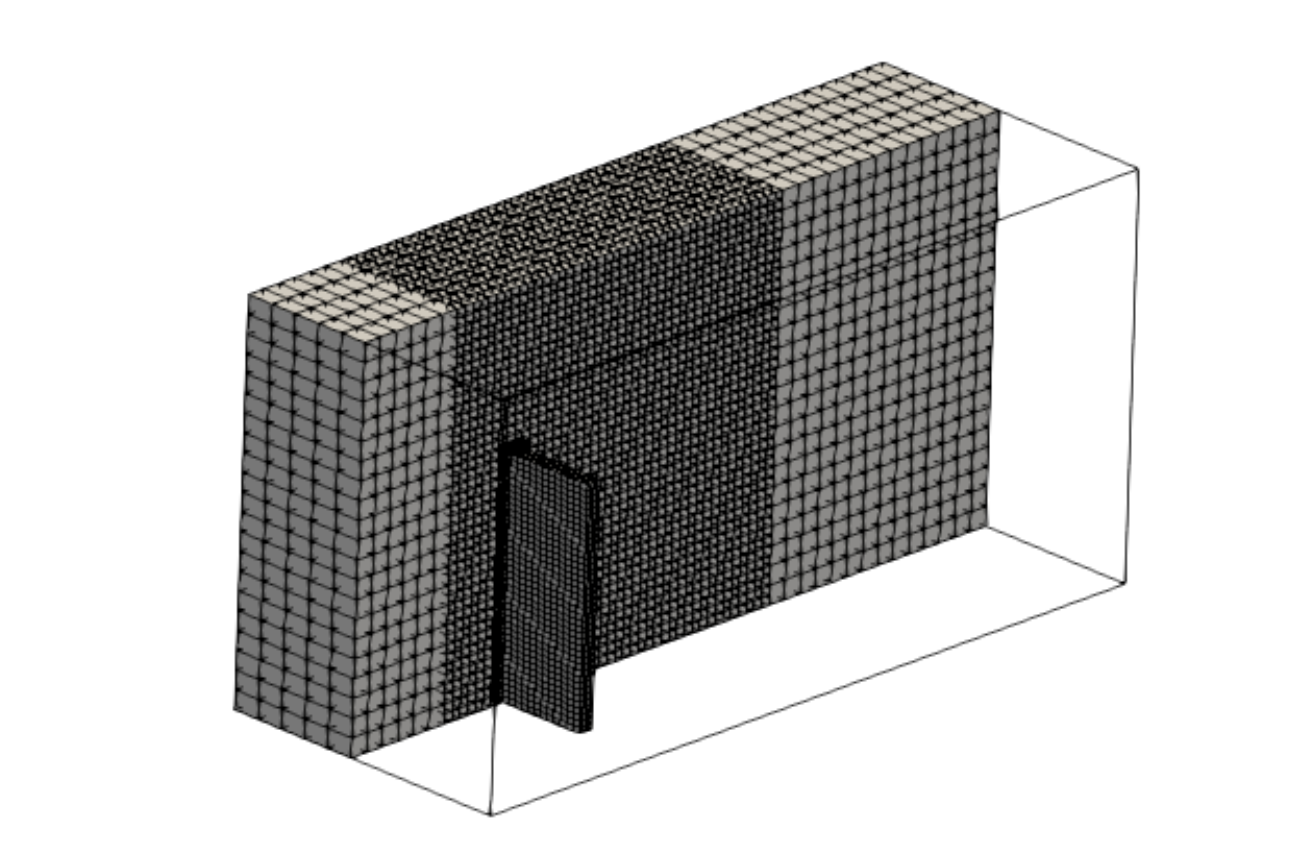}
  \caption{Flexible plate in cross-flow: meshes used for the fluid and solid domains.}
 \label{fig-thinplate-mesh}
 \end{center}
\end{figure}

In the last example, we study the dynamic fluid-structure interaction of thin, flexible plates in cross-flow. We first consider a single plate to assess the effect of adaptive integration on the forces and displacements. The setup of the problem is as illustrated in Fig. \ref{fig-thinplate-geom}, and the finite element mesh used for the analysis is shown in Fig. \ref{fig-thinplate-mesh}. The density and viscosity of the fluid are ${\rho^f} = {1} $ and ${\mu^f} = 0.01$, respectively. The material properties of the solid are: density, $\rho^s = 1.0$, Young’s modulus, $E=2000$, and Poisson’s ratio, $\nu^s  = 0.3$. A doubly-parabolic velocity profile with a maximum value of $U_m=2.25$ (and an average of $\bar{U}=1.0$) is imposed in the X-direction. Based on the length of the plate, the Reynolds number is $Re=60$.

\begin{figure}[H]
 \begin{center}
  \includegraphics[trim = 0mm 0mm 0mm 0mm, clip,scale=0.5]{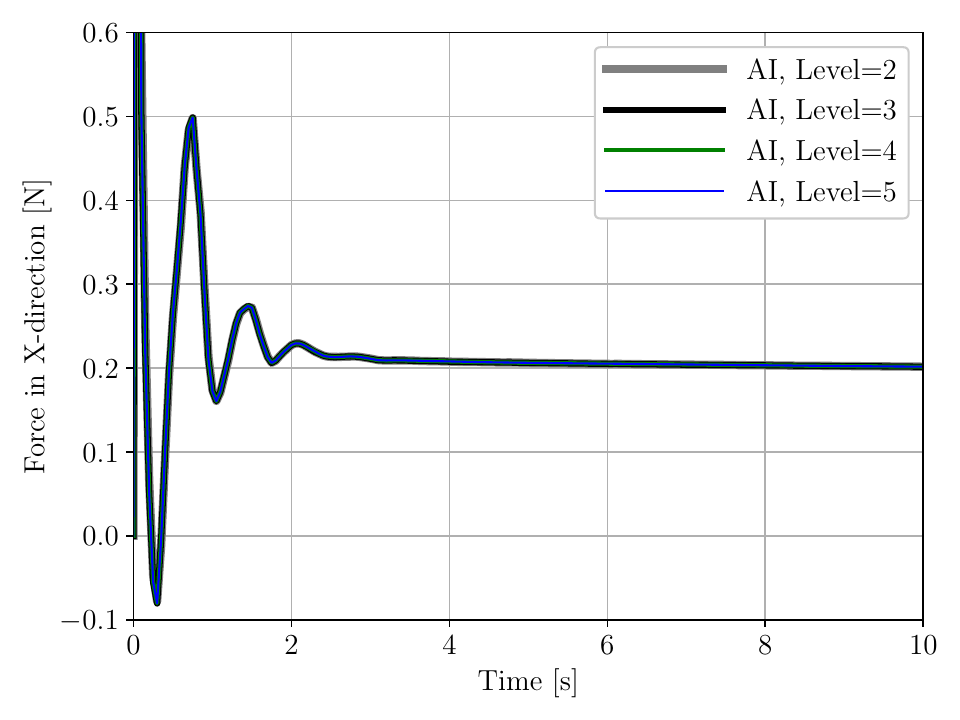}
  \includegraphics[trim = 0mm 0mm 0mm 0mm, clip,scale=0.5]{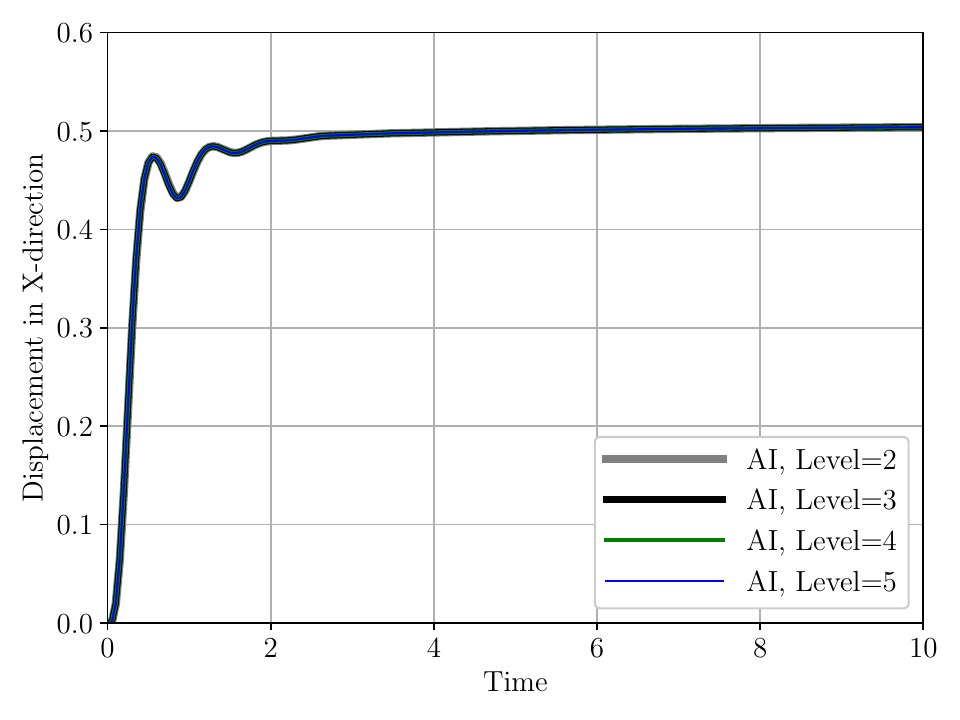}
  \caption{Flexible plate in cross-flow: evolution of horizontal displacement at point A for different levels adaptive integration.}
  \label{fig-thinplate-graphs}
 \end{center}
\end{figure}

\begin{figure}[H]
 \begin{center}
  \includegraphics[trim = 0mm 0mm 0mm 0mm, clip,scale=0.5]{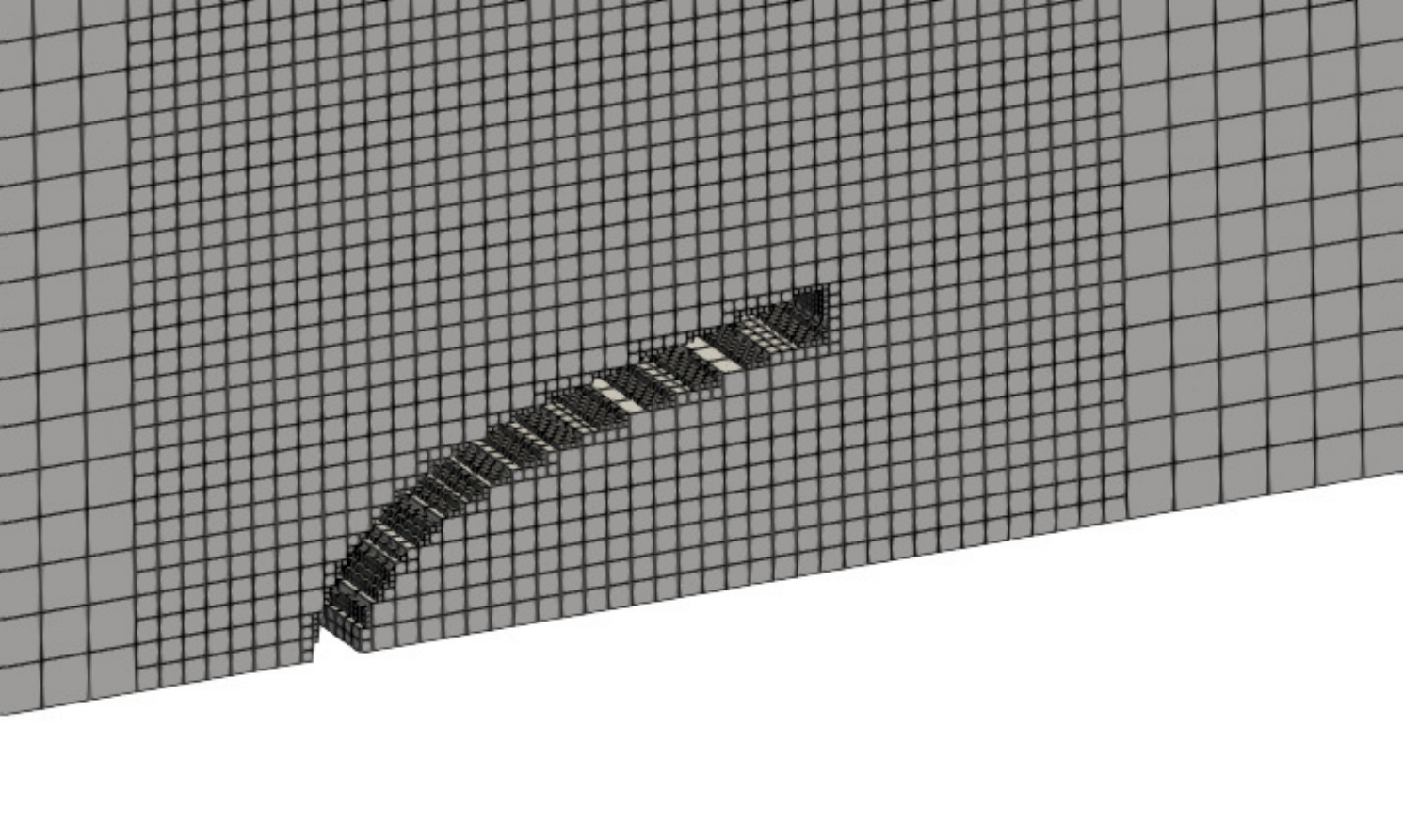}
  \caption{Flexible plate in cross-flow: cut view of the fluid mesh along with the sub-cells used for adaptive integration at $t=10$.}
  \label{fig-thinplate-defshapes}
 \end{center}
\end{figure}

Simulations are performed with $Q_1$ b-splines with different levels of adaptive integration using a constant time step $\Delta t=0.05$. Due to the presence of significant added-mass in this example, a first-order force predictor ($F^{s^P}_{n+1}=F^s_n$) with relaxation parameter $\beta=0.02$ is used. The evolution of the X-component of the total force on the plate and the horizontal displacement of point A (see Fig. \ref{fig-thinplate-geom}) is shown in Fig.\ref{fig-thinplate-graphs}. We can observe that there is a negligible difference in the force and displacement values obtained with different levels of adaptive integration. Moreover, the force response is free from spurious oscillations. The fluid mesh along with the sub-cells used for adaptive integration at the middle plane of the domain along the Y-axis at the final time instance is shown in Fig. \ref{fig-thinplate-defshapes}.

\begin{figure}[H]
 \begin{center}
  \subfloat[$t=0.30$]{\includegraphics[trim = 0mm 0mm 0mm 0mm, clip,scale=0.6]{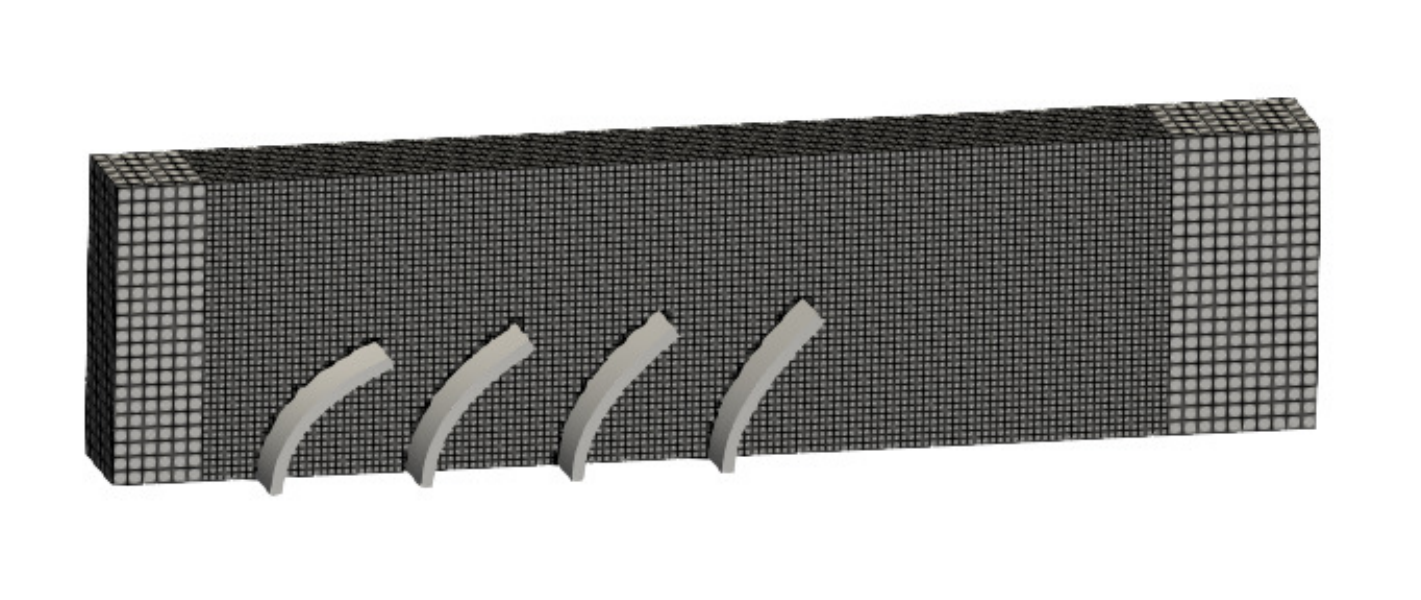}} \\
  \subfloat[$t=10.0$]{\includegraphics[trim = 0mm 0mm 0mm 0mm, clip,scale=0.6]{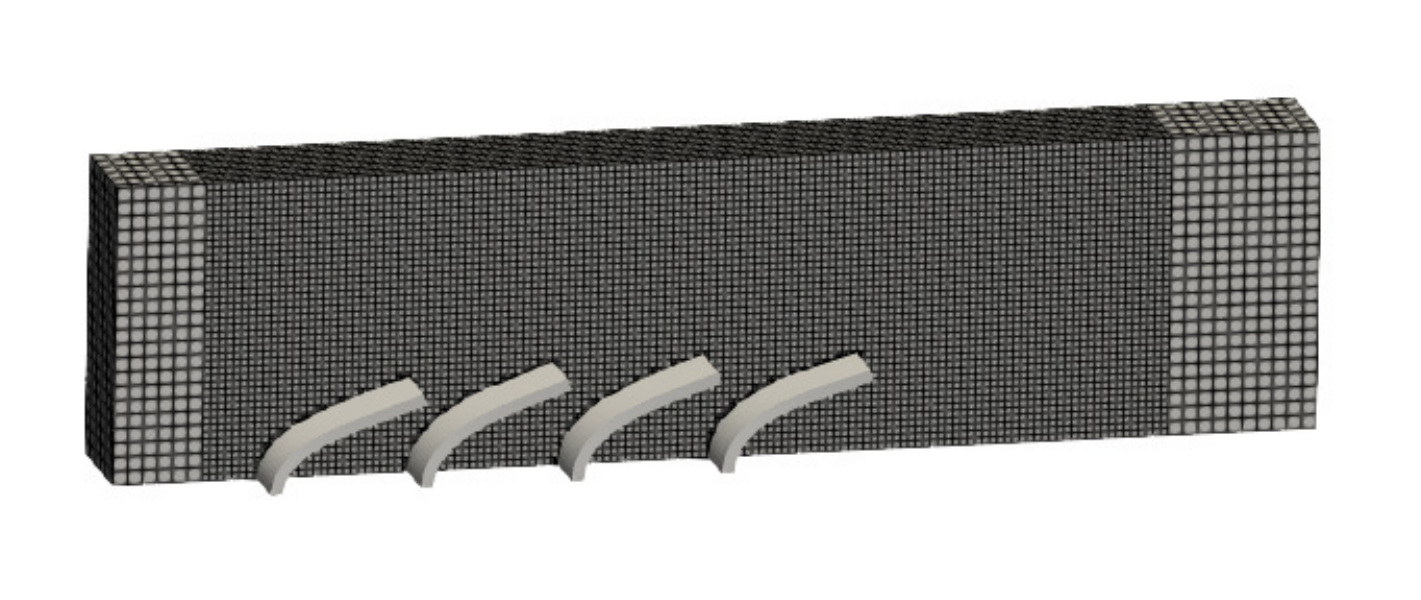}}
  \caption{Flexible plates in cross-flow: cut view of the fluid mesh along with the deformed shapes of the plates at time instants $t=0.3$ and $t=10.0$.}
  \label{fig-multiplates-defshapes}
 \end{center}
\end{figure}

\begin{figure}[H]
 \begin{center}
  \subfloat[$t=0.30$]{\includegraphics[trim = 0mm 0mm 0mm 0mm, clip,scale=0.6]{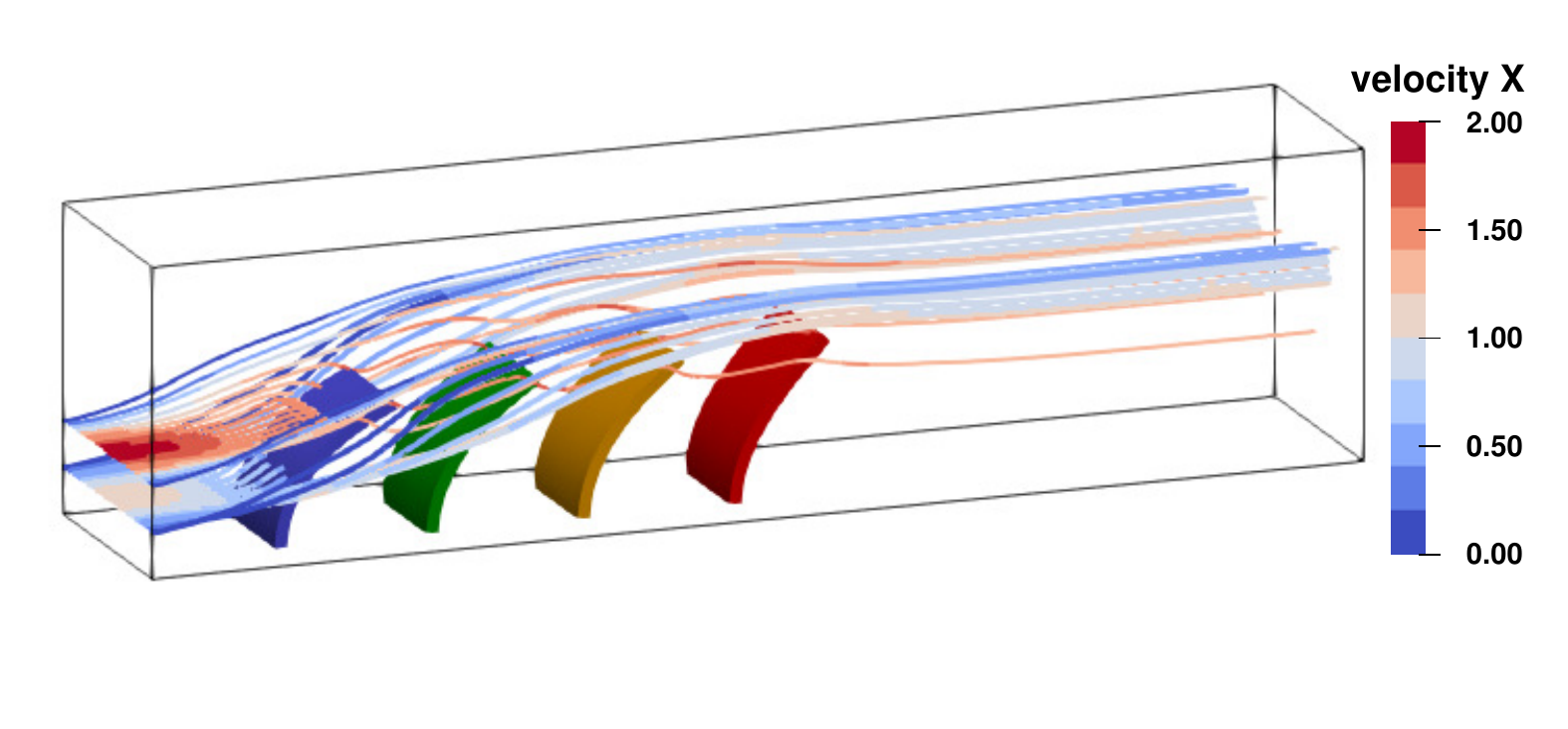}} \\
  \subfloat[$t=10.0$]{\includegraphics[trim = 0mm 0mm 0mm 0mm, clip,scale=0.6]{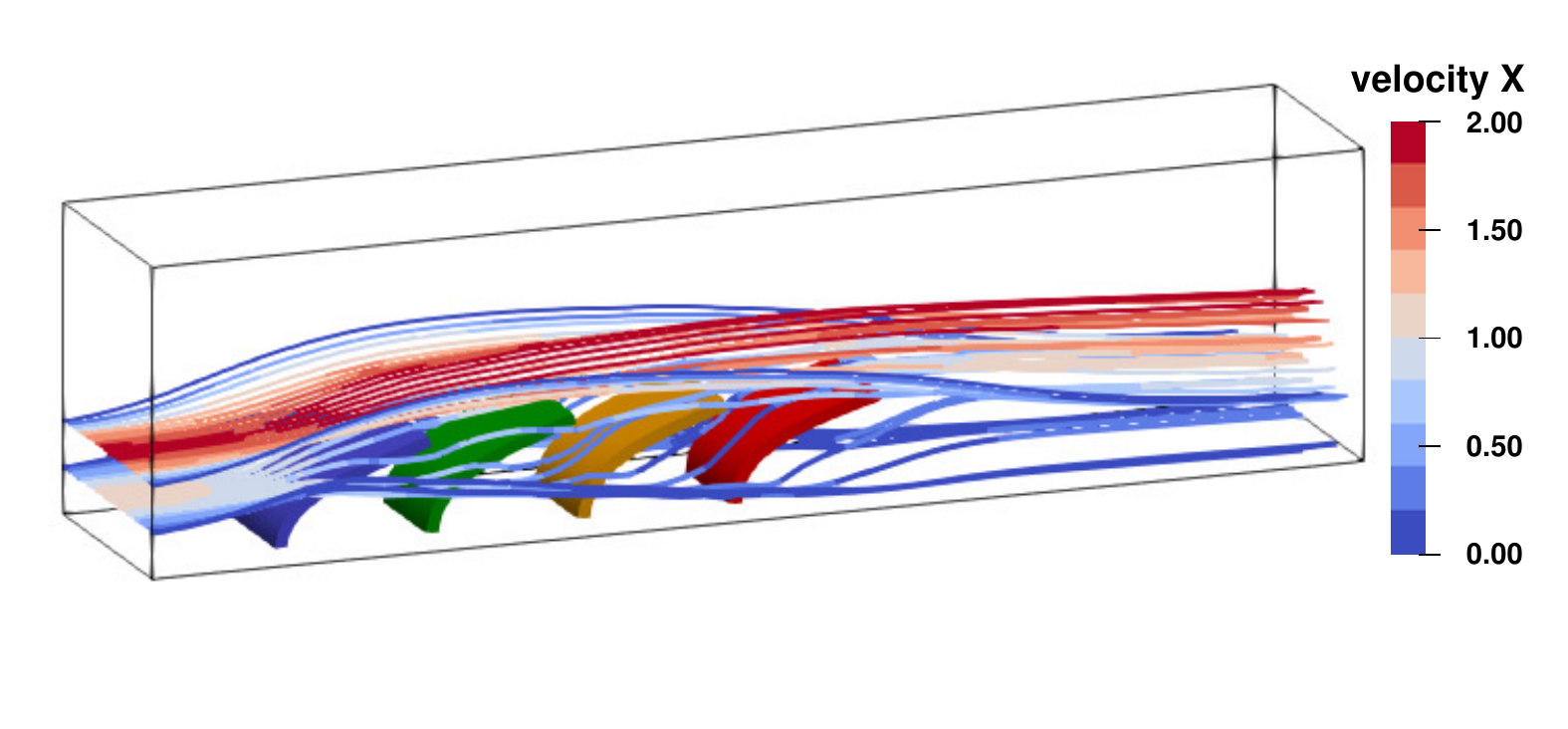}}
  \caption{Flexible plates in cross-flow: streamlines overlaid with velocity contours along with the deformed shapes of the plates at time instants $t=0.3$ and $t=10.0$.}
  \label{fig-multiplates-contours}
 \end{center}
\end{figure}

Figures \ref{fig-multiplates-defshapes} and \ref{fig-multiplates-contours} show deformed shapes and streamlines at two different time instances for the problem of four equally-spaced thin plates in an extended domain. Such simulations would have been quite challenging to perform using subtetrahedralisation. This example illustrates the clear advantage of using the adaptive integration technique for the integration of cut-cells. In accordance with previous examples, three levels of adaptive integration are sufficient to obtain forces and displacements of acceptable accuracy.

\section{Summary and conclusions} \label{section-conclusion}
In this paper, we comprehensively assessed the performance of adaptive integration of cut-cells in the context of laminar fluid-structure interaction problems in two and three dimensions. The present work offers some important insights into the performance of the adaptive integration technique for cut-cells for laminar FSI problems. First, the effect of adaptive integration on the convergence rates in velocity and pressure field is studied using the example of Kovasznay flow. It is demonstrated that excessive levels of adaptive integration are required to recover optimal convergence rates. Later, using the examples of unsteady flow past a fixed square in 2D, unsteady flow past a fixed cylinder in 3D, vortex-induced vibrations of a flexible plate behind a fixed cylinder in 2D and flexible plates in cross-flow in 3D, the accuracy of results obtained with different levels of the adaptive integration technique in computing forces and displacements for fluid-structure interaction problems are assessed.

The numerical results demonstrate that although excessive levels of adaptive integration are required for recovering optimal convergence rates, fewer levels are sufficient enough to obtain numerical results of comparable accuracy for FSI problems. While it certainly is possible to devise efficient alternatives to the standard quadtree/octree based recursive subdivision used in this work, for example, smart octrees \cite{KudelaCMAME2016}, present work shows that such sophisticated implementations are not necessary for simulating force and displacement response in laminar FSI problems. We conclude that when we take computational cost and accuracy of results into account, three levels of adaptive integration is an optimal choice for integrating cut-cells for laminar FSI problems.

\section*{Acknowledgement}
C Kadapa acknowledges the support of the Supercomputing Wales project, which is part-funded by
the European Regional Development Fund (ERDF) via the Welsh Government. Y Mei acknowledges the support of the National Natural Science Foundation of China (12002075) and the Fundamental Research Funds for the Central Universities (Grant No.DUT19RC(3)017).

\section*{References}

\end{document}